%Paper: hep-th/9312147
%From: peeters@orthe.physics.sunysb.edu (Bas Peeters)
%Date: Thu, 16 Dec 1993 11:59:04 -0500 (EST)

% File jytex.tex, for jyTeX version 2.6M (June 1992)
% Copyright (c) 1991, 1992 by Jonathan P. Yamron

\catcode`\@=11

%******************************************************************************

\message{Loading jyTeX fonts...}

%************************************************************
%*
%*             Available fonts
%*
%************************************************************

%************** 5-point fonts *******************************

\font\vptrm=cmr5
\font\vptmit=cmmi5
\font\vptsy=cmsy5
\font\vptbf=cmbx5

\skewchar\vptmit='177 \skewchar\vptsy='60
\fontdimen16 \vptsy=\the\fontdimen17 \vptsy

\def\vpt{\ifmmode\err@badsizechange\else
     \@mathfontinit
     \textfont0=\vptrm  \scriptfont0=\vptrm  \scriptscriptfont0=\vptrm
     \textfont1=\vptmit \scriptfont1=\vptmit \scriptscriptfont1=\vptmit
     \textfont2=\vptsy  \scriptfont2=\vptsy  \scriptscriptfont2=\vptsy
     \textfont3=\xptex  \scriptfont3=\xptex  \scriptscriptfont3=\xptex
     \textfont\bffam=\vptbf
     \scriptfont\bffam=\vptbf
     \scriptscriptfont\bffam=\vptbf
     \@fontstyleinit
     \def\rm{\vptrm\fam=\z@}%
     \def\bf{\vptbf\fam=\bffam}%
     \def\oldstyle{\vptmit\fam=\@ne}%
     \rm\fi}

%************** 6-point fonts *******************************

\font\viptrm=cmr6
\font\viptmit=cmmi6
\font\viptsy=cmsy6
\font\viptbf=cmbx6

\skewchar\viptmit='177 \skewchar\viptsy='60
\fontdimen16 \viptsy=\the\fontdimen17 \viptsy

\def\vipt{\ifmmode\err@badsizechange\else
     \@mathfontinit
     \textfont0=\viptrm  \scriptfont0=\vptrm  \scriptscriptfont0=\vptrm
     \textfont1=\viptmit \scriptfont1=\vptmit \scriptscriptfont1=\vptmit
     \textfont2=\viptsy  \scriptfont2=\vptsy  \scriptscriptfont2=\vptsy
     \textfont3=\xptex   \scriptfont3=\xptex  \scriptscriptfont3=\xptex
     \textfont\bffam=\viptbf
     \scriptfont\bffam=\vptbf
     \scriptscriptfont\bffam=\vptbf
     \@fontstyleinit
     \def\rm{\viptrm\fam=\z@}%
     \def\bf{\viptbf\fam=\bffam}%
     \def\oldstyle{\viptmit\fam=\@ne}%
     \rm\fi}

%************** 7-point fonts *******************************

\font\viiptrm=cmr7
\font\viiptmit=cmmi7
\font\viiptsy=cmsy7
\font\viiptit=cmti7
\font\viiptbf=cmbx7

\skewchar\viiptmit='177 \skewchar\viiptsy='60
\fontdimen16 \viiptsy=\the\fontdimen17 \viiptsy

\def\viipt{\ifmmode\err@badsizechange\else
     \@mathfontinit
     \textfont0=\viiptrm  \scriptfont0=\vptrm  \scriptscriptfont0=\vptrm
     \textfont1=\viiptmit \scriptfont1=\vptmit \scriptscriptfont1=\vptmit
     \textfont2=\viiptsy  \scriptfont2=\vptsy  \scriptscriptfont2=\vptsy
     \textfont3=\xptex    \scriptfont3=\xptex  \scriptscriptfont3=\xptex
     \textfont\itfam=\viiptit
     \scriptfont\itfam=\viiptit
     \scriptscriptfont\itfam=\viiptit
     \textfont\bffam=\viiptbf
     \scriptfont\bffam=\vptbf
     \scriptscriptfont\bffam=\vptbf
     \@fontstyleinit
     \def\rm{\viiptrm\fam=\z@}%
     \def\it{\viiptit\fam=\itfam}%
     \def\bf{\viiptbf\fam=\bffam}%
     \def\oldstyle{\viiptmit\fam=\@ne}%
     \rm\fi}

%************** 8-point fonts *******************************

\font\viiiptrm=cmr8
\font\viiiptmit=cmmi8
\font\viiiptsy=cmsy8
\font\viiiptit=cmti8
%\font\viiiptsl=cmsl8
\font\viiiptbf=cmbx8
%\font\viiipttt=cmtt8
%\font\viiiptss=cmss8

\skewchar\viiiptmit='177 \skewchar\viiiptsy='60
\fontdimen16 \viiiptsy=\the\fontdimen17 \viiiptsy

\def\viiipt{\ifmmode\err@badsizechange\else
     \@mathfontinit
     \textfont0=\viiiptrm  \scriptfont0=\viptrm  \scriptscriptfont0=\vptrm
     \textfont1=\viiiptmit \scriptfont1=\viptmit \scriptscriptfont1=\vptmit
     \textfont2=\viiiptsy  \scriptfont2=\viptsy  \scriptscriptfont2=\vptsy
     \textfont3=\xptex     \scriptfont3=\xptex   \scriptscriptfont3=\xptex
     \textfont\itfam=\viiiptit
     \scriptfont\itfam=\viiptit
     \scriptscriptfont\itfam=\viiptit
     \textfont\bffam=\viiiptbf
     \scriptfont\bffam=\viptbf
     \scriptscriptfont\bffam=\vptbf
     \@fontstyleinit
     \def\rm{\viiiptrm\fam=\z@}%
     \def\it{\viiiptit\fam=\itfam}%
     \def\bf{\viiiptbf\fam=\bffam}%
     \def\oldstyle{\viiiptmit\fam=\@ne}%
     \rm\fi}

%************** Optional 9-point fonts **********************

\def\getixpt{%
     \font\ixptrm=cmr9
     \font\ixptmit=cmmi9
     \font\ixptsy=cmsy9
     \font\ixptit=cmti9
%     \font\ixptsl=cmsl9
     \font\ixptbf=cmbx9
%     \font\ixpttt=cmtt9
%     \font\ixptss=cmss9
     \skewchar\ixptmit='177 \skewchar\ixptsy='60
     \fontdimen16 \ixptsy=\the\fontdimen17 \ixptsy}

\def\ixpt{\ifmmode\err@badsizechange\else
     \@mathfontinit
     \textfont0=\ixptrm  \scriptfont0=\viiptrm  \scriptscriptfont0=\vptrm
     \textfont1=\ixptmit \scriptfont1=\viiptmit \scriptscriptfont1=\vptmit
     \textfont2=\ixptsy  \scriptfont2=\viiptsy  \scriptscriptfont2=\vptsy
     \textfont3=\xptex   \scriptfont3=\xptex    \scriptscriptfont3=\xptex
     \textfont\itfam=\ixptit
     \scriptfont\itfam=\viiptit
     \scriptscriptfont\itfam=\viiptit
     \textfont\bffam=\ixptbf
     \scriptfont\bffam=\viiptbf
     \scriptscriptfont\bffam=\vptbf
     \@fontstyleinit
     \def\rm{\ixptrm\fam=\z@}%
     \def\it{\ixptit\fam=\itfam}%
     \def\bf{\ixptbf\fam=\bffam}%
     \def\oldstyle{\ixptmit\fam=\@ne}%
     \rm\fi}

%************** 10-point fonts ******************************

\font\xptrm=cmr10
\font\xptmit=cmmi10
\font\xptsy=cmsy10
\font\xptex=cmex10
\font\xptit=cmti10
\font\xptsl=cmsl10
\font\xptbf=cmbx10
\font\xpttt=cmtt10
\font\xptss=cmss10
\font\xptsc=cmcsc10
\font\xptbfs=cmbx10
\font\xptbmit=cmmib10

\skewchar\xptmit='177 \skewchar\xptbmit='177 \skewchar\xptsy='60
\fontdimen16 \xptsy=\the\fontdimen17 \xptsy

\def\xpt{\ifmmode\err@badsizechange\else
     \@mathfontinit
     \textfont0=\xptrm  \scriptfont0=\viiptrm  \scriptscriptfont0=\vptrm
     \textfont1=\xptmit \scriptfont1=\viiptmit \scriptscriptfont1=\vptmit
     \textfont2=\xptsy  \scriptfont2=\viiptsy  \scriptscriptfont2=\vptsy
     \textfont3=\xptex  \scriptfont3=\xptex    \scriptscriptfont3=\xptex
     \textfont\itfam=\xptit
     \scriptfont\itfam=\viiptit
     \scriptscriptfont\itfam=\viiptit
     \textfont\bffam=\xptbf
     \scriptfont\bffam=\viiptbf
     \scriptscriptfont\bffam=\vptbf
     \textfont\bfsfam=\xptbfs
     \scriptfont\bfsfam=\viiptbf
     \scriptscriptfont\bfsfam=\vptbf
     \textfont\bmitfam=\xptbmit
     \scriptfont\bmitfam=\viiptmit
     \scriptscriptfont\bmitfam=\vptmit
     \@fontstyleinit
     \def\rm{\xptrm\fam=\z@}%
     \def\it{\xptit\fam=\itfam}%
     \def\sl{\xptsl}%
     \def\bf{\xptbf\fam=\bffam}%
     \def\tt{\xpttt}%
     \def\ss{\xptss}%
     \def\sc{\xptsc}%
     \def\bfs{\xptbfs\fam=\bfsfam}%
     \def\bmit{\fam=\bmitfam}%
     \def\oldstyle{\xptmit\fam=\@ne}%
     \rm\fi}

%************** Optional 11-point fonts *********************

\def\getxipt{%
     \font\xiptrm=cmr10  scaled\magstephalf
     \font\xiptmit=cmmi10 scaled\magstephalf
     \font\xiptsy=cmsy10 scaled\magstephalf
     \font\xiptex=cmex10 scaled\magstephalf
     \font\xiptit=cmti10 scaled\magstephalf
     \font\xiptsl=cmsl10 scaled\magstephalf
     \font\xiptbf=cmbx10 scaled\magstephalf
     \font\xipttt=cmtt10 scaled\magstephalf
     \font\xiptss=cmss10 scaled\magstephalf
     \skewchar\xiptmit='177 \skewchar\xiptsy='60
     \fontdimen16 \xiptsy=\the\fontdimen17 \xiptsy}

\def\xipt{\ifmmode\err@badsizechange\else
     \@mathfontinit
     \textfont0=\xiptrm  \scriptfont0=\viiiptrm  \scriptscriptfont0=\viptrm
     \textfont1=\xiptmit \scriptfont1=\viiiptmit \scriptscriptfont1=\viptmit
     \textfont2=\xiptsy  \scriptfont2=\viiiptsy  \scriptscriptfont2=\viptsy
     \textfont3=\xiptex  \scriptfont3=\xptex     \scriptscriptfont3=\xptex
     \textfont\itfam=\xiptit
     \scriptfont\itfam=\viiiptit
     \scriptscriptfont\itfam=\viiptit
     \textfont\bffam=\xiptbf
     \scriptfont\bffam=\viiiptbf
     \scriptscriptfont\bffam=\viptbf
     \@fontstyleinit
     \def\rm{\xiptrm\fam=\z@}%
     \def\it{\xiptit\fam=\itfam}%
     \def\sl{\xiptsl}%
     \def\bf{\xiptbf\fam=\bffam}%
     \def\tt{\xipttt}%
     \def\ss{\xiptss}%
     \def\oldstyle{\xiptmit\fam=\@ne}%
     \rm\fi}

%************** 12-point fonts ******************************

\font\xiiptrm=cmr12
\font\xiiptmit=cmmi12
\font\xiiptsy=cmsy10  scaled\magstep1
\font\xiiptex=cmex10  scaled\magstep1
\font\xiiptit=cmti12
\font\xiiptsl=cmsl12
\font\xiiptbf=cmbx12
\font\xiipttt=cmtt12
\font\xiiptss=cmss12
\font\xiiptsc=cmcsc10 scaled\magstep1
\font\xiiptbfs=cmbx10  scaled\magstep1
\font\xiiptbmit=cmmib10 scaled\magstep1

\skewchar\xiiptmit='177 \skewchar\xiiptbmit='177 \skewchar\xiiptsy='60
\fontdimen16 \xiiptsy=\the\fontdimen17 \xiiptsy

\def\xiipt{\ifmmode\err@badsizechange\else
     \@mathfontinit
     \textfont0=\xiiptrm  \scriptfont0=\viiiptrm  \scriptscriptfont0=\viptrm
     \textfont1=\xiiptmit \scriptfont1=\viiiptmit \scriptscriptfont1=\viptmit
     \textfont2=\xiiptsy  \scriptfont2=\viiiptsy  \scriptscriptfont2=\viptsy
     \textfont3=\xiiptex  \scriptfont3=\xptex     \scriptscriptfont3=\xptex
     \textfont\itfam=\xiiptit
     \scriptfont\itfam=\viiiptit
     \scriptscriptfont\itfam=\viiptit
     \textfont\bffam=\xiiptbf
     \scriptfont\bffam=\viiiptbf
     \scriptscriptfont\bffam=\viptbf
     \textfont\bfsfam=\xiiptbfs
     \scriptfont\bfsfam=\viiiptbf
     \scriptscriptfont\bfsfam=\viptbf
     \textfont\bmitfam=\xiiptbmit
     \scriptfont\bmitfam=\viiiptmit
     \scriptscriptfont\bmitfam=\viptmit
     \@fontstyleinit
     \def\rm{\xiiptrm\fam=\z@}%
     \def\it{\xiiptit\fam=\itfam}%
     \def\sl{\xiiptsl}%
     \def\bf{\xiiptbf\fam=\bffam}%
     \def\tt{\xiipttt}%
     \def\ss{\xiiptss}%
     \def\sc{\xiiptsc}%
     \def\bfs{\xiiptbfs\fam=\bfsfam}%
     \def\bmit{\fam=\bmitfam}%
     \def\oldstyle{\xiiptmit\fam=\@ne}%
     \rm\fi}

%************** Optional 13-point fonts *********************

\def\getxiiipt{%
     \font\xiiiptrm=cmr12  scaled\magstephalf
     \font\xiiiptmit=cmmi12 scaled\magstephalf
     \font\xiiiptsy=cmsy9  scaled\magstep2
     \font\xiiiptit=cmti12 scaled\magstephalf
     \font\xiiiptsl=cmsl12 scaled\magstephalf
     \font\xiiiptbf=cmbx12 scaled\magstephalf
     \font\xiiipttt=cmtt12 scaled\magstephalf
     \font\xiiiptss=cmss12 scaled\magstephalf
     \skewchar\xiiiptmit='177 \skewchar\xiiiptsy='60
     \fontdimen16 \xiiiptsy=\the\fontdimen17 \xiiiptsy}

\def\xiiipt{\ifmmode\err@badsizechange\else
     \@mathfontinit
     \textfont0=\xiiiptrm  \scriptfont0=\xptrm  \scriptscriptfont0=\viiptrm
     \textfont1=\xiiiptmit \scriptfont1=\xptmit \scriptscriptfont1=\viiptmit
     \textfont2=\xiiiptsy  \scriptfont2=\xptsy  \scriptscriptfont2=\viiptsy
     \textfont3=\xivptex   \scriptfont3=\xptex  \scriptscriptfont3=\xptex
     \textfont\itfam=\xiiiptit
     \scriptfont\itfam=\xptit
     \scriptscriptfont\itfam=\viiptit
     \textfont\bffam=\xiiiptbf
     \scriptfont\bffam=\xptbf
     \scriptscriptfont\bffam=\viiptbf
     \@fontstyleinit
     \def\rm{\xiiiptrm\fam=\z@}%
     \def\it{\xiiiptit\fam=\itfam}%
     \def\sl{\xiiiptsl}%
     \def\bf{\xiiiptbf\fam=\bffam}%
     \def\tt{\xiiipttt}%
     \def\ss{\xiiiptss}%
     \def\oldstyle{\xiiiptmit\fam=\@ne}%
     \rm\fi}

%************** 14-point fonts ******************************

\font\xivptrm=cmr12   scaled\magstep1
\font\xivptmit=cmmi12  scaled\magstep1
\font\xivptsy=cmsy10  scaled\magstep2
\font\xivptex=cmex10  scaled\magstep2
\font\xivptit=cmti12  scaled\magstep1
\font\xivptsl=cmsl12  scaled\magstep1
\font\xivptbf=cmbx12  scaled\magstep1
\font\xivpttt=cmtt12  scaled\magstep1
\font\xivptss=cmss12  scaled\magstep1
\font\xivptsc=cmcsc10 scaled\magstep2
\font\xivptbfs=cmbx10  scaled\magstep2
\font\xivptbmit=cmmib10 scaled\magstep2

\skewchar\xivptmit='177 \skewchar\xivptbmit='177 \skewchar\xivptsy='60
\fontdimen16 \xivptsy=\the\fontdimen17 \xivptsy

\def\xivpt{\ifmmode\err@badsizechange\else
     \@mathfontinit
     \textfont0=\xivptrm  \scriptfont0=\xptrm  \scriptscriptfont0=\viiptrm
     \textfont1=\xivptmit \scriptfont1=\xptmit \scriptscriptfont1=\viiptmit
     \textfont2=\xivptsy  \scriptfont2=\xptsy  \scriptscriptfont2=\viiptsy
     \textfont3=\xivptex  \scriptfont3=\xptex  \scriptscriptfont3=\xptex
     \textfont\itfam=\xivptit
     \scriptfont\itfam=\xptit
     \scriptscriptfont\itfam=\viiptit
     \textfont\bffam=\xivptbf
     \scriptfont\bffam=\xptbf
     \scriptscriptfont\bffam=\viiptbf
     \textfont\bfsfam=\xivptbfs
     \scriptfont\bfsfam=\xptbfs
     \scriptscriptfont\bfsfam=\viiptbf
     \textfont\bmitfam=\xivptbmit
     \scriptfont\bmitfam=\xptbmit
     \scriptscriptfont\bmitfam=\viiptmit
     \@fontstyleinit
     \def\rm{\xivptrm\fam=\z@}%
     \def\it{\xivptit\fam=\itfam}%
     \def\sl{\xivptsl}%
     \def\bf{\xivptbf\fam=\bffam}%
     \def\tt{\xivpttt}%
     \def\ss{\xivptss}%
     \def\sc{\xivptsc}%
     \def\bfs{\xivptbfs\fam=\bfsfam}%
     \def\bmit{\fam=\bmitfam}%
     \def\oldstyle{\xivptmit\fam=\@ne}%
     \rm\fi}

%************** 17-point fonts ******************************

\font\xviiptrm=cmr17
\font\xviiptmit=cmmi12 scaled\magstep2
\font\xviiptsy=cmsy10 scaled\magstep3
\font\xviiptex=cmex10 scaled\magstep3
\font\xviiptit=cmti12 scaled\magstep2
\font\xviiptbf=cmbx12 scaled\magstep2
\font\xviiptbfs=cmbx10 scaled\magstep3

\skewchar\xviiptmit='177 \skewchar\xviiptsy='60
\fontdimen16 \xviiptsy=\the\fontdimen17 \xviiptsy

\def\xviipt{\ifmmode\err@badsizechange\else
     \@mathfontinit
     \textfont0=\xviiptrm  \scriptfont0=\xiiptrm  \scriptscriptfont0=\viiiptrm
     \textfont1=\xviiptmit \scriptfont1=\xiiptmit \scriptscriptfont1=\viiiptmit
     \textfont2=\xviiptsy  \scriptfont2=\xiiptsy  \scriptscriptfont2=\viiiptsy
     \textfont3=\xviiptex  \scriptfont3=\xiiptex  \scriptscriptfont3=\xptex
     \textfont\itfam=\xviiptit
     \scriptfont\itfam=\xiiptit
     \scriptscriptfont\itfam=\viiiptit
     \textfont\bffam=\xviiptbf
     \scriptfont\bffam=\xiiptbf
     \scriptscriptfont\bffam=\viiiptbf
     \textfont\bfsfam=\xviiptbfs
     \scriptfont\bfsfam=\xiiptbfs
     \scriptscriptfont\bfsfam=\viiiptbf
     \@fontstyleinit
     \def\rm{\xviiptrm\fam=\z@}%
     \def\it{\xviiptit\fam=\itfam}%
     \def\bf{\xviiptbf\fam=\bffam}%
     \def\bfs{\xviiptbfs\fam=\bfsfam}%
     \def\oldstyle{\xviiptmit\fam=\@ne}%
     \rm\fi}

%************** 21-point fonts ******************************

\font\xxiptrm=cmr17  scaled\magstep1
%\font\xxiptmit=cmmi12 scaled\magstep3
%\font\xxiptsy=cmsy10 scaled\magstep4
%\font\xxiptex=cmex10 scaled\magstep4
%\font\xxiptbf=cmbx12 scaled\magstep3

%\skewchar\xxiptmit='177 \skewchar\xxiptsy='60
%\fontdimen16 \xxiptsy=\the\fontdimen17 \xxiptsy

\def\xxipt{\ifmmode\err@badsizechange\else
     \@mathfontinit
%     \textfont0=\xxiptrm  \scriptfont0=\xivptrm  \scriptscriptfont0=\xptrm
%     \textfont1=\xxiptmit \scriptfont1=\xivptmit \scriptscriptfont1=\xptmit
%     \textfont2=\xxiptsy  \scriptfont2=\xivptsy  \scriptscriptfont2=\xptsy
%     \textfont3=\xxiptex  \scriptfont3=\xivptex  \scriptscriptfont3=\xptex
%     \textfont\bffam=\xxiptbf
%     \scriptfont\bffam=\xivptbf
%     \scriptscriptfont\bffam=\xptbf
     \@fontstyleinit
     \def\rm{\xxiptrm\fam=\z@}%
     \rm\fi}

%************** 25-point fonts ******************************

\font\xxvptrm=cmr17  scaled\magstep2
%\font\xxvptmit=cmmi12 scaled\magstep4
%\font\xxvptsy=cmsy10 scaled\magstep5
%\font\xxvptex=cmex10 scaled\magstep5
%\font\xxvptbf=cmbx12 scaled\magstep4

%\skewchar\xxvptmit='177 \skewchar\xxvptsy='60
%\fontdimen16 \xxvptsy=\the\fontdimen17 \xxvptsy

\def\xxvpt{\ifmmode\err@badsizechange\else
     \@mathfontinit
%     \textfont0=\xxvptrm  \scriptfont0=\xviiptrm  \scriptscriptfont0=\xiiptrm
%     \textfont1=\xxvptmit \scriptfont1=\xviiptmit \scriptscriptfont1=\xiiptmit
%     \textfont2=\xxvptsy  \scriptfont2=\xviiptsy  \scriptscriptfont2=\xiiptsy
%     \textfont3=\xxvptex  \scriptfont3=\xviiptex  \scriptscriptfont3=\xiiptex
%     \textfont\bffam=\xxvptbf
%     \scriptfont\bffam=\xviiptbf
%     \scriptscriptfont\bffam=\xiiptbf
     \@fontstyleinit
     \def\rm{\xxvptrm\fam=\z@}%
     \rm\fi}

%************** Other fonts *********************************

%\font\dummy=dummy

%******************************************************************************

\message{Loading jyTeX macros...}

%************************************************************
%*
%*              Simple modifications to plain
%*
%************************************************************
\message{modifications to plain.tex,}

% The "\outer" qualifier is removed from the definitions of \newcount through
% \newif so that they may be used in definitions.  \newif is also changed to
% make \if commands globally defined.

\def\newcount{\alloc@0\count\countdef\insc@unt}
\def\newdimen{\alloc@1\dimen\dimendef\insc@unt}
\def\newskip{\alloc@2\skip\skipdef\insc@unt}
\def\newmuskip{\alloc@3\muskip\muskipdef\@cclvi}
\def\newbox{\alloc@4\box\chardef\insc@unt}
\def\newtoks{\alloc@5\toks\toksdef\@cclvi}
\def\newhelp#1#2{\newtoks#1\global#1\expandafter{\csname#2\endcsname}}
\def\newread{\alloc@6\read\chardef\sixt@@n}
\def\newwrite{\alloc@7\write\chardef\sixt@@n}
\def\newfam{\alloc@8\fam\chardef\sixt@@n}
\def\newinsert#1{\global\advance\insc@unt by\m@ne
     \ch@ck0\insc@unt\count
     \ch@ck1\insc@unt\dimen
     \ch@ck2\insc@unt\skip
     \ch@ck4\insc@unt\box
     \allocationnumber=\insc@unt
     \global\chardef#1=\allocationnumber
     \wlog{\string#1=\string\insert\the\allocationnumber}}
\def\newif#1{\count@\escapechar \escapechar\m@ne
     \expandafter\expandafter\expandafter
          \xdef\@if#1{true}{\let\noexpand#1=\noexpand\iftrue}%
     \expandafter\expandafter\expandafter
          \xdef\@if#1{false}{\let\noexpand#1=\noexpand\iffalse}%
     \global\@if#1{false}\escapechar=\count@}

%************** Some parameter changes **********************

\newlinechar=`\^^J
\overfullrule=0pt

%************** Font-related modifications ******************

% The plain fonts are mapped onto the corresponding jyTeX fonts

% Some control sequences are disabled.

\let\itfam=\undefined

\let\bffam=\undefined

\count18=3

% German sharp s is given a new name (\ss is already taken)

\chardef\sharps="19

% The mathcode assignments of characters in the math italic font are changed to
% allow for switching to boldface.

\mathchardef\alpha="710B
\mathchardef\beta="710C
\mathchardef\gamma="710D
\mathchardef\delta="710E
\mathchardef\epsilon="710F
\mathchardef\zeta="7110
\mathchardef\eta="7111
\mathchardef\theta="7112
\mathchardef\iota="7113
\mathchardef\kappa="7114
\mathchardef\lambda="7115
\mathchardef\mu="7116
\mathchardef\nu="7117
\mathchardef\xi="7118
\mathchardef\pi="7119
\mathchardef\rho="711A
\mathchardef\sigma="711B
\mathchardef\tau="711C
\mathchardef\upsilon="711D
\mathchardef\phi="711E
\mathchardef\chi="711F
\mathchardef\psi="7120
\mathchardef\omega="7121
\mathchardef\varepsilon="7122
\mathchardef\vartheta="7123
\mathchardef\varpi="7124
\mathchardef\varrho="7125
\mathchardef\varsigma="7126
\mathchardef\varphi="7127
\mathchardef\imath="717B
\mathchardef\jmath="717C
\mathchardef\ell="7160
\mathchardef\wp="717D
\mathchardef\partial="7140
\mathchardef\flat="715B
\mathchardef\natural="715C
\mathchardef\sharp="715D

%************** Miscellaneous changes ***********************

% The dimension \p@ (1pt) is replaced with \rp@ (relative pt, defined below),
% whose size is determined by the base type size of the document.

\def\angle{{\vbox{\ialign{$\m@th\scriptstyle##$\crcr
     \not\mathrel{\mkern14mu}\crcr
     \noalign{\nointerlineskip}
     \mkern2.5mu\leaders\hrule height.34\rp@\hfill\mkern2.5mu\crcr}}}}
\def\vdots{\vbox{\baselineskip4\rp@ \lineskiplimit\z@
     \kern6\rp@\hbox{.}\hbox{.}\hbox{.}}}
\def\ddots{\mathinner{\mkern1mu\raise7\rp@\vbox{\kern7\rp@\hbox{.}}\mkern2mu
     \raise4\rp@\hbox{.}\mkern2mu\raise\rp@\hbox{.}\mkern1mu}}
\def\overrightarrow#1{\vbox{\ialign{##\crcr
     \rightarrowfill\crcr
     \noalign{\kern-\rp@\nointerlineskip}
     $\hfil\displaystyle{#1}\hfil$\crcr}}}
\def\overleftarrow#1{\vbox{\ialign{##\crcr
     \leftarrowfill\crcr
     \noalign{\kern-\rp@\nointerlineskip}
     $\hfil\displaystyle{#1}\hfil$\crcr}}}
\def\overbrace#1{\mathop{\vbox{\ialign{##\crcr
     \noalign{\kern3\rp@}
     \downbracefill\crcr
     \noalign{\kern3\rp@\nointerlineskip}
     $\hfil\displaystyle{#1}\hfil$\crcr}}}\limits}
\def\underbrace#1{\mathop{\vtop{\ialign{##\crcr
     $\hfil\displaystyle{#1}\hfil$\crcr
     \noalign{\kern3\rp@\nointerlineskip}
     \upbracefill\crcr
     \noalign{\kern3\rp@}}}}\limits}
\def\big#1{{\hbox{$\left#1\vbox to8.5\rp@ {}\right.\n@space$}}}
\def\Big#1{{\hbox{$\left#1\vbox to11.5\rp@ {}\right.\n@space$}}}
\def\bigg#1{{\hbox{$\left#1\vbox to14.5\rp@ {}\right.\n@space$}}}
\def\Bigg#1{{\hbox{$\left#1\vbox to17.5\rp@ {}\right.\n@space$}}}
\def\@vereq#1#2{\lower.5\rp@\vbox{\baselineskip\z@skip\lineskip-.5\rp@
     \ialign{$\m@th#1\hfil##\hfil$\crcr#2\crcr=\crcr}}}
\def\rlh@#1{\vcenter{\hbox{\ooalign{\raise2\rp@
     \hbox{$#1\rightharpoonup$}\crcr
     $#1\leftharpoondown$}}}}
\def\bordermatrix#1{\begingroup\m@th
     \setbox\z@\vbox{%
          \def\cr{\crcr\noalign{\kern2\rp@\global\let\cr\endline}}%
          \ialign{$##$\hfil\kern2\rp@\kern\p@renwd
               &\thinspace\hfil$##$\hfil&&\quad\hfil$##$\hfil\crcr
               \omit\strut\hfil\crcr
               \noalign{\kern-\baselineskip}%
               #1\crcr\omit\strut\cr}}%
     \setbox\tw@\vbox{\unvcopy\z@\global\setbox\@ne\lastbox}%
     \setbox\tw@\hbox{\unhbox\@ne\unskip\global\setbox\@ne\lastbox}%
     \setbox\tw@\hbox{$\kern\wd\@ne\kern-\p@renwd\left(\kern-\wd\@ne
          \global\setbox\@ne\vbox{\box\@ne\kern2\rp@}%
          \vcenter{\kern-\ht\@ne\unvbox\z@\kern-\baselineskip}%
          \,\right)$}%
     \null\;\vbox{\kern\ht\@ne\box\tw@}\endgroup}
\def\endinsert{\egroup
     \if@mid\dimen@\ht\z@
          \advance\dimen@\dp\z@
          \advance\dimen@12\rp@
          \advance\dimen@\pagetotal
          \ifdim\dimen@>\pagegoal\@midfalse\p@gefalse\fi
     \fi
     \if@mid\bigskip\box\z@
          \bigbreak
     \else\insert\topins{\penalty100 \splittopskip\z@skip
               \splitmaxdepth\maxdimen\floatingpenalty\z@
               \ifp@ge\dimen@\dp\z@
                    \vbox to\vsize{\unvbox\z@\kern-\dimen@}%
               \else\box\z@\nobreak\bigskip
               \fi}%
     \fi
     \endgroup}

% \normalbaselines is removed from \cases and \matrix.

\def\cases#1{\left\{\,\vcenter{\m@th
     \ialign{$##\hfil$&\quad##\hfil\crcr#1\crcr}}\right.}
\def\matrix#1{\null\,\vcenter{\m@th
     \ialign{\hfil$##$\hfil&&\quad\hfil$##$\hfil\crcr
          \mathstrut\crcr
          \noalign{\kern-\baselineskip}
          #1\crcr
          \mathstrut\crcr
          \noalign{\kern-\baselineskip}}}\,}

% \raggedbottom modified slightly

\newif\ifraggedbottom

\def\raggedbottom{\ifraggedbottom\else
     \advance\topskip by\z@ plus60pt \raggedbottomtrue\fi}%
\def\normalbottom{\ifraggedbottom
     \advance\topskip by\z@ plus-60pt \raggedbottomfalse\fi}

%************************************************************
%*
%*              Miscellaneous definitions
%*
%************************************************************
\message{hacks,}

%************** Hack registers ******************************

\toksdef\toks@i=1
\toksdef\toks@ii=2

%************** Basic macros ********************************

\def\TeX{T\kern-.1667em \lower.5ex \hbox{E}\kern-.125em X\null}
\def\jyTeX{{\leavevmode
     \raise.587ex \hbox{\it\j}\kern-.1em \lower.048ex \hbox{\it y}\kern-.12em
     \TeX}}

\let\then=\iftrue
\def\ifnoarg#1\then{\def\hack@{#1}\ifx\hack@\empty}
\def\ifundefined#1\then{%
     \expandafter\ifx\csname\expandafter\blank\string#1\endcsname\relax}
\def\useif#1\then{\csname#1\endcsname}
\def\usename#1{\csname#1\endcsname}
\def\useafter#1#2{\expandafter#1\csname#2\endcsname}

% Modify so that I can have \loop's within \loop's?
\long\def\loop#1\repeat{\def\@iterate{#1\expandafter\@iterate\fi}\@iterate
     \let\@iterate=\relax}
%\long\def\loop#1\repeat{\def\@loopbody{#1}\@iterate}
%\def\@iterate{\@loopbody\let\next=\@iterate\else\let\next=\relax\fi\next}

\let\TeXend=\end
\def\begin#1{\begingroup\def\@@blockname{#1}\usename{begin#1}}
\def\end#1{\usename{end#1}\def\hack@{#1}%
     \ifx\@@blockname\hack@
          \endgroup
     \else\err@badgroup\hack@\@@blockname
     \fi}
\def\@@blockname{}

\def\defaultoption[#1]#2{%
     \def\hack@{\ifx\hack@ii[\toks@={#2}\else\toks@={#2[#1]}\fi\the\toks@}%
     \futurelet\hack@ii\hack@}

\def\markup#1{\let\@@marksf=\empty
     \ifhmode\edef\@@marksf{\spacefactor=\the\spacefactor\relax}\/\fi
     ${}^{\hbox{\subscriptfonts#1}}$\@@marksf}

%************** Time registers ******************************

\newtoks\shortyear
\newtoks\militaryhour
\newtoks\standardhour
\newtoks\minute
\newtoks\amorpm

\def\settime{\count@=\time\divide\count@ by60
     \militaryhour=\expandafter{\number\count@}%
     {\multiply\count@ by-60 \advance\count@ by\time
          \xdef\hack@{\ifnum\count@<10 0\fi\number\count@}}%
     \minute=\expandafter{\hack@}%
     \ifnum\count@<12
          \amorpm={am}
     \else\amorpm={pm}
          \ifnum\count@>12 \advance\count@ by-12 \fi
     \fi
     \standardhour=\expandafter{\number\count@}%
     \def\hack@19##1##2{\shortyear={##1##2}}%
          \expandafter\hack@\the\year}

\def\monthword#1{%
     \ifcase#1
          $\bullet$\err@badcountervalue{monthword}%
          \or January\or February\or March\or April\or May\or June%
          \or July\or August\or September\or October\or November\or December%
     \else$\bullet$\err@badcountervalue{monthword}%
     \fi}

\def\monthabbr#1{%
     \ifcase#1
          $\bullet$\err@badcountervalue{monthabbr}%
          \or Jan\or Feb\or Mar\or Apr\or May\or Jun%
          \or Jul\or Aug\or Sep\or Oct\or Nov\or Dec%
     \else$\bullet$\err@badcountervalue{monthabbr}%
     \fi}

\def\militarytime{\the\militaryhour:\the\minute}
\def\standardtime{\the\standardhour:\the\minute}

%************** Number styles *******************************

\def\@setnumstyle#1#2{\expandafter\global\expandafter\expandafter
     \expandafter\let\expandafter\expandafter
     \csname @\expandafter\blank\string#1style\endcsname
     \csname#2\endcsname}
\def\numstyle#1{\usename{@\expandafter\blank\string#1style}#1}
\def\ifblank#1\then{\useafter\ifx{@\expandafter\blank\string#1}\blank}

\def\blank#1{}

\def\Roman#1{\expandafter\uppercase\expandafter{\romannumeral#1}}
\def\alphabetic#1{%
     \ifcase#1
          $\bullet$\err@badcountervalue{alphabetic}%
          \or a\or b\or c\or d\or e\or f\or g\or h\or i\or j\or k\or l\or m%
          \or n\or o\or p\or q\or r\or s\or t\or u\or v\or w\or x\or y\or z%
     \else$\bullet$\err@badcountervalue{alphabetic}%
     \fi}
\def\Alphabetic#1{\expandafter\uppercase\expandafter{\alphabetic{#1}}}
\def\symbols#1{%
     \ifcase#1
          $\bullet$\err@badcountervalue{symbols}%
          \or*\or\dag\or\ddag\or\S\or$\|$%
          \or**\or\dag\dag\or\ddag\ddag\or\S\S\or$\|\|$%
     \else$\bullet$\err@badcountervalue{symbols}%
     \fi}

%************** String macros *******************************

\catcode`\^^?=13 \def^^?{\relax}

\def\trimleading#1\to#2{\edef#2{#1}%
     \expandafter\@trimleading\expandafter#2#2^^?^^?}
\def\@trimleading#1#2#3^^?{\ifx#2^^?\def#1{}\else\def#1{#2#3}\fi}

\def\trimtrailing#1\to#2{\edef#2{#1}%
     \expandafter\@trimtrailing\expandafter#2#2^^? ^^?\relax}
\def\@trimtrailing#1#2 ^^?#3{\ifx#3\relax\toks@={}%
     \else\def#1{#2}\toks@={\trimtrailing#1\to#1}\fi
     \the\toks@}

\def\trim#1\to#2{\trimleading#1\to#2\trimtrailing#2\to#2}

\catcode`\^^?=15

%************** List macros *********************************

\long\def\additemL#1\to#2{\toks@={\^^\{#1}}\toks@ii=\expandafter{#2}%
     \xdef#2{\the\toks@\the\toks@ii}}

\long\def\additemR#1\to#2{\toks@={\^^\{#1}}\toks@ii=\expandafter{#2}%
     \xdef#2{\the\toks@ii\the\toks@}}

\def\getitemL#1\to#2{\expandafter\@getitemL#1\hack@#1#2}
\def\@getitemL\^^\#1#2\hack@#3#4{\def#4{#1}\def#3{#2}}

%************************************************************
%*
%*             Font-related macros
%*
%************************************************************
\message{font macros,}

%************** Font set-up *********************************

\newdimen\rp@
\newcount\@@sizeindex \@@sizeindex=0
\newcount\@@factori
\newcount\@@factorii
\newcount\@@factoriii
\newcount\@@factoriv

\countdef\maxfam=18
\newfam\itfam
\newfam\bffam
\newfam\bfsfam
\newfam\bmitfam

\def\@mathfontinit{\count@=4
     \loop\textfont\count@=\nullfont
          \scriptfont\count@=\nullfont
          \scriptscriptfont\count@=\nullfont
          \ifnum\count@<\maxfam\advance\count@ by\@ne
     \repeat}

\def\@fontstyleinit{%
     \def\it{\err@fontnotavailable\it}%
     \def\bf{\err@fontnotavailable\bf}%
     \def\bfs{\err@bfstobf}%
     \def\bmit{\err@fontnotavailable\bmit}%
     \def\sc{\err@fontnotavailable\sc}%
     \def\sl{\err@sltoit}%
     \def\ss{\err@fontnotavailable\ss}%
     \def\tt{\err@fontnotavailable\tt}}

\def\@parameterinit#1{\rm\rp@=.1em \@getscaling{#1}%
     \let\^^\=\@doscaling\scalingskipslist
     \setbox\strutbox=\hbox{\vrule
          height.708\baselineskip depth.292\baselineskip width\z@}}

\def\@getfactor#1#2#3#4{\@@factori=#1 \@@factorii=#2
     \@@factoriii=#3 \@@factoriv=#4}

\def\@getscaling#1{\count@=#1 \advance\count@ by-\@@sizeindex\@@sizeindex=#1
     \ifnum\count@<0
          \let\@mulordiv=\divide
          \let\@divormul=\multiply
          \multiply\count@ by\m@ne
     \else\let\@mulordiv=\multiply
          \let\@divormul=\divide
     \fi
     \edef\@@scratcha{\ifcase\count@                {1}{1}{1}{1}\or
          {1}{7}{23}{3}\or     {2}{5}{3}{1}\or      {9}{89}{13}{1}\or
          {6}{25}{6}{1}\or     {8}{71}{14}{1}\or    {6}{25}{36}{5}\or
          {1}{7}{53}{4}\or     {12}{125}{108}{5}\or {3}{14}{53}{5}\or
          {6}{41}{17}{1}\or    {13}{31}{13}{2}\or   {9}{107}{71}{2}\or
          {11}{139}{124}{3}\or {1}{6}{43}{2}\or     {10}{107}{42}{1}\or
          {1}{5}{43}{2}\or     {5}{69}{65}{1}\or    {11}{97}{91}{2}\fi}%
     \expandafter\@getfactor\@@scratcha}

\def\@doscaling#1{\@mulordiv#1by\@@factori\@divormul#1by\@@factorii
     \@mulordiv#1by\@@factoriii\@divormul#1by\@@factoriv}

%************* Size-changing commands ***********************

\newskip\headskip
\newskip\footskip

\def\typesize=#1pt{\count@=#1 \advance\count@ by-10
     \ifcase\count@
          \@setsizex\or\err@badtypesize\or
          \@setsizexii\or\err@badtypesize\or
          \@setsizexiv
     \else\err@badtypesize
     \fi}

\def\@setsizex{\getixpt
     \def\subsubscriptfonts{\vpt}%
          \def\subsubscriptsize{\vpt\@parameterinit{-8}}%
     \def\subscriptfonts{\viipt}\def\subscriptsize{\viipt\@parameterinit{-4}}%
     \def\footnotefonts{\viiipt}\def\footnotesize{\viiipt\@parameterinit{-2}}%
     \def\smallfonts{\ixpt}\def\smallsize{\ixpt\@parameterinit{-1}}%
     \def\normalfonts{\xpt}\def\normalsize{\xpt\@parameterinit{0}}%
     \def\bigfonts{\xiipt}\def\bigsize{\xiipt\@parameterinit{2}}%
     \def\Bigfonts{\xivpt}\def\Bigsize{\xivpt\@parameterinit{4}}%
     \def\biggfonts{\xviipt}\def\biggsize{\xviipt\@parameterinit{6}}%
     \def\Biggfonts{\xxipt}\def\Biggsize{\xxipt\@parameterinit{8}}%
     \def\tinyfonts{\vpt}\def\tinysize{\vpt\@parameterinit{-8}}%
     \def\HUGEFONTS{\xxvpt}\def\HUGESIZE{\xxvpt\@parameterinit{10}}%
     \normalsize\fixedskipslist}

\def\@setsizexii{\getxipt
     \def\subsubscriptfonts{\vipt}%
          \def\subsubscriptsize{\vipt\@parameterinit{-6}}%
     \def\subscriptfonts{\viiipt}%
          \def\subscriptsize{\viiipt\@parameterinit{-2}}%
     \def\footnotefonts{\xpt}\def\footnotesize{\xpt\@parameterinit{0}}%
     \def\smallfonts{\xipt}\def\smallsize{\xipt\@parameterinit{1}}%
     \def\normalfonts{\xiipt}\def\normalsize{\xiipt\@parameterinit{2}}%
     \def\bigfonts{\xivpt}\def\bigsize{\xivpt\@parameterinit{4}}%
     \def\Bigfonts{\xviipt}\def\Bigsize{\xviipt\@parameterinit{6}}%
     \def\biggfonts{\xxipt}\def\biggsize{\xxipt\@parameterinit{8}}%
     \def\Biggfonts{\xxvpt}\def\Biggsize{\xxvpt\@parameterinit{10}}%
     \def\tinyfonts{\vpt}\def\tinysize{\vpt\@parameterinit{-8}}%
     \def\HUGEFONTS{\xxvpt}\def\HUGESIZE{\xxvpt\@parameterinit{10}}%
     \normalsize\fixedskipslist}

\def\@setsizexiv{\getxiiipt
     \def\subsubscriptfonts{\viipt}%
          \def\subsubscriptsize{\viipt\@parameterinit{-4}}%
     \def\subscriptfonts{\xpt}\def\subscriptsize{\xpt\@parameterinit{0}}%
     \def\footnotefonts{\xiipt}\def\footnotesize{\xiipt\@parameterinit{2}}%
     \def\smallfonts{\xiiipt}\def\smallsize{\xiiipt\@parameterinit{3}}%
     \def\normalfonts{\xivpt}\def\normalsize{\xivpt\@parameterinit{4}}%
     \def\bigfonts{\xviipt}\def\bigsize{\xviipt\@parameterinit{6}}%
     \def\Bigfonts{\xxipt}\def\Bigsize{\xxipt\@parameterinit{8}}%
     \def\biggfonts{\xxvpt}\def\biggsize{\xxvpt\@parameterinit{10}}%
     \def\Biggfonts{\err@sizetoolarge\Biggfonts\HUGEFONTS}%
          \def\Biggsize{\err@sizetoolarge\Biggsize\HUGESIZE}%
     \def\tinyfonts{\vpt}\def\tinysize{\vpt\@parameterinit{-8}}%
     \def\HUGEFONTS{\xxvpt}\def\HUGESIZE{\xxvpt\@parameterinit{10}}%
     \normalsize\fixedskipslist}

\def\subsubscriptfonts{\vpt} \def\subsubscriptsize{\vpt\@parameterinit{-8}}
\def\subscriptfonts{\viipt}  \def\subscriptsize{\viipt\@parameterinit{-4}}
\def\footnotefonts{\viiipt}  \def\footnotesize{\viiipt\@parameterinit{-2}}
\def\smallfonts{\err@sizenotavailable\smallfonts}
                             \def\smallsize{\ixpt\@parameterinit{-1}}
\def\normalfonts{\xpt}       \def\normalsize{\xpt\@parameterinit{0}}
\def\bigfonts{\xiipt}        \def\bigsize{\xiipt\@parameterinit{2}}
\def\Bigfonts{\xivpt}        \def\Bigsize{\xivpt\@parameterinit{4}}
\def\biggfonts{\xviipt}      \def\biggsize{\xviipt\@parameterinit{6}}
\def\Biggfonts{\xxipt}       \def\Biggsize{\xxipt\@parameterinit{8}}
\def\tinyfonts{\vpt}         \def\tinysize{\vpt\@parameterinit{-8}}
\def\HUGEFONTS{\xxvpt}       \def\HUGESIZE{\xxvpt\@parameterinit{10}}

%************************************************************
%*
%*             Document layout
%*
%************************************************************
\message{document layout,}

%************** Page format *********************************

\newtoks\everyoutput \everyoutput={}
\newdimen\depthofpage
\newcount\pagenum \pagenum=0

\newdimen\oddtopmargin  \newdimen\eventopmargin
\newdimen\oddleftmargin \newdimen\evenleftmargin
\newtoks\oddhead        \newtoks\evenhead
\newtoks\oddfoot        \newtoks\evenfoot

\def\topmargin{\afterassignment\@seteventop\oddtopmargin}
\def\leftmargin{\afterassignment\@setevenleft\oddleftmargin}
\def\head{\afterassignment\@setevenhead\oddhead}
\def\foot{\afterassignment\@setevenfoot\oddfoot}

\def\@seteventop{\eventopmargin=\oddtopmargin}
\def\@setevenleft{\evenleftmargin=\oddleftmargin}
\def\@setevenhead{\evenhead=\oddhead}
\def\@setevenfoot{\evenfoot=\oddfoot}

\def\pagenumstyle#1{\@setnumstyle\pagenum{#1}}

\newif\ifdraft
\def\draft{\drafttrue\leftmargin=.5in \overfullrule=5pt }

\def\outputstyle#1{\global\expandafter\let\expandafter
          \@outputstyle\csname#1output\endcsname
     \usename{#1setup}}

\output={\@outputstyle}

\def\normaloutput{\the\everyoutput
     \global\advance\pagenum by\@ne
     \ifodd\pagenum
          \voffset=\oddtopmargin \hoffset=\oddleftmargin
     \else\voffset=\eventopmargin \hoffset=\evenleftmargin
     \fi
     \advance\voffset by-1in  \advance\hoffset by-1in
     \count0=\pagenum
     \expandafter\shipout\pagebox
     \ifnum\outputpenalty>-\@MM\else\dosupereject\fi}

\newdimen\fullhsize
\newbox\leftpage
\newcount\leftpagenum
\newcount\outputpagenum \outputpagenum=0
\let\leftorright=L

\def\twoupoutput{\the\everyoutput
     \global\advance\pagenum by\@ne
     \if L\leftorright
          \global\setbox\leftpage=\leftline{\pagebox}%
          \global\leftpagenum=\pagenum
          \global\let\leftorright=R%
     \else\global\advance\outputpagenum by\@ne
          \ifodd\outputpagenum
               \voffset=\oddtopmargin \hoffset=\oddleftmargin
          \else\voffset=\eventopmargin \hoffset=\evenleftmargin
          \fi
          \advance\voffset by-1in  \advance\hoffset by-1in
          \count0=\leftpagenum \count1=\pagenum
          \shipout\vbox{\hbox to\fullhsize
               {\box\leftpage\hfil\leftline{\pagebox}}}%
          \global\let\leftorright=L%
     \fi
     \ifnum\outputpenalty>-\@MM
     \else\dosupereject
          \if R\leftorright
               \globaldefs=\@ne\head={\hfil}\foot={\hfil}\globaldefs=\z@
               \null\newpage
          \fi
     \fi}

\def\pagebox{\vbox{\makeheadline\pagebody\makefootline}}

\def\makeheadline{%
     \vbox to\z@{\baselinestretch=\@m
          \vskip\topskip\vskip-.708\baselineskip\vskip-\headskip
          \line{\vbox to\ht\strutbox{}%
               \ifodd\pagenum\the\oddhead\else\the\evenhead\fi}%
          \vss}%
     \nointerlineskip}

\def\pagebody{\vbox to\vsize{%
     \boxmaxdepth\maxdepth
     \ifvoid\topins\else\unvbox\topins\fi
     \depthofpage=\dp255
     \unvbox255
     \ifraggedbottom\kern-\depthofpage\vfil\fi
     \ifvoid\footins
     \else\vskip\skip\footins
          \footnoterule
          \unvbox\footins
          \vskip-\footnoteskip
     \fi}}

\def\makefootline{\baselineskip=\footskip
     \line{\ifodd\pagenum\the\oddfoot\else\the\evenfoot\fi}}

%************** Sectioning commands *************************

\newskip\abovechapterskip
\newskip\belowchapterskip
\newskip\abovesectionskip
\newskip\belowsectionskip
\newskip\abovesubsectionskip
\newskip\belowsubsectionskip

\def\chapterstyle#1{\global\expandafter\let\expandafter\@chapterstyle
     \csname#1text\endcsname}
\def\sectionstyle#1{\global\expandafter\let\expandafter\@sectionstyle
     \csname#1text\endcsname}
\def\subsectionstyle#1{\global\expandafter\let\expandafter\@subsectionstyle
     \csname#1text\endcsname}

\def\chapter#1{%
     \ifdim\lastskip=17sp \else\chapterbreak\vskip\abovechapterskip\fi
     \@chapterstyle{\ifblank\chapternumstyle\then
          \else\newchapternum=\next\chapternumformat\ \fi#1}%
     \nobreak\vskip\belowchapterskip\vskip17sp }

\def\section#1{%
     \ifdim\lastskip=17sp \else\sectionbreak\vskip\abovesectionskip\fi
     \@sectionstyle{\ifblank\sectionnumstyle\then
          \else\newsectionnum=\next\sectionnumformat\ \fi#1}%
     \nobreak\vskip\belowsectionskip\vskip17sp }

\def\subsection#1{%
     \ifdim\lastskip=17sp \else\subsectionbreak\vskip\abovesubsectionskip\fi
     \@subsectionstyle{\ifblank\subsectionnumstyle\then
          \else\newsubsectionnum=\next\subsectionnumformat\ \fi#1}%
     \nobreak\vskip\belowsubsectionskip\vskip17sp }

%************** Text formatting commands ********************

\let\TeXunderline=\underline
\let\TeXoverline=\overline
\def\underline#1{\relax\ifmmode\TeXunderline{#1}\else
     $\TeXunderline{\hbox{#1}}$\fi}
\def\overline#1{\relax\ifmmode\TeXoverline{#1}\else
     $\TeXoverline{\hbox{#1}}$\fi}

\def\baselinestretch{\afterassignment\@baselinestretch\count@}
\def\@baselinestretch{\baselineskip=\normalbaselineskip
     \divide\baselineskip by\@m\baselineskip=\count@\baselineskip
     \setbox\strutbox=\hbox{\vrule
          height.708\baselineskip depth.292\baselineskip width\z@}%
     \bigskipamount=\the\baselineskip
          plus.25\baselineskip minus.25\baselineskip
     \medskipamount=.5\baselineskip
          plus.125\baselineskip minus.125\baselineskip
     \smallskipamount=.25\baselineskip
          plus.0625\baselineskip minus.0625\baselineskip}

\def\\{\ifhmode\ifnum\lastpenalty=-\@M\else\hfil\penalty-\@M\fi\fi
     \ignorespaces}
\def\newpage{\vfil\break}

\def\lefttext#1{\par{\@text\leftskip=\z@\rightskip=\centering
     \noindent#1\par}}
\def\righttext#1{\par{\@text\leftskip=\centering\rightskip=\z@
     \noindent#1\par}}
\def\centertext#1{\par{\@text\leftskip=\centering\rightskip=\centering
     \noindent#1\par}}
\def\@text{\parindent=\z@ \parfillskip=\z@ \everypar={}%
     \spaceskip=.3333em \xspaceskip=.5em
     \def\\{\ifhmode\ifnum\lastpenalty=-\@M\else\penalty-\@M\fi\fi
          \ignorespaces}}

\def\beginleft{\par\@text\leftskip=\z@ \rightskip=\centering}
     
\def\beginright{\par\@text\leftskip=\centering\rightskip=\z@ }
     
\def\begincenter{\par\@text\leftskip=\centering\rightskip=\centering}

\def\beginnarrow{\defaultoption[\parindent]\@beginnarrow}
\def\@beginnarrow[#1]{\par\advance\leftskip by#1\advance\rightskip by#1}

\begingroup
\catcode`\[=1 \catcode`\{=11
\gdef\beginignore[\endgroup\bgroup
     \catcode`\e=0 \catcode`\\=12 \catcode`\{=11 \catcode`\f=12 \let\or=\relax
     \let\nd{ignor=\fi \let\}=\egroup
     \iffalse}
\endgroup

\long\def\marginnote#1{\leavevmode
     \edef\@marginsf{\spacefactor=\the\spacefactor\relax}%
     \ifdraft\strut\vadjust{%
          \hbox to\z@{\hskip\hsize\hskip.1in
               \vbox to\z@{\vskip-\dp\strutbox
                    \marginnoteformat
                    \vskip-\ht\strutbox
                    \noindent\strut#1\par
                    \vss}%
               \hss}}%
     \fi
     \@marginsf}

%************** The \bye command ****************************

\newtoks\everybye \everybye={\par\vfil}
\outer\def\bye{\the\everybye
     \footnotecheck
     \prelabelcheck
     \streamcheck
     \supereject
     \TeXend}

%************************************************************
%*
%*             Footnotes
%*
%************************************************************
\message{footnotes,}

\newcount\footnotenum \footnotenum=0
\newskip\footnoteskip
\let\@footnotelist=\empty

\def\footnotenumstyle#1{\@setnumstyle\footnotenum{#1}%
     \useafter\ifx{@footnotenumstyle}\symbols
          \global\let\@footup=\empty
     \else\global\let\@footup=\markup
     \fi}

\def\footnote{\footnotecheck\defaultoption[]\@footnote}
\def\@footnote[#1]{\@footnotemark[#1]\@footnotetext}

\def\footnotemark{\defaultoption[]\@footnotemark}
\def\@footnotemark[#1]{\let\@footsf=\empty
     \ifhmode\edef\@footsf{\spacefactor=\the\spacefactor\relax}\/\fi
     \ifnoarg#1\then
          \global\advance\footnotenum by\@ne
          \@footup{\footnotenumformat}%
          \edef\@@foota{\footnotenum=\the\footnotenum\relax}%
          \expandafter\additemR\expandafter\@footup\expandafter
               {\@@foota\footnotenumformat}\to\@footnotelist
          \global\let\@footnotelist=\@footnotelist
     \else\markup{#1}%
          \additemR\markup{#1}\to\@footnotelist
          \global\let\@footnotelist=\@footnotelist
     \fi
     \@footsf}

\def\footnotetext{%
     \ifx\@footnotelist\empty\err@extrafootnotetext\else\@footnotetext\fi}
\def\@footnotetext{%
     \getitemL\@footnotelist\to\@@foota
     \global\let\@footnotelist=\@footnotelist
     \insert\footins\bgroup
     \footnoteformat
     \splittopskip=\ht\strutbox\splitmaxdepth=\dp\strutbox
     \interlinepenalty=\interfootnotelinepenalty\floatingpenalty=\@MM
     \noindent\llap{\@@foota}\strut
     \bgroup\aftergroup\@footnoteend
     \let\@@scratcha=}
\def\@footnoteend{\strut\par\vskip\footnoteskip\egroup}

\def\footnoterule{\normalfonts
     \kern-.3em \hrule width2in height.04em \kern .26em }

\def\footnotecheck{%
     \ifx\@footnotelist\empty
     \else\err@extrafootnotemark
          \global\let\@footnotelist=\empty
     \fi}

%************************************************************
%*
%*             Labelling macros
%*
%************************************************************
\message{labels,}

\let\@@labeldef=\xdef
\newif\if@labelfile
\newwrite\@labelfile
\let\@prelabellist=\empty

\def\label#1#2{\trim#1\to\@@labarg\edef\@@labtext{#2}%
     \edef\@@labname{lab@\@@labarg}%
     \useafter\ifundefined\@@labname\then\else\@yeslab\fi
     \useafter\@@labeldef\@@labname{#2}%
     \ifstreaming
          \expandafter\toks@\expandafter\expandafter\expandafter
               {\csname\@@labname\endcsname}%
          \immediate\write\streamout{\noexpand\label{\@@labarg}{\the\toks@}}%
     \fi}
\def\@yeslab{%
     \useafter\ifundefined{if\@@labname}\then
          \err@labelredef\@@labarg
     \else\useif{if\@@labname}\then
               \err@labelredef\@@labarg
          \else\global\usename{\@@labname true}%
               \useafter\ifundefined{pre\@@labname}\then
               \else\useafter\ifx{pre\@@labname}\@@labtext
                    \else\err@badlabelmatch\@@labarg
                    \fi
               \fi
               \if@labelfile
               \else\global\@labelfiletrue
                    \immediate\write\sixt@@n{--> Creating file \jobname.lab}%
                    \immediate\openout\@labelfile=\jobname.lab
               \fi
               \immediate\write\@labelfile
                    {\noexpand\prelabel{\@@labarg}{\@@labtext}}%
          \fi
     \fi}

\def\putlab#1{\trim#1\to\@@labarg\edef\@@labname{lab@\@@labarg}%
     \useafter\ifundefined\@@labname\then\@nolab\else\usename\@@labname\fi}
\def\@nolab{%
     \useafter\ifundefined{pre\@@labname}\then
          \undefinedlabelformat
          \err@needlabel\@@labarg
          \useafter\xdef\@@labname{\undefinedlabelformat}%
     \else\usename{pre\@@labname}%
          \useafter\xdef\@@labname{\usename{pre\@@labname}}%
     \fi
     \useafter\newif{if\@@labname}%
     \expandafter\additemR\@@labarg\to\@prelabellist}

\def\prelabel#1{\useafter\gdef{prelab@#1}}

\def\ifundefinedlabel#1\then{%
     \expandafter\ifx\csname lab@#1\endcsname\relax}
\def\useiflab#1\then{\csname iflab@#1\endcsname}

\def\prelabelcheck{{%
     \def\^^\##1{\useiflab{##1}\then\else\err@undefinedlabel{##1}\fi}%
     \@prelabellist}}

%************************************************************
%*
%*             Equation numbering
%*
%************************************************************
\message{equation numbering,}

\newcount\chapternum
\newcount\sectionnum
\newcount\subsectionnum
\newcount\equationnum
\newcount\subequationnum
\newcount\figurenum
\newcount\subfigurenum
\newcount\tablenum
\newcount\subtablenum

\newif\if@subeqncount
\newif\if@subfigcount
\newif\if@subtblcount

\def\newchapternum{\newsectionnum=\z@\@resetnum\chapternum}
\def\newsectionnum{\newsubsectionnum=\z@\@resetnum\sectionnum}
\def\newsubsectionnum{\newequationnum=\z@\newfigurenum=\z@\newtablenum=\z@
     \@resetnum\subsectionnum}
\def\newequationnum{\newsubequationnum=\z@\@resetnum\equationnum}
\def\newsubequationnum{\@resetnum\subequationnum}
\def\newfigurenum{\newsubfigurenum=\z@\@resetnum\figurenum}
\def\newsubfigurenum{\@resetnum\subfigurenum}
\def\newtablenum{\newsubtablenum=\z@\@resetnum\tablenum}
\def\newsubtablenum{\@resetnum\subtablenum}

\def\@resetnum#1{\global\advance#1by1 \edef\next{\the#1\relax}\global#1}

\newchapternum=0

\def\chapternumstyle#1{\@setnumstyle\chapternum{#1}}
\def\sectionnumstyle#1{\@setnumstyle\sectionnum{#1}}
\def\subsectionnumstyle#1{\@setnumstyle\subsectionnum{#1}}
\def\equationnumstyle#1{\@setnumstyle\equationnum{#1}}
\def\subequationnumstyle#1{\@setnumstyle\subequationnum{#1}%
     \ifblank\subequationnumstyle\then\global\@subeqncountfalse\fi
     \ignorespaces}
\def\figurenumstyle#1{\@setnumstyle\figurenum{#1}}
\def\subfigurenumstyle#1{\@setnumstyle\subfigurenum{#1}%
     \ifblank\subfigurenumstyle\then\global\@subfigcountfalse\fi
     \ignorespaces}
\def\tablenumstyle#1{\@setnumstyle\tablenum{#1}}
\def\subtablenumstyle#1{\@setnumstyle\subtablenum{#1}%
     \ifblank\subtablenumstyle\then\global\@subtblcountfalse\fi
     \ignorespaces}

\def\eqnlabel#1{%
     \if@subeqncount
          \newsubequationnum=\next
     \else\newequationnum=\next
          \ifblank\subequationnumstyle\then
          \else\global\@subeqncounttrue
               \newsubequationnum=\@ne
          \fi
     \fi
     \label{#1}{\puteqnformat}(\puteqn{#1})%
     \ifdraft\rlap{\hskip.1in{\tt#1}}\fi}

\let\puteqn=\putlab

\def\equation#1#2{\useafter\gdef{eqn@#1}{#2\eqno\eqnlabel{#1}}}
\def\Equation#1{\useafter\gdef{eqn@#1}}

\def\putequation#1{\useafter\ifundefined{eqn@#1}\then
     \err@undefinedeqn{#1}\else\usename{eqn@#1}\fi}

\def\eqnseriesstyle#1{\gdef\@eqnseriesstyle{#1}}
\def\begineqnseries{\subequationnumstyle{\@eqnseriesstyle}%
     \defaultoption[]\@begineqnseries}
\def\@begineqnseries[#1]{\edef\@@eqnname{#1}}
\def\endeqnseries{\subequationnumstyle{blank}%
     \expandafter\ifnoarg\@@eqnname\then
     \else\label\@@eqnname{\puteqnformat}%
     \fi
     \aftergroup\ignorespaces}

\def\figlabel#1{%
     \if@subfigcount
          \newsubfigurenum=\next
     \else\newfigurenum=\next
          \ifblank\subfigurenumstyle\then
          \else\global\@subfigcounttrue
               \newsubfigurenum=\@ne
          \fi
     \fi
     \label{#1}{\putfigformat}\putfig{#1}%
     {\def\marginnoteformat{\tt}\marginnote{#1}}}

\let\putfig=\putlab

\def\figseriesstyle#1{\gdef\@figseriesstyle{#1}}
\def\beginfigseries{\subfigurenumstyle{\@figseriesstyle}%
     \defaultoption[]\@beginfigseries}
\def\@beginfigseries[#1]{\edef\@@figname{#1}}
\def\endfigseries{\subfigurenumstyle{blank}%
     \expandafter\ifnoarg\@@figname\then
     \else\label\@@figname{\putfigformat}%
     \fi
     \aftergroup\ignorespaces}

\def\tbllabel#1{%
     \if@subtblcount
          \newsubtablenum=\next
     \else\newtablenum=\next
          \ifblank\subtablenumstyle\then
          \else\global\@subtblcounttrue
               \newsubtablenum=\@ne
          \fi
     \fi
     \label{#1}{\puttblformat}\puttbl{#1}%
     {\def\marginnoteformat{\tt}\marginnote{#1}}}

\let\puttbl=\putlab

\def\tblseriesstyle#1{\gdef\@tblseriesstyle{#1}}
\def\begintblseries{\subtablenumstyle{\@tblseriesstyle}%
     \defaultoption[]\@begintblseries}
\def\@begintblseries[#1]{\edef\@@tblname{#1}}
\def\endtblseries{\subtablenumstyle{blank}%
     \expandafter\ifnoarg\@@tblname\then
     \else\label\@@tblname{\puttblformat}%
     \fi
     \aftergroup\ignorespaces}

%************************************************************
%*
%*             Reference numbering
%*
%************************************************************
\message{reference numbering,}

\newcount\referencenum \referencenum=0
\newcount\@@prerefcount \@@prerefcount=0
\newcount\@@thisref
\newcount\@@lastref
\newcount\@@loopref
\newcount\@@refseq
\newdimen\refnumindent
\let\@undefreflist=\empty

\def\referencenumstyle#1{\@setnumstyle\referencenum{#1}}

\def\referencestyle#1{\usename{@ref#1}}

\def\@refsequential{%
     \gdef\@refpredef##1{\global\advance\referencenum by\@ne
          \let\^^\=0\label{##1}{\^^\{\the\referencenum}}%
          \useafter\gdef{ref@\the\referencenum}{{##1}{\undefinedlabelformat}}}%
     \gdef\@reference##1##2{%
          \ifundefinedlabel##1\then
          \else\def\^^\####1{\global\@@thisref=####1\relax}\putlab{##1}%
               \useafter\gdef{ref@\the\@@thisref}{{##1}{##2}}%
          \fi}%
     \gdef\endputreferences{%
          \loop\ifnum\@@loopref<\referencenum
                    \advance\@@loopref by\@ne
                    \expandafter\expandafter\expandafter\@printreference
                         \csname ref@\the\@@loopref\endcsname
          \repeat
          \par}}

\def\@refpreordered{%
     \gdef\@refpredef##1{\global\advance\referencenum by\@ne
          \additemR##1\to\@undefreflist}%
     \gdef\@reference##1##2{%
          \ifundefinedlabel##1\then
          \else\global\advance\@@loopref by\@ne
               {\let\^^\=0\label{##1}{\^^\{\the\@@loopref}}}%
               \@printreference{##1}{##2}%
          \fi}
     \gdef\endputreferences{%
          \def\^^\####1{\useiflab{####1}\then
               \else\reference{####1}{\undefinedlabelformat}\fi}%
          \@undefreflist
          \par}}

\def\beginprereferences{\par
     \def\reference##1##2{\global\advance\referencenum by1\@ne
          \let\^^\=0\label{##1}{\^^\{\the\referencenum}}%
          \useafter\gdef{ref@\the\referencenum}{{##1}{##2}}}}
\def\endprereferences{\global\@@prerefcount=\the\referencenum\par}

\def\beginputreferences{\par
     \refnumindent=\z@\@@loopref=\z@
     \loop\ifnum\@@loopref<\referencenum
               \advance\@@loopref by\@ne
               \setbox\z@=\hbox{\referencenum=\@@loopref
                    \referencenumformat\enskip}%
               \ifdim\wd\z@>\refnumindent\refnumindent=\wd\z@\fi
     \repeat
     \putreferenceformat
     \@@loopref=\z@
     \loop\ifnum\@@loopref<\@@prerefcount
               \advance\@@loopref by\@ne
               \expandafter\expandafter\expandafter\@printreference
                    \csname ref@\the\@@loopref\endcsname
     \repeat
     \let\reference=\@reference}

\def\@printreference#1#2{\ifx#2\undefinedlabelformat\err@undefinedref{#1}\fi
     \noindent\ifdraft\rlap{\hskip\hsize\hskip.1in \tt#1}\fi
     \llap{\referencenum=\@@loopref\referencenumformat\enskip}#2\par}

\def\reference#1#2{{\par\refnumindent=\z@\putreferenceformat\noindent#2\par}}

\def\putref#1{\trim#1\to\@@refarg
     \expandafter\ifnoarg\@@refarg\then
          \toks@={\relax}%
     \else\@@lastref=-\@m\def\@@refsep{}\def\@more{\@nextref}%
          \toks@={\@nextref#1,,}%
     \fi\the\toks@}
\def\@nextref#1,{\trim#1\to\@@refarg
     \expandafter\ifnoarg\@@refarg\then
          \let\@more=\relax
     \else\ifundefinedlabel\@@refarg\then
               \expandafter\@refpredef\expandafter{\@@refarg}%
          \fi
          \def\^^\##1{\global\@@thisref=##1\relax}%
          \global\@@thisref=\m@ne
          \setbox\z@=\hbox{\putlab\@@refarg}%
     \fi
     \advance\@@lastref by\@ne
     \ifnum\@@lastref=\@@thisref\advance\@@refseq by\@ne\else\@@refseq=\@ne\fi
     \ifnum\@@lastref<\z@
     \else\ifnum\@@refseq<\thr@@
               \@@refsep\def\@@refsep{,}%
               \ifnum\@@lastref>\z@
                    \advance\@@lastref by\m@ne
                    {\referencenum=\@@lastref\putrefformat}%
               \else\undefinedlabelformat
               \fi
          \else\def\@@refsep{--}%
          \fi
     \fi
     \@@lastref=\@@thisref
     \@more}

%************************************************************
%*
%*             Job streaming
%*
%************************************************************
\message{streaming,}

\newif\ifstreaming

\def\streamto{\defaultoption[\jobname]\@streamto}
\def\@streamto[#1]{\global\streamingtrue
     \immediate\write\sixt@@n{--> Streaming to #1.str}%
     \newwrite\streamout\immediate\openout\streamout=#1.str }

\def\streamfrom{\defaultoption[\jobname]\@streamfrom}
\def\@streamfrom[#1]{\newread\streamin\openin\streamin=#1.str
     \ifeof\streamin
          \expandafter\err@nostream\expandafter{#1.str}%
     \else\immediate\write\sixt@@n{--> Streaming from #1.str}%
          \let\@@labeldef=\gdef
          \ifstreaming
               \edef\@elc{\endlinechar=\the\endlinechar}%
               \endlinechar=\m@ne
               \loop\read\streamin to\@@scratcha
                    \ifeof\streamin
                         \streamingfalse
                    \else\toks@=\expandafter{\@@scratcha}%
                         \immediate\write\streamout{\the\toks@}%
                    \fi
                    \ifstreaming
               \repeat
               \@elc
               \input #1.str
               \streamingtrue
          \else\input #1.str
          \fi
          \let\@@labeldef=\xdef
     \fi}

\def\streamcheck{\ifstreaming
     \immediate\write\streamout{\pagenum=\the\pagenum}%
     \immediate\write\streamout{\footnotenum=\the\footnotenum}%
     \immediate\write\streamout{\referencenum=\the\referencenum}%
     \immediate\write\streamout{\chapternum=\the\chapternum}%
     \immediate\write\streamout{\sectionnum=\the\sectionnum}%
     \immediate\write\streamout{\subsectionnum=\the\subsectionnum}%
     \immediate\write\streamout{\equationnum=\the\equationnum}%
     \immediate\write\streamout{\subequationnum=\the\subequationnum}%
     \immediate\write\streamout{\figurenum=\the\figurenum}%
     \immediate\write\streamout{\subfigurenum=\the\subfigurenum}%
     \immediate\write\streamout{\tablenum=\the\tablenum}%
     \immediate\write\streamout{\subtablenum=\the\subtablenum}%
     \immediate\closeout\streamout
     \fi}

%************************************************************
%*
%*             Error messages
%*
%************************************************************

\def\err@badtypesize{%
     \errhelp={The limited availability of certain fonts requires^^J%
          that the base type size be 10pt, 12pt, or 14pt.^^J}%
     \errmessage{--> Illegal base type size}}

\def\err@badsizechange{\immediate\write\sixt@@n
     {--> Size change not allowed in math mode, ignored}}

\def\err@sizetoolarge#1{\immediate\write\sixt@@n
     {--> \noexpand#1 too big, substituting HUGE}}

\def\err@sizenotavailable#1{\immediate\write\sixt@@n
     {--> Size not available, \noexpand#1 ignored}}

\def\err@fontnotavailable#1{\immediate\write\sixt@@n
     {--> Font not available, \noexpand#1 ignored}}

\def\err@sltoit{\immediate\write\sixt@@n
     {--> Style \noexpand\sl not available, substituting \noexpand\it}%
     \it}

\def\err@bfstobf{\immediate\write\sixt@@n
     {--> Style \noexpand\bfs not available, substituting \noexpand\bf}%
     \bf}

\def\err@badgroup#1#2{%
     \errhelp={The block you have just tried to close was not the one^^J%
          most recently opened.^^J}%
     \errmessage{--> \noexpand\end{#1} doesn't match \noexpand\begin{#2}}}

\def\err@badcountervalue#1{\immediate\write\sixt@@n
     {--> Counter (#1) out of bounds}}

\def\err@extrafootnotemark{\immediate\write\sixt@@n
     {--> \noexpand\footnotemark command
          has no corresponding \noexpand\footnotetext}}

\def\err@extrafootnotetext{%
     \errhelp{You have given a \noexpand\footnotetext command without first
          specifying^^Ja \noexpand\footnotemark.^^J}%
     \errmessage{--> \noexpand\footnotetext command has no corresponding
          \noexpand\footnotemark}}

\def\err@labelredef#1{\immediate\write\sixt@@n
     {--> Label "#1" redefined}}

\def\err@badlabelmatch#1{\immediate\write\sixt@@n
     {--> Definition of label "#1" doesn't match value in \jobname.lab}}

\def\err@needlabel#1{\immediate\write\sixt@@n
     {--> Label "#1" cited before its definition}}

\def\err@undefinedlabel#1{\immediate\write\sixt@@n
     {--> Label "#1" cited but never defined}}

\def\err@undefinedeqn#1{\immediate\write\sixt@@n
     {--> Equation "#1" not defined}}

\def\err@undefinedref#1{\immediate\write\sixt@@n
     {--> Reference "#1" not defined}}

\def\err@nostream#1{%
     \errhelp={You have tried to input a stream file that doesn't exist.^^J}%
     \errmessage{--> Stream file #1 not found}}

%************************************************************
%*
%*             Initialization
%*
%************************************************************
\message{jyTeX initialization}

\everyjob{\immediate\write16{--> jyTeX version \fmtversion}%
     \edef\@@jobname{\jobname}%
%     \openin0=\inputpath jysupp
%     \ifeof0
%     \else\closein0
%          \immediate\write16{--> Additional macros loaded from jysupp.tex}%
%          \jyinput jysupp
%     \fi
%     \openin0=\inputpath jylocal
%     \ifeof0
%     \else\closein0
%          \immediate\write16{--> Additional macros loaded from jylocal.tex}%
%          \jyinput jylocal
%     \fi
     \edef\jobname{\@@jobname}%
     \settime
     \openin0=\jobname.lab
     \ifeof0
     \else\closein0
          \immediate\write16{--> Getting labels from file \jobname.lab}%
          \input\jobname.lab
     \fi}

%************** Spacing *************************************

\def\fixedskipslist{%
     \^^\{\topskip}%
     \^^\{\splittopskip}%
     \^^\{\maxdepth}%
     \^^\{\skip\topins}%
     \^^\{\skip\footins}%
     \^^\{\headskip}%
     \^^\{\footskip}}

\def\scalingskipslist{%
     \^^\{\p@renwd}%
     \^^\{\delimitershortfall}%
     \^^\{\nulldelimiterspace}%
     \^^\{\scriptspace}%
     \^^\{\jot}%
     \^^\{\normalbaselineskip}%
     \^^\{\normallineskip}%
     \^^\{\normallineskiplimit}%
     \^^\{\baselineskip}%
     \^^\{\lineskip}%
     \^^\{\lineskiplimit}%
     \^^\{\bigskipamount}%
     \^^\{\medskipamount}%
     \^^\{\smallskipamount}%
     \^^\{\parskip}%
     \^^\{\parindent}%
     \^^\{\abovedisplayskip}%
     \^^\{\belowdisplayskip}%
     \^^\{\abovedisplayshortskip}%
     \^^\{\belowdisplayshortskip}%
     \^^\{\abovechapterskip}%
     \^^\{\belowchapterskip}%
     \^^\{\abovesectionskip}%
     \^^\{\belowsectionskip}%
     \^^\{\abovesubsectionskip}%
     \^^\{\belowsubsectionskip}}

%************** Document layout *****************************

\def\twoupsetup{%                                % setup for twoup style
     \topmargin=.75in
     \leftmargin=.5in
     \vsize=6.9in
     \hsize=4.75in
     \fullhsize=10in
     \let\draft=\relax}

\outputstyle{normal}                             % page style

\def\marginnoteformat{\subscriptsize             % paragraphing of margin notes
     \hsize=1in \baselinestretch=1000 \everypar={}%
     \tolerance=5000 \hbadness=5000 \parskip=0pt \parindent=0pt
     \leftskip=0pt \rightskip=0pt \raggedright}

\head={\ifdraft\normalfonts\it\hfil DRAFT\hfil   % format of headline
     \llap{\number\day\ \monthword\month\ \militarytime}\else\hfil\fi}
\foot={\hfil\normalfonts\numstyle\pagenum\hfil}  % format of footline

\normalbaselineskip=12pt                         % usual \baselineskip
\normallineskip=0pt                              % usual \lineskip
\normallineskiplimit=0pt                         % usual \lineskiplimit
\normalbaselines                                 % set \baselineskip

\topskip=.85\baselineskip
\splittopskip=\topskip
\headskip=2\baselineskip
\footskip=\headskip

\pagenumstyle{arabic}                            % counter style

\parskip=0pt                                     % no skip between paragraphs
\parindent=20pt                                  % usual \parindent

\baselinestretch=1000                            % set \big-, \med-, \smallskip

%************** Sectioning **********************************

\chapterstyle{left}                              % position of heading
\chapternumstyle{blank}                          % counter style
\def\chapterbreak{\newpage}                      % break before heading
\abovechapterskip=0pt                            % space before heading
\belowchapterskip=1.5\baselineskip               % space after heading
     plus.38\baselineskip minus.38\baselineskip
\def\chapternumformat{\numstyle\chapternum.}     % format of heading counter

\sectionstyle{left}                              % position of heading
\sectionnumstyle{blank}                          % counter style
\def\sectionbreak{\vskip0pt plus4\baselineskip\penalty-100
     \vskip0pt plus-4\baselineskip}              % break before heading
\abovesectionskip=1.5\baselineskip               % space before heading
     plus.38\baselineskip minus.38\baselineskip
\belowsectionskip=\the\baselineskip              % space after heading
     plus.25\baselineskip minus.25\baselineskip
\def\sectionnumformat{%                          % format of heading counter
     \ifblank\chapternumstyle\then\else\numstyle\chapternum.\fi
     \numstyle\sectionnum.}

\subsectionstyle{left}                           % position of heading
\subsectionnumstyle{blank}                       % counter style
\def\subsectionbreak{\vskip0pt plus4\baselineskip\penalty-100
     \vskip0pt plus-4\baselineskip}              % break before heading
\abovesubsectionskip=\the\baselineskip           % space before heading
     plus.25\baselineskip minus.25\baselineskip
\belowsubsectionskip=.75\baselineskip            % space after heading
     plus.19\baselineskip minus.19\baselineskip
\def\subsectionnumformat{%                       % format of heading counter
     \ifblank\chapternumstyle\then\else\numstyle\chapternum.\fi
     \ifblank\sectionnumstyle\then\else\numstyle\sectionnum.\fi
     \numstyle\subsectionnum.}

%************** Footnotes ***********************************

\footnotenumstyle{symbols}                       % counter style
\footnoteskip=0pt                                % jyTeX spacing parameter
\def\footnotenumformat{\numstyle\footnotenum}    % \footnotemark format
\def\footnoteformat{\footnotesize                % paragraphing of text
     \everypar={}\parskip=0pt \parfillskip=0pt plus1fil
     \leftskip=1em \rightskip=0pt
     \spaceskip=0pt \xspaceskip=0pt
     \def\\{\ifhmode\ifnum\lastpenalty=-10000
          \else\hfil\penalty-10000 \fi\fi\ignorespaces}}

%************** Labels **************************************

\def\undefinedlabelformat{$\bullet$}             % mark for undefined label

%************** Equation numbering **************************

\equationnumstyle{arabic}                        % counter style
\subequationnumstyle{blank}                      % counter style
\figurenumstyle{arabic}                          % counter style
\subfigurenumstyle{blank}                        % counter style
\tablenumstyle{arabic}                           % counter style
\subtablenumstyle{blank}                         % counter style

\eqnseriesstyle{alphabetic}                      % sub-counter style for series
\figseriesstyle{alphabetic}                      % sub-counter style for series
\tblseriesstyle{alphabetic}                      % sub-counter style for series

\def\puteqnformat{\hbox{%                        % equation number format
     \ifblank\chapternumstyle\then\else\numstyle\chapternum.\fi
     \ifblank\sectionnumstyle\then\else\numstyle\sectionnum.\fi
     \ifblank\subsectionnumstyle\then\else\numstyle\subsectionnum.\fi
     \numstyle\equationnum
     \numstyle\subequationnum}}
\def\putfigformat{\hbox{%                        % figure number format
     \ifblank\chapternumstyle\then\else\numstyle\chapternum.\fi
     \ifblank\sectionnumstyle\then\else\numstyle\sectionnum.\fi
     \ifblank\subsectionnumstyle\then\else\numstyle\subsectionnum.\fi
     \numstyle\figurenum
     \numstyle\subfigurenum}}
\def\puttblformat{\hbox{%                        % table number format
     \ifblank\chapternumstyle\then\else\numstyle\chapternum.\fi
     \ifblank\sectionnumstyle\then\else\numstyle\sectionnum.\fi
     \ifblank\subsectionnumstyle\then\else\numstyle\subsectionnum.\fi
     \numstyle\tablenum
     \numstyle\subtablenum}}

%************** Reference numbering *************************

\referencestyle{sequential}                      % referencing method
\referencenumstyle{arabic}                       % counter style
\def\putrefformat{\numstyle\referencenum}        % format of reference citation
\def\referencenumformat{\numstyle\referencenum.} % format of number in list
\def\putreferenceformat{%                        % paragraphing of list
     \everypar={\hangindent=1em \hangafter=1 }%
     \def\\{\hfil\break\null\hskip-1em \ignorespaces}%
     \leftskip=\refnumindent\parindent=0pt \interlinepenalty=1000 }

%************** Font initialization *************************

\normalsize

%******************************************************************************

\def\fmtversion{2.6M (June 1992)}

\catcode`\@=12
\def\upref#1/{\markup{[\putref{#1}]}}

\typesize=12pt
%\sectionnumstyle{arabic}
%\subsectionnumstyle{arabic}

\footnoteskip=2pt
\footnotenumstyle{arabic}
%%%%%%%%%%%%%%%%%%%%%%%%%%%%%%%%%%%%%%%%%%%%%%%%%%%%%%%%%%%%%%%%%%%%%
%
% Finally some useful macros for journals, ...
%

\typesize=12pt
\smallsize
\sectionnumstyle{arabic}
%\draft
\def\csprop{\langle x, t_2 | y, t_1 \rangle }
\def\psprop{\langle x | \exp \left( - {i\over\hbar} \hat H \Delta t
\right) | p \rangle }
\def\Sco{S_{\rm conf}}
\def\Scl{S_{\rm cl}}
\def\Sxy{\Scl (x,y;t)}
\def\cspro0{\langle x,t | y, t{=}0 \rangle }
\def\unit{1 \hskip-.3em \raise2pt\hbox{$ \scriptstyle |$ } }
\def\procoh{\langle w^*, t{=}T | z, t{=}0 \rangle }
\def\Tr{{\rm Tr}}
\def\tr{{\rm tr}}
\def\Ampl{\langle x | \exp \left( - {\epsilon\over\hbar} \hat H \right)
| y \rangle }
\def\MatElt{ \langle x | \exp \left( - { \epsilon \over \hbar } \hat H
\right) | p \rangle }
\def\Dsl{D \hskip-.6em \raise1pt\hbox{$ / $ } }
\def\SAmpl{ \langle x,\bar\eta | \exp \left( - { \epsilon \over
\hbar } \hat H \right) | y,\chi \rangle }
\def\Samhis1{ \langle x,\bar\eta | \exp \left( -  \epsilon \hat H
\right) | y,\chi \rangle }
\def\SMat{ \langle x,\bar\eta | \exp \left( - { \epsilon \over
\hbar } \hat H \right) | p,\xi \rangle }
\def\Dsl{D \hskip-.6em \raise1pt\hbox{$ / $ } }

\equation{hameq}{{\partial\over\partial x} \Sxy = p(t) \quad ; \quad
{\partial\over\partial y} \Sxy = - p(0) }

\equation{defU}{\cspro0 = \exp \left( {i\over\hbar} U(x,y;t) \right)}

\equation{schr}{\eqalign{\langle x,t | \hat p(t) | y, t{=}0 \rangle &=
\int\! dx^\prime \, \langle x,t | \hat p (t) | x^\prime ,t \rangle \,
\langle x^\prime, t | y, t{=}0 \rangle = - i \hbar
{\partial\over\partial x } \cspro0 \cr &= \left(
{\partial\over\partial x} U(x,y;t) \right) \cspro0 }}

\equation{defUhat}{ \langle x,t | \hat U \bigl( \hat x (t), \hat x(0)
\bigr) | y,t{=}0 \rangle = U(x,y;t) }

\equation{hamopeq}{\hat p(t) = {\partial\over\partial \hat x(t)} \hat
U \bigl(\hat x(t),\hat x(0) \bigr) \quad ; \quad \hat p(0) = -
{\partial \over \partial \hat x(0)} \hat U \bigl( \hat x(t), \hat x(0)
\bigr) }

\equation{compl}{\cspro0 = \int \! dx_1 \ldots dx_{N-1} \, \langle x,t
| x_{N-1}, t_{N-1} \rangle \langle x_{N-1},t_{N-1} | x_{N-2},t_{N-2}
\rangle \ldots \langle x_1,t_1 | y, t{=}0 \rangle }

\equation{Hferm}{\hat H = {1\over 2} g^{-1/4}(\hat x) \hat\pi_i
g^{1/2}(\hat x)
g^{ij}(\hat x) \hat \pi_j g^{-1/4}(\hat x) - {1\over 8}
R_{abcd}(\omega(\hat x)) \bigl(
\hat\psi^a_\alpha \hat\psi^b_\alpha \bigr) \bigl( \hat\psi^c_\beta
\hat\psi^d_\beta \bigr) }

\equation{defcovD}{\hat \pi_i = \hat p_i + {1\over 2i}
\omega_{iab}(\hat x)
\hat\psi^a_\alpha \hat\psi^b_\alpha \quad ; \quad \bigl\{
\hat\psi^a_\alpha , \hat\psi^b_\beta \bigr\} = \hbar \delta^{ab}
\delta_{\alpha\beta} \quad ; \quad \bigl[\hat p_i, \hat x^j \bigr] = - i\hbar
\delta_i^j }

\equation{regan}{{\cal A} = \lim_{M^2\rightarrow\infty} \Tr \hat J
\exp \left( {\hat {\cal R} \over M^2 } \right) }

\equation{ratfunc}{{F\bigl[ V[x] \bigr] \over F\bigl[ V=0 \bigr]} =
\lim_{N\rightarrow\infty} {F\bigr[ V,N \bigr] \over F \bigl[ 0,N
\bigr] }}

\equation{tdisc}{F\bigl[ V,N \bigr] = \int \! dx_1\ldots dx_N \exp
\left( \sum_{j=0}^{N-1} \left[ - {1\over 2} m \bigl( x_{j+1} - x_j
\bigr)^2 \epsilon^{-1} + V(x_j) \epsilon \right] \right) }

\equation{Trotf}{\exp\left( -{i\over\hbar} H \Delta t \right) =
\exp\left( -{i\over\hbar} T \Delta t \right) \exp\left( -{i\over\hbar}
V \Delta t \right) + {\cal O} (\Delta t)^2 }

\equation{intpj}{\eqalign{\langle x_{j+1}, t_{j+1} | x_j,t_j \rangle
&= \sum_{p_{j+1}} \langle x_{j+1} | \exp \left( -{i\over\hbar}
T \Delta t \right) | p_{j+1} \rangle \langle p_{j+1} | \exp \left(
-{i\over\hbar} V \Delta t \right) | x_j \rangle \cr &= \int
{d^n p_{j+1}\over (2\pi\hbar)^n} \, \exp \left( {i\over\hbar}
(x_{j+1}-x_j)p_{j+1} \right) \exp \left( -{i\over\hbar}
T(p_{j+1}) \Delta t \right) \exp \left( -{i\over\hbar}
V(x_j) \Delta t \right) \cr }}

\equation{cspi}{\langle x_{j+1},t_{j+1} | x_j,t_j \rangle = \left(
{m\over 2\pi i \hbar \Delta t} \right)^{n/2} \exp \left( {i\over\hbar}
S(x_{j+1},x_j;t_{j+1}-t_j) \right) + {\cal O}(\Delta t)^2 }

\equation{pspi}{\langle x_j | \exp \left( - {i\over\hbar} \hat H
\Delta t \right) | p_j \rangle = \langle x_j | 1 -
{i\over\hbar} \hat H \Delta t | p_j \rangle }

\equation{rescpi}{\eqalign{\langle x | \exp \left( - {i\over\hbar}
\hat H \Delta t \right) | y \rangle &= (2\pi\hbar)^{-n}
g^{-1/4}(x) g^{-1/4}(y)
\left( {\hbar\over\epsilon} \right)^{n/2} \int\!d^nq \, \exp \left(
{i\over (\epsilon\hbar)^{1/2}} q_j (x^j-y^j) \right) \cr & \qquad \exp
\left( - {i\over 2m} q_i g^{ij}(x) q_j \right) \left[ 1 +
\sqrt{\epsilon\hbar} A(q,x) + \epsilon\hbar B(q,x) + \ldots \right]
\cr}}

\equation{inpr}{\langle x | p \rangle = (2\pi \hbar)^{-n/2} g^{-1/4}(x)
\exp \left( {i\over\hbar} p_i x^i \right) }

\equation{phsppi}{\cspro0 = \lim_{N\rightarrow\infty}
\int \! {dp_N dx_{N-1} \ldots dx_1 dp_1\over (2\pi\hbar)^{N}} \, \exp
\left( {i\over\hbar}
\sum_{j=0}^{N-1} \left[ p_{j+1}(x_{j+1}-x_j)\epsilon^{-1} - H(x_j,p_j)
\epsilon \right] \right) }

\equation{defcoh}{|z\rangle = e^{a^\dagger z} |0\rangle \quad ;
\quad \langle z^* | = \langle 0 | e^{z^* a} \quad ; \quad
[a,a^\dagger]=1 }

\equation{inpcoh}{\langle w^* | z \rangle = e^{w^* z} \quad ;
\quad \unit = \int\! {dz dz^* \over 2\pi i} \, |z\rangle
e^{- z^* z} \langle z^* | }

\equation{defprocoh}{\eqalign{\procoh &= \int \! \prod_{j=1}^{N-1} {dz_j
dz_j^* \over 2\pi i} \, \langle w^* | \exp\left(-{i\over\hbar} \hat H
\Delta t \right) | z_{N-1} \rangle \, e^{-z_{N-1}^* z_{N-1}} \cr &
\qquad \langle
z_{N-1}^* | \exp\left( -{i\over\hbar} \hat H \Delta t \right) | z_{N-2}
\rangle \ldots
e^{-z_1^* z_1}\, \langle z_1^* | \exp\left( -{i\over\hbar} \hat H
\Delta t \right) | z \rangle }}

\equation{defscrham}{\langle z_{j+1}^* | \exp \left( - {i\over\hbar}
\hat H
\Delta t \right) | z_j \rangle = \exp \left( - {i\over\hbar}
h(z_{j+1}^*,z_j) \Delta t \right) \langle z^*_{j+1} | z_j \rangle }

\equation{intprop}{\exp\left( w^* z_{N-1} - z_{N-1}^* z_{N-1} +
z_{N-1}^* z_{N-2} \ldots + z_1^* z - {i\over\hbar} \sum_{j=0}^{N-1}
h(z_{j+1}^*,z_j) \Delta t \right) }

\equation{trace}{\Tr \exp\left( - {i\over\hbar} \hat H \Delta t \right)
= \int\! {dw^* dz\over 2\pi i} \, e^{-w^*z} \,\langle
w^*,\Delta t | z,t{=}0 \rangle }

\equation{pipbc}{\exp\left( \int\! dt\, \left[ -z^* \dot z -
{i\over\hbar} h(z^*,z) \right] \right) }

\equation{pathcoh}{e^{w^* z_{\rm cl}(T)} \, \int \! {Dz^*
Dz\over 2\pi i} \,
\exp \left( \int\! dt \left[ - z^* \dot z - {i\over\hbar}
h(z^*,z) \right] \right) }

\equation{pathcohsym}{e^{{1\over 2}\left[ w^* z_{\rm cl}(T) +
w^*_{\rm cl}(0) z \right]}
\, \int \! {Dz^* Dz\over 2\pi i} \,
\exp \left( \int\! dt \left[ - {1\over 2} \left( z^* \dot z - \dot z^*
z \right) - {i\over\hbar} h(z^*,z) \right] \right) }

\equation{qufi}{z_{\rm qu}(0)= z_{\rm qu}(T)= z^*_{\rm qu}(0)=
z^*_{\rm qu}(T)= 0 }

\equation{xpcl}{w^*_{\rm cl}(t) = (x_{\rm cl}(t) - i p_{\rm cl}(t)
)/\sqrt{2\hbar} \quad ; \quad z_{\rm cl}(t) = ( x_{\rm cl}(t) + i p_{\rm
cl}(t) )/\sqrt{2\hbar} }

\equation{defxpqu}{x_{\rm qu}(t) = x(t) + {1\over 2} \left( z_{\rm
cl}(t) + w^*_{\rm cl}(t) \right) \quad ; \quad p_{\rm qu}(t) = p(t) +
{1\over 2i} \left( z_{\rm cl}(t) - w^*_{\rm cl}(t) \right) }

\equation{meas}{\eqalign{ Dx_{\rm qu} Dp_{\rm qu} \quad ; \qquad &
x_{\rm qu}(t_j) \in [ - \infty + a_j, + \infty + a_j ] \cr & p_{\rm
qu}(t_j) \in [ -\infty + b_j , + \infty + b_j ] \cr }}

\equation{newmeas}{ Dx_{\rm qu} Dp_{\rm qu} \quad ; \qquad - \infty
\leq x_{\rm qu} \leq \infty, \quad -\infty \leq p_{\rm qu} \leq \infty
}

\equation{cohfer}{|\eta\rangle = e^{\hat\psi^\dagger \eta} |0\rangle \quad
; \quad \langle \bar\eta | = \langle 0 | e^{\bar\eta \hat\psi } }

\equation{inpcofer}{\langle \bar\eta | \xi \rangle = e^{\bar\eta \xi}
\quad ; \quad \unit = \int \! d \bar\eta d\xi \, | \xi\rangle
e^{-\bar\eta \xi} \langle \bar\eta| }

\equation{propcohfer}{e^{\bar\eta \xi_{\rm cl}(T)} \int\! D\bar\eta D\xi \,
\exp\left( \int\limits_0^T\! dt \left[ - \bar\eta \dot\xi -
{i\over\hbar} h(\bar\eta,\xi) \right] \right) }

\equation{prchfersym}{e^{{1\over 2}\left[ \bar\eta \xi_{\rm cl}(T) +
\bar\eta_{\rm cl}(0) \xi \right]} \int\! D\bar\eta D\xi \, \exp\left(
\int\limits_0^T\! dt
\left[ - {1\over 2} (\bar\eta \dot\xi - \dot{\bar\eta} \xi ) -
{i\over\hbar} h(\bar\eta,\xi) \right] \right) }

\equation{fertra}{\Tr \hat A = \langle 0 | \hat A | 0 \rangle +
\langle 1 | \hat A | 1 \rangle = \int\! d\xi d\bar\eta \, e^{\bar\eta
\xi} \langle \bar\eta | \hat A | \xi \rangle }

\equation{traprofer}{\int\! d\xi d\bar\eta \prod_{j=1}^{N-1}
d\bar\eta_j d\xi_j \, \exp\left( \bar\eta \xi + \bar\eta \xi_{N-1} -
\bar\eta_{N-1} \xi_{N-1} \ldots - \bar\eta_1 \xi_1 + \bar\eta_1 \xi -
{i\over\hbar} \sum_{j=0}^{N-1} h(\bar\eta_{j+1},\xi_j) \Delta t \right) }

\equation{Matrelt}{\MatElt}

\equation{defham}{\hat H = {1\over 2}g^{-1/4}\hat p_i g^{1/2}g^{ij} \hat
p_j g^{-1/4} - {1\over 2} \xi \hbar^2 R }

\equation{defAkl}{ \langle x | \left( \hat H \right)^k | p \rangle
\equiv \sum_{l=0}^{2k} A_l^k(x) \, p^l \, \langle x | p \rangle }

\equation{A2k}{ A^k_{2k}(x) p^{2k} = \left( {1\over 2} p^2 \right)^k }

\equation{A2kmin1}{\eqalign{ A^k_{2k-1}(x) p^{2k-1} &= - i \hbar k
{1\over 2} \left( {1\over 2} p^2
\right)^{k-1} \left( \partial_i g^{ij} \right) p_j \cr & \qquad
\qquad- i \hbar {k\choose 2}
\left( {1\over 2} p^2 \right)^{k-2} {1\over 2} g^{ij} \left(
\partial_i g^{kl} \right) p_j p_k p_l \cr }}

\equation{A2kmin2}{\eqalign{A^k_{2k-2}&(x) p^{2k-2} = \cr &\hbar^2 k
\left( {1\over 2} p^2
\right)^{k-1} \left[ {1\over 32} g^{ij} \left( \partial_i \log g
\right) \left( \partial_j \log g \right) + {1\over 8} g^{ij} \left(
\partial_i \partial_j \log g \right) + {1\over 8} \left( \partial_i
g^{ij} \right) \left( \partial_j \log g \right) \right] \cr - &
\hbar^2 {k\choose 2} \left( {1\over 2} p^2 \right)^{k-2} \left[
{1\over 2} g^{ij} \left(
\partial_i \partial_k g^{kl} \right) + {1\over 4}\Bigl( \partial_i
g^{ij}
\Bigr) \left( \partial_k g^{kl} \right) \right. \cr & \left. \qquad
\qquad \qquad \qquad \qquad + {1\over 4}\left( \partial_i g^{ik}
\right) \left( \partial_k g^{jl} \right) + {1\over 4}g^{ik} \left( \partial_i
\partial_k g^{jl} \right) \right] p_j p_l \cr  - &
\hbar^2 {k\choose 3} \left( {1\over 2}p^2 \right)^{k-3} \left[ {1\over
2} g^{ik} g^{jl}
\Bigl( \partial_i \partial_j g^{mn}  \Bigr) + {3\over 4} g^{im} \left(
\partial_i g^{kl} \right) \Bigl( \partial_j g^{jn} \Bigr) \right. \cr
& \left. \qquad \qquad \qquad \qquad \qquad + {1\over 2} g^{jl}
\left( \partial_j g^{ik} \right) \Bigl( \partial_i g^{mn} \Bigr) +
{1\over 4} g^{ij} \left( \partial_i g^{kl} \right) \Bigl( \partial_j
g^{mn} \Bigr) \right] p_k p_l p_m p_n \cr
-& \hbar^2 {k\choose 4} \left( {1\over 2}p^2 \right)^{k-4} \left[ {3\over 4}
g^{ij}
g^{mn} \left( \partial_i g^{kl} \right) \Bigl( \partial_n g^{pq}
\Bigr) \right] p_j p_k p_l p_m p_p p_q \quad
- \quad {1\over 2} \xi \hbar^2 k \left( {1\over 2} p^2 \right)^{k-1} R \cr}}

\equation{ordEps}{\eqalign{\Ampl &= \int\! d^n p\,\langle x | \exp \left( -
{\epsilon\over\hbar} \hat H \right) | p \rangle \langle p | y \rangle
\cr & \hskip-3em
= g^{-1/4}(x)g^{-1/4}(y) (2\pi\hbar)^{-n} \left( {\hbar\over\epsilon}
\right)^{n/2} \int\! d^n q \,\exp \left( i { q_i (x-y)^i \over
\sqrt{\epsilon\hbar} } \right) \cr & \qquad \qquad
\sum_{k=0}^{\infty} { (-1)^k \over k! }
\left( {\epsilon \over \hbar } \right)^k \sum_{l=0}^{2k} A^k_l(x) q^l
\left( {\epsilon \over \hbar } \right)^{-l/2} \cr }}

\equation{momint}{\eqalign{\langle x | \exp \Bigl(  &-
{\epsilon\over\hbar} \hat H \Bigr) | y \rangle
 = g^{-1/4}(x) g^{-1/4}(y) \left( 4 \pi^2 \hbar
\epsilon \right )^{-n/2} \int\! d^n q\,\exp \left( -  {1\over 2}
g^{ij} q_i q_j
+ i { q_i (x-y)^i \over \sqrt{ \epsilon\hbar } } \right) \cr
 \Biggl[  1 \quad &+ \quad
i \sqrt{\epsilon\hbar } \left\{  {1\over 2} \Bigl( \partial_i g^{ij}
\Bigr) q_j -  {1\over 4}
g^{ij} \left( \partial_i g^{kl} \right) q_j q_k q_l \right\}
\cr \noalign{\break}  +
\epsilon \hbar \Biggl\{ &\left[ {1\over 2} \xi R - {1\over 32} g^{ij}
\Bigl(
\partial_i \log g  \Bigr) \Bigl( \partial_j \log g \Bigr) - {1\over
8} g^{ij} \Bigl( \partial_i \partial_j \log g \Bigr) - {1\over 8}
\Bigl( \partial_i  g^{ij} \Bigr) \Bigl( \partial_j \log g \Bigr)
\right]    \cr
& - \left[ {1\over 4} g^{ij} \left(  \partial_i \partial_k g^{kl} \right)
+ {1\over 8} \Bigl( \partial_i g^{ij}  \Bigr) \left( \partial_k
g^{kl} \right) + {1\over 8} \left( \partial_i g^{ik}  \right) \left(
\partial_k g^{jl} \right) + {1\over 8} g^{ik} \left( \partial_i
\partial_k g^{jl} \right) \right] q_j q_l   \cr  & +
 \left[ {1\over 12} g^{ik} g^{jl}
\Bigl( \partial_i \partial_j g^{mn}  \Bigr) +  {1\over 8}g^{im} \left(
\partial_i g^{kl} \right) \Bigl( \partial_j g^{jn} \Bigr)
 \right. \cr & \qquad \qquad \qquad \qquad
+ \left. {1\over 12} g^{jl}
\left( \partial_j g^{ik} \right) \Bigl( \partial_i g^{mn} \Bigr) +
{1\over 24} g^{ij} \left( \partial_i g^{kl} \right) \Bigl( \partial_j
g^{mn} \Bigr) \right] q_k q_l q_m q_n  \cr
 &- \left[ {1\over 32} g^{ij}
g^{mn} \left( \partial_i g^{kl} \right) \Bigl( \partial_n g^{pq}
\Bigr) \right] q_j q_k q_l q_m q_p q_q  \Biggr\} \quad + \quad {\cal
O} \bigl( \epsilon^{3/2} \bigr) \quad \Biggr] \cr}}

\equation{factTra}{\eqalign{\langle x | \exp \Bigl(  &-
{\epsilon\over\hbar} \hat H \Bigr) | y \rangle
 = ( 2 \pi \hbar \epsilon )^{-n/2} g^{-1/4}(x)
g^{-1/4}(y) \cr &\left[ g^{1/2}(x) + g^{1/4}(x) (y-x)^i \left( \partial_i
g^{1/4}(x) \right) + {1\over 2} g^{1/4}(x) (y-x)^i (y-x)^j \left(
\partial_i \partial_j g^{1/4}(x) \right) \right] \cr &
\exp \left( - {1 \over 2 \epsilon \hbar } g_{ij}(x) (y-x)^i (y-x)^j
\right)
\Biggl[ 1 - {1\over 4} {1\over \epsilon\hbar } \left( \partial_k
 g_{ij}(x) \right) (y-x)^i (y-x)^j (y-x)^k \cr &+ {1 \over 2 } \left(
{1\over 4} {1\over \epsilon\hbar } \left( \partial_k
g_{ij}(x) \right) (y-x)^i (y-x)^j (y-x)^k \right)^2 \cr &
- {1\over 12 } {1\over \epsilon \hbar } \left( \partial_k \partial_l
g_{ij}(x) - {1\over 2} g_{mn}(x) \Gamma^m_{ij}(x) \Gamma^n_{kl}(x)
\right) (y-x)^i (y-x)^j (y-x)^k (y-x)^l \cr &+ \left(  {1\over 2} \xi
- {1 \over 12}
\right) \epsilon \hbar R(x) - {1\over 12} R_{ij}(x) (y-x)^i (y-x)^j
 \quad + \quad {\cal
O} \bigl( \epsilon^{3/2} \bigr) \quad \Biggr] \cr}}

\equation{infampl}{\eqalign{\Ampl &= (2\pi\hbar\epsilon )^{-n/2}
\exp\left\{ -
{\epsilon\over\hbar} \left[ {1\over 2} g_{ij}(x) + {1\over 4}
\partial_k g_{ij}(x) (y-x)^k \right. \right. \cr & \left. \left.
\hskip-4em + {1\over 12} \left( \partial_k
\partial_l g_{ij}(x) - {1\over 2} g_{mn}(x) \Gamma^m_{ij}(x)
\Gamma^n_{kl}(x) \right) (y-x)^k (y-x)^l \right] { (y-x)^i \over
\epsilon } { (y-x)^j \over \epsilon } \right. \cr & \left.
 \hskip-2em +
\epsilon\hbar \left( {1\over 2} \xi - {1\over 12} \right) R(x) -
{1\over 12} R_{ij}(x) (y-x)^i (y-x)^j  \quad + \quad {\cal
O} \bigl( \epsilon^{3/2} \bigr) \quad \right\} \cr }}

\equation{expAct}{\eqalign{S_{{\rm cl}}&(x,y;\epsilon) =
\int_{t=-\epsilon}^{t=0} \! dt \, {1\over 2} g_{ij}(x(t))\dot x^i(t)
\dot x^j(t)  \cr &= {1\over\epsilon} \Biggl[
{1\over 2} g_{ij}(x)(y-x)^i(y-x)^j + {1\over 4} \partial_k g_{ij}(x)
(y-x)^i(y-x)^j(y-x)^k \cr & \qquad +
{1\over 12} \Bigl( \partial_k \partial_l g_{ij}(x) - {1\over 2}
g_{mn}(x) \Gamma^m_{ij}(x) \Gamma^n_{kl}(x)
\Bigr)(y-x)^i(y-x)^j(y-x)^k (y-x)^l \Biggr] +
{1\over \epsilon} {\cal O}(y-x)^5 \cr }}

\equation{final}{\eqalign{\Ampl &= (2\pi\hbar\epsilon)^{-n/2} \exp \left( -
{1 \over \hbar} S_{{\rm cl}} \right) \biggl[ 1 + {1\over 2}
\epsilon\hbar \left( {1\over 2} \xi - {1\over 12} \right) \bigl( R(x)
+ R(y) \bigr) \cr & \qquad - {1\over 24} \left( R_{ij}(x) +
R_{ij}(y) \right) (x-y)^i (x-y)^j + {\cal O} \bigl( \epsilon^{3/2}
\bigr) \biggr] \cr }}

\equation{comp}{\int\!dz\, \sqrt{g(z)} \langle x | \exp \left( -
{\epsilon\over\hbar} \hat H \right) | z \rangle \langle z | \exp
\left( - {\epsilon\over\hbar} \hat H \right) | y \rangle = \langle x |
\exp \left( - {2\epsilon\over\hbar} \hat H \right) | y \rangle}

\equation{HWeyl}{\hat H_{\rm Weyl} = \hat H - {1\over 8} \hbar^2 R }

\equation{defD}{D^{1/2}(x,y;\epsilon) = \epsilon^{-n/2} g^{1/4}(x)
g^{1/4}(y) \left[ 1 - {1\over 12} R_{ij} (y-x)^i (y-x)^j \right] }

\equation{defDij}{D_{ij}(x,y;\epsilon) \equiv - {\partial\over
\partial x^i} {\partial \over \partial y^j} S_{{\rm cl}}(x,y;\epsilon)
}

\equation{rewr}{\Ampl = (2\pi\hbar )^{-n/2} \tilde D^{1/2} \exp\left( -
{1\over\hbar} S_{{\rm cl}} \right) \left[ 1 + {1\over 2} \epsilon
\hbar \left( {1\over 2} \xi - {1\over 12} \right) \bigl( R(x) + R(y)
\bigr) \right] }

\equation{Schrod}{\left[ \hat H(y) + \hbar {\partial\over
\partial\epsilon} \right] \theta(\epsilon) \langle x | \exp \left( -
{\epsilon\over\hbar} \hat H \right) | y \rangle = \hbar g^{-1/2}(x)
\delta(\epsilon) \delta^n(x-y) }

\equation{defHamy}{\hat H(y) = - {1\over 2} \hbar^2 g^{-1/2}(y) {\partial
\over \partial y^i } g^{1/2}(y) g^{ij}(y) {\partial\over \partial y^j}
- {1\over 2} \xi \hbar^2 R(y)}

\equation{ddeps}{-\hbar{\partial\over\partial
\epsilon} \langle x| \exp\bigl( -{\epsilon\over\hbar} \hat H \bigr)
y\rangle = \int\! dz \, \sqrt{g(z)} \langle x | \hat H | z\rangle
\langle z| \exp\bigl( -{\epsilon\over\hbar} \hat H \bigr)
y\rangle = \hat H(x) \langle x| \exp\bigl( -{\epsilon\over\hbar} \hat
H \bigr) y\rangle }

\equation{An2}{{\cal A}_2^{scalar} = - {\hbar \over 24 \pi } R }

\equation{Lsusy}{\eqalign{L &= {1\over 2} g_{ij}(\phi)
\dot\phi^i \dot\phi^j + {i\over 2} g_{ij}(\phi) \chi^i_\alpha D_t
\chi^j_\alpha + {1\over 8} R_{ijkl}(\Gamma) \bigl( \chi^i_\alpha
\chi^j_\alpha \bigr) \bigl( \chi^k_\beta \chi_\beta^l \bigr) \cr
& \qquad D_t \chi^j_\alpha = \partial_t \chi_\alpha^j + \dot\phi^k
\Gamma^j_{kl} \chi^l_\alpha \quad ; \qquad \alpha , \beta = 1,2 \cr
&\quad R_{ijk}{}^l(\Gamma) = -\partial_i \Gamma_{jk}^l - \Gamma^l_{in}
\Gamma^n_{jk} - ( i \leftrightarrow j ) \cr }}

\equation{susytra}{\eqalign{ \delta \phi^i &= i \left( \epsilon_1
\chi^i_2 - \epsilon_2 \chi^i_1 \right) = \epsilon^T \sigma_2 \chi^i
\cr \delta \chi^i &= - i \dot \phi^i \sigma_2 \epsilon -
\Gamma^i{}_{jk} \delta \phi^j \chi^k \cr }}

\equation{Qcl}{Q_\alpha^{({\rm cl})} = g_{ij}(\phi) \chi^i_\alpha
\dot\phi^j }

\equation{Lsusyflat}{\eqalign{L &= {1\over 2} g_{ij}(\phi)
\dot\phi^i \dot\phi^j + {i\over 2} \psi^a_\alpha D_t
\psi^a_\alpha + {1\over 8} R_{abcd}(\omega) \bigl( \psi^a_\alpha
\psi^b_\alpha \bigr) \bigl( \psi^c_\beta \psi^d_\beta \bigr) \cr
& \qquad D_t \psi^a_\alpha = \partial_t \psi_\alpha^a + \dot\phi^j
\omega_j{}^a{}_b \psi^b_\alpha \cr &\quad R_{ijab}(\omega) =
\partial_i \omega_{jab} + \omega_{ia}{}^c \omega_{jcb} - (i
\leftrightarrow j ) = R_{ijkl}(\Gamma) e^k_a e^l_b \cr & \qquad
\partial_i e_a^j + \Gamma^j_{ik} e_a^k + \omega_{ia}{}^b e_b^j = 0 \cr
}}

\equation{Qclflat}{Q_\alpha^{({\rm cl})}= e_{ia}(\phi) \psi^a_\alpha
\dot\phi^i }

\equation{constr}{C^a_\alpha = p(\psi^a_\alpha) + {i\over 2} \psi^a_\alpha}

\equation{DiracB}{\eqalign{\left\{ p(\psi^a_\alpha), \psi^b_\beta \right\}_D &=
\left\{ p(\psi^a_\alpha), \psi^b_\beta \right\}_P - \Bigl\{
p(\psi^a_\alpha), C^c_\gamma \Bigr\}_P i \left\{ C^c_\gamma,
\psi^b_\beta \right\}_P \cr &= - {1\over 2} \delta^{ab}
\delta_{\alpha\beta} \cr }}

\equation{Anticom}{\left\{ \psi^a_\alpha , \psi^b_\beta \right\} =
\delta^{ab} \delta_{\alpha\beta} }

\equation{momentum}{p_i = g_{ij} \dot\phi^j + {i\over 2} \omega_{iab}
\psi^a_\alpha \psi^b_\alpha \quad ; \quad \left[ p_i , \phi^j \right]
= -i \delta^j_i }

\equation{Qqu}{Q_\alpha^{({\rm qu})} = e^i_a(\phi) \psi^a_\alpha
g^{1/4} \pi_i g^{-1/4}
\quad ; \quad \pi_i \equiv p_i - {i\over 2} \omega_{iab} \psi^a_\alpha
\psi^b_\alpha }

\equation{commut}{\left[ \pi_i , \psi^a_\alpha \right] = i
\omega_i{}^a{}_b \psi^b_\alpha \quad ; \quad \left[ \pi_i , e_a^j
\right] = -i \left( \partial_i e_a^j \right) }

\equation{QQ}{\left\{ Q_\alpha^{({\rm qu})} , Q_\beta^{({\rm qu})}
\right\}
= g^{1/4} \left[ \delta_{\alpha\beta} \delta^{ab} e^i_a \pi_i e^j_b \pi_j -
\psi^j_\beta \psi^i_\alpha \left[ \pi_i , \pi_j \right] - i \left(
\psi^j_\alpha \psi^b_\beta + \psi^b_\beta \psi^j_\alpha \right)
\omega_{jb}{}^k \pi_k \right] g^{-1/4} }

\equation{pipi}{\left[ \pi_i , \pi_j \right] = - {1\over 2} R_{ijab}
\psi^a_\alpha \psi^b_\alpha }

\equation{cycl}{\eqalign{ & \psi^a_2 \left( \psi^b_1 \psi^c_1 \psi^d_1
+ \psi^c_1 \psi^d_1 \psi^b_1
+ \psi^d_1 \psi^b_1  \psi^c_1 \right) R_{abcd} = 0 \cr & \qquad = 3
\psi^a_2 \left( \psi^b_1 \psi^c_1\psi^d_1 -\delta^{bc} \psi^d_1
\right) R_{abcd} = 3 \psi^a_2 \psi^b_1 \psi^c_1 \psi^d_1 R_{abcd} }}

\equation{QQisHqu}{\eqalign{ & \left\{ Q_\alpha^{({\rm qu})} ,
Q_\beta^{({\rm qu})} \right\} = 2 \delta_{\alpha\beta} H^{({\rm qu})}
\cr &
\qquad H^{({\rm qu})}= {1\over 2} \delta^{ab} g^{1/4} \left(
e^i_a \pi_i
e_b^j \pi_j + e_a^i \omega_{ib}{}^c e_c^j \pi_j \right) g^{-1/4} -
{1\over 8}
R_{abcd}(\omega) \left( \psi^a_\alpha \psi^b_\alpha \right) \left(
\psi^c_\beta \psi^d_\beta \right) \cr }}

\equation{HSWeyl}{{1\over 2} R_{abcd} \left[ \theta(t_a,t_b)
\theta(t_b,t_c) \theta(t_c,t_d) \bar\psi^a \psi^b \bar\psi^c \psi^d -
\theta(t_a,t_b)
\theta(t_b,t_d) \theta(t_d,t_c) \bar\psi^a \psi^b \psi^d \bar\psi^c
+ 22 \; {\rm more} \>\right] }

\equation{1isth}{{1\over 2} R_{abcd}\bar\psi^a \psi^b \bar\psi^c
\psi^d = {1\over 2} R_{abcd}\bar\psi^a \psi^b \bar\psi^c
\psi^d \Bigl[ \theta(t_a,t_b) \theta(t_b,t_c) \theta(t_c,t_d) +
\theta(t_a,t_b) \theta(t_b,t_d) \theta(t_d,t_c) + 22 \; {\rm more} \>
\Bigr] }

\equation{thiden}{\eqalign{\theta(t_a,t_d) \theta(t_c,t_b) &=
\theta(t_a,t_d)
\theta(t_d,t_c) \theta(t_c,t_b) + \theta(t_a,t_c) \theta(t_c,t_d)
\theta(t_d,t_b) + \theta(t_a,t_c) \theta(t_c,t_b) \theta(t_b,t_d)\cr &
\qquad +
\theta(t_c,t_a) \theta(t_a,t_b) \theta(t_b,t_d) + \theta(t_c,t_a)
\theta(t_a,t_d) \theta(t_d,t_b) + \theta(t_c,t_b) \theta(t_b,t_a)
\theta(t_a,t_d) }}

\equation{SMatelt}{\SMat}

\equation{defSham}{\eqalign{\hat H &= {1\over 2} g^{-1/4}
\pi_i g^{1/2} g^{ij} \pi_j g^{-1/4} - {1\over 8}
\hbar^2 R_{abcd} \left( \psi^a_\alpha \psi^b_\alpha \right) \left(
\psi^c_\beta \psi^d_\beta \right) \cr \pi_i &= p_i - {i\hbar\over 2}
\omega_{iab}
\psi^a_\alpha \psi^b_\alpha \quad ; \qquad \alpha = 1,2 \cr }}

\equation{defferm}{\eqalign{& \psi^a = {1\over\sqrt{2}}\bigl( \psi^a_1
+ i \psi^a_2 \bigr) \quad ; \quad \bar\psi^a = {1\over\sqrt{2}}\bigl(
\psi^a_1 - i \psi^a_2 \bigr) \cr & \bigl\{ \psi^a, \bar\psi^b \bigr\}
= \delta^{ab} \quad ; \quad \bigl\{ \psi^a, \psi^b \bigr\} = 0 =
\bigl\{ \bar\psi^a, \bar\psi^b \bigr\} \cr }}

\equation{defBkl}{\langle x,\bar\eta | \bigl( \hat H \bigr)^k | p,\xi
\rangle \equiv \sum_{l=0}^{2k} B^k_l(x,\bar\eta,\xi ) \, p^l \, \langle
x,\bar\eta | p,\xi \rangle }

\equation{Hparts}{\eqalign{\hat\alpha &= {1\over 2} g^{-1/4} \hat p_i
g^{1/2} g^{ij} \hat p_j g^{-1/4} = \hat H_{\rm bos} \cr
\hat\beta &= - i \hbar g^{ij} \omega_{iab} \bar\psi^a \psi^b
g^{1/4} \hat p_j g^{-1/4} \cr
\hat\gamma &= - {1\over 2} \hbar^2 g^{-1/2} \partial_i \bigl( g^{1/2}
g^{ij} \omega_{jab} \bigr) \bar\psi^a \psi^b - {1\over 2} \hbar^2
g^{ij} \omega_{iab} \omega_{jcd} \bar\psi^a \psi^b \bar\psi^c \psi^d -
{1\over 2} \hbar^2 R_{abcd}\bar\psi^a \psi^b \bar\psi^c \psi^d \cr }}

\equation{B2k}{B^k_{2k}(x,\bar\eta,\xi) p^{2k} = A^k_{2k}(x) p^{2k} =
\left( {1\over 2} p^2 \right)^k }

\equation{B2kmin1}{B^k_{2k-1}(x,\bar\eta,\xi) p^{2k-1} = A^k_{2k-1}(x)
p^{2k-1} - i k \hbar \left( {1\over 2} p^2 \right)^{k-1}
g^{ij} \omega_{iab} \bar\eta^a \xi^b p_j }

\equation{B2kmin2}{\eqalign{B^k_{2k-2}(&x,\bar\eta,\xi) p^{2k-2} =
A^k_{2k-2}(x) p^{2k-2} \cr &- i k \hbar A^{k-1}_{2k-3}(x) p^{2k-3}
g^{ij} \omega_{iab} \bar\eta^a \xi^b p_j - {1\over 2} k(k-1)
\hbar^2 \left( {1\over 2} p^2 \right)^{k-2} g^{ij} \partial_i \bigl(
g^{kl} \omega_{kab} \bigr) \bar\eta^a \xi^b p_j p_l \cr & + {1\over 4}
k \hbar^2 \left( {1\over 2} p^2 \right)^{k-1} g^{ij} \bigl( \partial_i
\log g \bigr) \omega_{jab} \bar\eta^a \xi^b - {1\over 4} k(k-1)
\hbar^2 \left( {1\over 2} p^2 \right)^{k-2} g^{ij} \bigl( \partial_i
g^{kl} \bigr) \omega_{jab} \bar\eta^a \xi^b p_k p_l \cr & - {1\over 2}
k \hbar^2 \left( {1\over 2} p^2 \right)^{k-1} \left[ g^{-1/2}
\partial_i \bigl( g^{ij} g^{1/2} \omega_{jab} \bigr) \bar\eta^a \xi^b
+ \bigl( g^{ij} \omega_{iab} \omega_{jcd} + R_{abcd} \bigr) \bigl(
\bar\eta^a \xi^d \delta^{bc} - \bar\eta^a \bar\eta^c \xi^b \xi^d \bigr)
\right]
\cr & - {1\over 2} k(k-1) \hbar^2 \left( {1\over 2} p^2 \right)^{k-2}
g^{ij} \omega_{iab} g^{kl} \omega_{kcd} \bigl( \bar\eta^a \xi^d
\delta^{bc} - \bar\eta^a \bar\eta^c \xi^b \xi^d \bigr) p_j p_l \cr }}

\equation{Smomint}{\SAmpl = \int\! d^np d^n\bar\xi d^n\xi \,
e^{-\bar\xi \xi} \SMat \langle p,\bar\xi | y,\chi \rangle }

\equation{defSinp}{\langle x,\bar\eta | p,\xi \rangle = \langle x | p
\rangle \langle \bar\eta | \xi \rangle = (2\pi\hbar)^{-n/2}
g^{-1/4}(x) e^{{i\over\hbar} px} e^{\bar\eta\xi} }

\equation{STraAmp}{\eqalign{\SAmpl &= (2\pi\epsilon\hbar)^{-n/2} \exp
\left( - {1\over\hbar} S_B - {1\over\hbar} \tilde S_F \right) \cr &
\qquad \left[ 1 - {1\over 12} \epsilon\hbar R(x) - {1\over 12}
R_{ij}(x) (y-x)^i (y-x)^j + {1\over 2} \epsilon\hbar R_{ab}(x)
\bar\eta^a \chi^b \right] \cr }}

\equation{SBos}{\eqalign{S_B &= {1\over 2\epsilon} g_{ij}(x) (y-x)^i
(y-x)^j + {1\over 4\epsilon} \partial_k g_{ij}(x) (y-x)^i (y-x)^j
(y-x)^k \cr & \quad + {1\over 12\epsilon} \Bigl( \partial_k\partial_l
g_{ij}(x) - {1\over 2} g_{mn}(x) \Gamma^m_{ij}(x) \Gamma^n_{kl}(x)
\Bigr) (y-x)^i (y-x)^j (y-x)^k (y-x)^l }}

\equation{SFer}{\eqalign{\tilde S_F &= - \hbar \delta_{ab} \bar\eta^a
\chi^b - \hbar (y-x)^i \omega_{iab}(x) \bar\eta^a \chi^b
 - {1\over2} \hbar (y-x)^i (y-x)^j \Bigl( \partial_i
\omega_{jab}(x) + \omega_{ia}{}^c(x) \omega_{jcb}(x) \Bigr) \bar\eta^a
\chi^b \cr & \qquad - {1\over 2} \epsilon\hbar^2 R_{abcd}(x) \bar\eta^a \chi^b
\bar\eta^c \chi^d \cr }}

\equation{SFercon}{\tilde S_F = S_F - \hbar\delta_{ab} \bar\eta^a
\psi^b_{\rm cl}(0) \quad ; \qquad S_F = \hbar \int\limits_{-\epsilon}^0\!
dt\,\left( \delta_{ab}
\bar\psi^a \dot\psi^b + \dot x^i \omega_{iab} \bar\psi^a \psi^b -
{1\over 2} \hbar R_{abcd} \bar\psi^a \psi^b \bar\psi^c \psi^d \right) }

\equation{eom}{\eqalign{&\ddot x^i + \Gamma^i{}_{jk} \dot x^j \dot x^k
- R^i{}_{jab} \dot x^j \bar\psi^a \psi^b + {1\over 2} \hbar g^{ij}
\left( \partial_j R_{abcd} \right)\bar\psi^a \psi^b \bar\psi^c \psi^d=
0 \cr & \dot\psi^a + \dot x^i
\omega_i{}^a{}_b \psi^b - R^a{}_{bcd} \psi^b \bar\psi^c \psi^d = 0 \cr
& \dot{\bar\psi^a} + \dot x^i
\omega_i{}^a{}_b \bar\psi^b - R^a{}_{bcd} \bar\psi^b \bar\psi^c \psi^d
= 0 \cr }}

\equation{TayL}{S= \epsilon L(0) - {1\over 2} \epsilon^2 \dot L(0) + \ldots }

\equation{expter}{\eqalign{& \dot x^i(0) = {(x-y)^i \over \epsilon}
+ {1\over 2} (x-y)^j R^i{}_{jab} \bar\eta^a
\chi^b  \cr & \bar\psi^a(0) = \bar\eta^a \cr & \psi^a(0) =
\chi^a - (x-y)^i \omega_i{}^a{}_b \chi^b \cr &
\dot\psi^a(0) = {\psi^a(0) -\chi^a \over \epsilon} - {1\over 2}
{(x-y)^i (x-y)^j \over \epsilon}\left( \partial_i \omega_j{}^a{}_b -
\omega_i{}^a{}_c \omega_j{}^c{}_b \right) \chi^b \cr }}

\equation{Scompo}{ \langle x,\bar\eta | \exp \left( - {2\epsilon
\over \hbar } \hat H \right) | y,\chi \rangle = \int\!d^nz d^n\bar\xi
d^n\xi \sqrt{g(z)} e^{-\bar\xi \xi} \langle x,\bar\eta | \exp \left( -
{ \epsilon \over \hbar } \hat H \right) | z,\xi \rangle \langle
z,\bar\xi | \exp \left( - { \epsilon \over \hbar } \hat H \right) |
y,\chi \rangle }

\equation{SconFer}{\Sco = \int\! dt \, \left( {1\over 2} g_{ij} \dot
x^i \dot x^j + \delta_{ab} \bar\psi^a \dot\psi^b +
\dot x^i \omega_{iab} \bar\psi^a \psi^b - {1\over 2}
R_{abcd} \bar\psi^a \psi^b \bar\psi^c \psi^d \right) - \delta_{ab}
\bar\eta^a \psi^b_{\rm cl}(0) }

\equation{defDS}{D_S = {\rm sdet} D_{AB} \quad ; \quad D_{AB} \equiv -
{\partial\over \partial \Phi^A}  \left( S_B + \tilde S_F \right)
{\overleftarrow{\partial}\over \partial \Phi^B} }

\equation{DisABCD}{D_{AB} = \left( \eqalign{ &A_{ij} \quad B_{ib} \cr
&C_{aj} \quad D_{ab} \cr } \right) }

\equation{ABCD}{\eqalign{A_{ij} &= {1\over\epsilon} g_{ij}(x) +
\partial_i \omega_{jab}(x) \bar\eta^a \chi^b \cr & \qquad
- {1\over 2} \Bigl( \partial_i \omega_{jab}(x) + \partial_j
\omega_{iab}(x) + \omega_{ia}{}^c(x) \omega_{jcb}(x) +
\omega_{ja}{}^c(x) \omega_{icb}(x) \Bigr) \bar\eta^a
\chi^b \cr B_{ib} &= - \omega_{iab}(x) \bar\eta^a \cr
C_{aj} &= \omega_{jab}(x) \chi^b \cr D_{ab} &=
\delta_{ab} + (y-x)^i \omega_{iab}(x) + {1\over
2} (y-x)^i (y-x)^j \Bigl( \partial_i \omega_{jab}(x) +
\omega_{ia}{}^c(x) \omega_{jcb}(x) \Bigr) \cr & \qquad + \epsilon
\Bigl( R_{abcd}(x) - R_{adcb}(x) \Bigr) \bar\eta^c \chi^d \cr }}

\equation{expSdet}{ D_S^{1/2} = (\epsilon)^{-n/2} g^{1/2}(x)
\Bigl[ 1 + {1\over 2} \epsilon\,\tr a + {1\over 2} \epsilon\,\tr CB -
{1\over 2} \tr d + {1\over 4} \tr d^2 \Bigr] }

\equation{DSis}{\tilde D_S^{1/2} = g^{-1/4}(x) D_S^{1/2} g^{-1/4}(y) =
(\epsilon)^{-n/2} \left[ 1 - {1\over 12} R_{ij}(x) (y-x)^i
(y-x)^j + {1\over 2} \epsilon R_{ab} \bar\eta^a \chi^b \right] }

\equation{proprew}{\SAmpl = (2\pi\hbar)^{-n/2} \tilde D_S^{1/2} \exp
\left( - {1\over \hbar} \bigl( S_B + \tilde S_F \bigr) \right) \left[
1 - {1\over 12} \epsilon \hbar R \right] }

\equation{Htrun}{\eqalign{H^{({\rm qu})}_{N=2,{\rm truncated}} &=
{1\over 2} g^{-1/4}
\pi_i g^{1/2} g^{ij} \pi_j g^{-1/4} - {1\over 16}
\hbar^2 R \cr \pi_i &= p_i - {i\hbar\over 2} \omega_{iab}
\psi^a_I \psi^b_I \cr }}

\equation{FerReg}{\eqalign{\hat H &= {1\over 2} g^{-1/4}
\pi_i g^{1/2} g^{ij} \pi_j g^{-1/4} - {1\over 8}
\hbar^2 R \cr \pi_i &= p_i - {i\hbar\over 2} \omega_{iab}
\psi^a_I \psi^b_I \cr }}

\equation{FerTrans}{\eqalign{\SAmpl &= (2\pi\epsilon\hbar)^{-n/2}
\exp\left( -{1\over\hbar} \bigl( S_B + \tilde S_F \bigr) \right)
\biggl[ 1 + {1\over 24}
\epsilon\hbar R(x) \cr & \qquad - {1\over 12} R_{ij}(x) (y-x)^i
(y-x)^j + {1\over 16} (y-x)^i (y-x)^j \omega_{ia}{}^b(x)
\omega_{jb}{}^a(x) \biggr] \cr }}

\equation{Soverh}{\eqalign{{1\over\hbar}\tilde S_F &=  -
\delta_{ab} \bar\eta^a \chi^b - {1\over 4} (y-x)^i \omega_{iab}(x)
(\bar\eta^a + \chi^a) (\bar\eta^b + \chi^b)  \cr & \qquad - {1\over 8}
(y-x)^i (y-x)^j \partial_i \omega_{jab}(x) (\bar\eta^a + \chi^a)
(\bar\eta^b + \chi^b) \cr }}

\equation{1ABCD}{\eqalign{A_{ij} &= {1\over\epsilon} g_{ij}(x) +
{1\over 8} \Bigl( \partial_i \omega_{jab}(x) - \partial_j
\omega_{iab}(x) \Bigr) (\bar\eta^a + \chi^a ) (\bar\eta^b + \chi^b )
\cr B_{ib} &= - {1\over 2} \omega_{iab}(x) ( \bar\eta^a + \chi^a ) \cr
C_{aj} &= {1\over 2} \omega_{jab}(x) (\bar\eta^b + \chi^b ) \cr D_{ab}
&= \delta_{ab} + {1\over 2} (y-x)^i \omega_{iab}(x) + {1\over 4}
(y-x)^i (y-x)^j \partial_i \omega_{jab}(x) \cr }}

\equation{DS1is}{\tilde D_S^{1/2} = (\epsilon)^{-n/2} \Bigl[ 1 -
{1\over 12} R_{ij}(x) (y-x)^i (y-x)^j + {1\over 16} (y-x)^i (y-x)^j
\omega_{ia}{}^b (x) \omega_{jb}{}^a (x) \Bigr] }

\equation{prop1re}{\SAmpl = (2\pi\hbar)^{-n/2} \tilde D_S^{1/2} \exp
\left( - {1\over\hbar} \bigl( S_B + \tilde S_F \bigr) \right) \Bigl[ 1
+ {1\over 24} \epsilon \hbar R \Bigr] }

\equation{SferAn}{\Sco = \int\! dt\, \left( {1\over2} g_{ij} \dot x^i
\dot x^j + {1\over2} \delta_{ab} \psi^a_1 \dot\psi^b_1 + {1\over2}
\delta_{ab} \psi^a_2 \dot\psi^b_2 + {1\over2} \dot x^i \omega_{iab}
\psi^a_1 \psi^b_1 \right) - \delta_{ab} \bar\eta^a \psi^b_{\rm cl}(0) }

\equation{SAnn}{{\cal A}_n^{spin-{1\over 2}} = -{\hbar\over
2^{n/2}} \lim_{\epsilon \rightarrow 0} \int \! d^a\chi d^a\bar\eta \,
e^{\bar\eta \chi} \langle x,
\bar\eta | \exp \left( - {\epsilon\over\hbar} \hat H \right) | x,
\chi \rangle \quad ; \qquad a = 1\ldots n }

\equation{partcs}{\hat H = {1\over 2} g^{-1/4}(\hat x) \hat p_i
g^{ij}(\hat x) g^{1/2}(\hat x) \hat p_j g^{-1/4}(\hat x) }

\equation{ppMaxw}{\hat H = {1\over 2} \left( \hat p_i - {e\over c}
A_i(\hat x) \right) \left( \hat p^i - {e\over c} A^i(\hat x) \right) }

\equation{respro}{\eqalign{\langle x| \exp\left( -
{\epsilon\over\hbar} \hat
H \right) | y\rangle &= (2\pi\hbar)^{-n/2} \exp\left( -{1\over\hbar}
\Scl (x,y;\epsilon) \right) \epsilon^{-n/2} \left( 1 - {1\over 12}
R_{ij}(x)(x-y)^i(x-y)^j \right) \cr & \qquad \exp \left( - {1\over 12}
\hbar R(x) \epsilon \right) \cr }}

\equation{Scova}{L_{\rm conf} = {1\over 2} g_{ij} \dot x^i \dot x^j +
{\hbar^2\over 8} R }

\equation{Hconf}{H_{\rm conf} = {1\over 2} g^{ij} p_i p_j - {1\over 8}
\hbar^2 R - {1\over 8} \hbar^2 \Gamma^i{}_{jk} \Gamma^j{}_{il} g^{kl}
}

\equation{Spluspro}{S =
{1\over\epsilon}\int\limits_{-1}^0\! d\tau\, {1\over 2} g_{ij}\bigl(
x_0(\tau) + x_{\rm qu}(\tau) \bigr) \left[ \bigl( \dot x^i_0(\tau) +
\dot x^i_{\rm qu}(\tau) \bigr) \bigl( \dot x^j_0(\tau) + \dot x^j_{\rm
qu}(\tau) \bigr) + b^i(\tau) c^j(\tau) + a^i(\tau) a^j(\tau) \right] }

\equation{onlypro}{\eqalign{ < x^i_{\rm qu}(\sigma) x^j_{\rm qu}(\tau)
> &= -\epsilon \hbar g^{ij} \Delta(\sigma,\tau) \cr < b^i(\sigma)
c^j(\tau) > &= -2 \epsilon \hbar g^{ij} \partial^2_\sigma
\Delta(\sigma,\tau) \cr < a^i(\sigma) a^j(\tau) > &= \epsilon\hbar
g^{ij} \partial^2_\sigma \Delta(\sigma,\tau) \cr }}

\equation{propmod}{
\Delta(\sigma,\tau) = -2 \sum\limits_{n=1}^\infty
{\sin (n\pi\sigma) \sin (n\pi\tau) \over n^2 \pi^2 } }

\equation{restpath}{- {1\over 32} \epsilon\hbar g^{ij}g^{kl}g^{mn}
\left(\partial_i g_{km} \right) \left( \partial_j g_{ln} \right) +
{1\over 48} \epsilon\hbar g^{ij}g^{kl}g^{mn} \left( \partial_i
g_{km}\right) \left( \partial_l g_{jn} \right) = -{1\over 24}
\epsilon\hbar g_{kl} \Gamma^{k}{}_i{}^j \Gamma^{l}{}_j{}^i }

\equation{pathres}{- {1\over 24} \epsilon\hbar g^{ij} g^{kl} g^{mn}
\left( \partial_i g_{km} \right) \left( \partial_l g_{jn} \right) }

\equation{Nis2r}{- {1\over 24} \epsilon\hbar g_{kl} \Gamma^k{}_i{}^j
\Gamma^l{}_j{}^i -{1\over 12} \epsilon\hbar g^{ij} \omega_{ia}{}^b
\omega_{jb}{}^a }

\equation{Hamgen}{\hat H = g^{-1/4}(\hat x)
\bigl( \hat p_i - A_i(\hat x) \bigr) g^{1/2}(\hat x) g^{ij}(\hat x)
\bigl( \hat p_j - A_j(\hat x) \bigr) g^{-1/4}(\hat x) + V(\hat x) }

\equation{Scofigu}{\Sco = \int\limits_{-\Delta t}^0 \left[
{1\over 2} g_{ij}(x) \dot x^i \dot x^j + A_i(x) \dot x^i + V(x) -
{1\over 8} \hbar^2 R(x) \right] \, dt }

\equation{Scogen}{\Sco = {1\over\epsilon} \int\limits_{-1}^0 \! d\tau
\, {1\over 2} g_{ij}(x(\tau)) \left[ \dot x^i(\tau) \dot x^j(\tau) +
b^i(\tau) c^j(\tau) + a^i(\tau) a^j(\tau) \right] }

\equation{bgdec}{x^i(\tau) = x^i_0(\tau) + x^i_{\rm qu}(\tau) \quad ;
\quad x^i_0(\tau) = x^i + (x-y)^i \tau }

\equation{skin}{S^{kin} = {1\over\epsilon} \int\limits_{-1}^0 \!
d\tau\, {1\over 2} g_{ij} \left[ \dot x^i \dot x^j + b^i c^j + a^i a^j
\right] }

\equation{sint}{S^{int} = {1\over\epsilon} \int\limits_{-1}^0 \!
d\tau\, \left( {1\over 2} \partial_k g_{ij} x^k + {1\over 4}
\partial_k \partial_l g_{ij} x^k x^l \right) \left[ \dot x^i \dot x^j
+ b^i c^j + a^i a^j \right] }

\equation{canprop}{\eqalign{\Delta^{ij}(s,t) & \equiv \langle
0 | T \hat x^i(s) \hat x^j(t) |0\rangle \cr & = \theta(s
-t) \langle 0| \bigl( \hat x^i + \hat p^i s \bigr) \bigl( \hat
x^j + \hat p^j t \bigr) | 0\rangle \quad + \quad (s
\leftrightarrow t) \cr & = \theta (s-t) \langle 0| \hat p^i
s \, \hat x^j \bigl( 1 + t/\epsilon \bigr) |0\rangle \quad +
\quad (s\leftrightarrow t) \cr & = \theta(s-t)
{\hbar\over i} g^{ij} s \bigl( 1 + t/\epsilon \bigr)
\quad + \quad (s\leftrightarrow t) \cr }}

\equation{GreenBos}{ \Delta^{ij}(\sigma,\tau) = - \epsilon\hbar
g^{ij} \bigl[ \sigma(\tau+1) \theta(\sigma-\tau) + \tau(\sigma+1)
\theta(\tau-\sigma) \bigr] }

\equation{modpro}{\Delta(\sigma,\tau) = -2 \sum\limits_{n=1}^\infty
{\sin (n\pi\sigma) \sin (n\pi\tau) \over n^2 \pi^2 }}

\equation{Sbackg}{S_0 = {1\over\epsilon} \left[ {1\over 2} g_{ij} +
{1\over 4} \partial_k g_{ij} (y-x)^k + {1\over 12} \partial_k
\partial_l g_{ij} (y-x)^k (y-x)^l \right] (y-x)^i (y-x)^j }

\equation{lS1}{-{1\over 12} \partial_k \partial_l g_{ij} \left[
{1\over 2} g^{ij} (x-y)^k (x-y)^l - g^{jl} (x-y)^i (x-y)^k + {1\over
2} g^{kl} (x-y)^i (x-y)^j \right] }

\equation{qS1}{{1\over\epsilon} {1\over 24} g_{mn} \Gamma^m_{ij}
\Gamma^n_{kl} (x-y)^i (x-y)^j (x-y)^k (x-y)^l }

\equation{lS2}{{1\over 24} \epsilon\hbar \partial_k \partial_l g_{ij}
\left[ g^{kl} g^{ij} - g^{ik} g^{jl} \right] }

\equation{qS3}{- {1\over 32} \epsilon\hbar g^{ij}g^{kl}g^{mn}
\left(\partial_i g_{km} \right) \left( \partial_j g_{ln} \right) +
{1\over 48} \epsilon\hbar g^{ij}g^{kl}g^{mn} \left( \partial_i
g_{km}\right) \left( \partial_l g_{jn} \right) }

\equation{qS3p}{-{1\over 24} \epsilon\hbar g_{kl} \Gamma^k{}_i{}^j
\Gamma^l{}_j{}^i }

\equation{Ssplit}{S = S_0 + S^{kin}_{\rm fer} + S^{int}_{\rm fer} }

\equation{Skinfer}{S^{kin}_{\rm fer} = \int\limits^0_{-1} \! d\tau \,
\delta_{ab} \bar\psi^a \dot\psi^b }

\equation{Sintfer2}{S^{int}_{\rm fer} = \int\limits^0_{-1}
\! d\tau \, \left[ \dot x^i \omega_{iab} \bar\psi^a \psi^b - {1\over 2}
\epsilon R_{abcd} \bar\psi^a \psi^b \bar\psi^c \psi^d \right] }

\equation{Sintfer1}{S^{int}_{\rm fer} = \int\limits^0_{-1}
\! d\tau \, \left[ {1\over 2} \dot x^i \omega_{iab} \psi^a_1 \psi^b_1
\right] }

\equation{ferGreen}{G^{ab}_{\alpha\beta}(\sigma,\tau) = < \psi^a_{{\rm
qu},\alpha}(\sigma)
\psi^b_{{\rm qu},\beta}(\tau) > = {1\over 2} \hbar \delta^{ab}
\delta_{\alpha\beta}
\bigl( \theta(\sigma-\tau) - \theta(\tau-\sigma) \bigr) + {i\over 2}
\hbar \delta^{ab}
\epsilon_{\alpha\beta} }

\equation{Greenfer}{\eqalign{ & < \bar\psi^a_{\rm qu}(\sigma) \psi^b_{\rm
qu}(\tau) > = - \hbar \delta^{ab} \theta(\tau-\sigma) \cr & <
\psi^a_{\rm qu}(\sigma)
\psi^b_{\rm qu}(\tau) > = 0 = < \bar\psi^a_{\rm qu}(\sigma) \bar\psi^b_{\rm
qu}(\tau) > \cr }}

\equation{modexp}{\eqalign{\psi(t) &= \sum\limits_n b_n \cos \bigl(
(n+{1\over 2}) \pi t/\epsilon \bigr) \cr \bar\psi(t) &= \sum\limits_n
\bar b_n \sin \bigl( (n+{1\over 2}) \pi t/\epsilon \bigr) \cr }}

\equation{femopr}{< T \bar\psi^a(s) \psi^b(t) > = {2\over\pi}
\delta^{ab} \sum\limits_{n=0}^\infty {1\over n+{1\over 2}} \sin \bigl(
(n+{1\over 2}) \pi s/\epsilon \bigr) \cos \bigl( (n+{1\over 2})\pi
t/\epsilon \bigr) }

\equation{Sintcl1}{-{1\over\hbar} \int\limits_{-1}^{0} \! d\tau \,
\left[ {1\over 4} \dot x^i_0(\tau) \omega_{iab}(x_0(\tau)) \bigl(
\bar\eta^a +\chi^a \bigr) \bigl( \bar\eta^b + \chi^b \bigr) \right] }

%******************** begin titlepage ********************

\pagenumstyle{blank}
\footnoteskip=2pt
\line{\it November 1993\hfil ITP-SB-93-51}

%\vskip4em
\vskip1em
\baselineskip=32pt
\begin{center}{\bigsize{\sc The Hamiltonian approach and phase space
path integration\\ for nonlinear sigma models with and without
fermions }}
\end{center}
\vskip1em
\vfil
\baselineskip=16pt

\begin{center}
{\bigsize
Bas Peeters\footnote[$\ \bowtie$]{\sc e-mail:
peeters@max.physics.sunysb.edu}\vskip .5cm
Peter van Nieuwenhuizen\footnote[$\ \ddagger$]{\sc e-mail:
vannieu@max.physics.sunysb.edu}}
%\vskip .5cm
\vskip .3cm
{\it Institute for Theoretical Physics\vskip 4pt
State University of New York at Stony Brook\vskip 4pt
Stony Brook, NY~~11794-3840}
\vskip .5cm

\end{center}

\vfil

\centertext{\bfs \bigsize ABSTRACT}
\vskip\belowsectionskip

\begin{narrow}[4em]

Instead of imposing the Schr\"{o}dinger equation to obtain the
configuration space propagator $\csprop$ for a quantum mechanical
nonlinear sigma model, we directly evaluate the phase space propagator
$\psprop$ by expanding the exponent and pulling all operators $\hat p$
to the right and $\hat x$ to the left. Contrary to the widespread
belief that it is sufficient to keep only terms linear in $\Delta t$
in the expansion if one is only interested in the final result through
order $\Delta t$, we find that all terms in the expansion must be
retained. We solve the combinatorical problem of summing the infinite
series in closed form through order $\Delta t$. Our results
straightforwardly generalize to higher orders in $\Delta t$. We then
include fermions for which we use coherent states in phase space. For
supersymmetric $N{=}1$ and $N{=}2$ quantum mechanics, we
find that if the super Van Vleck determinant replaces the original Van
Vleck determinant the propagator factorizes into a classical part,
this super determinant and the extra scalar curvature term which was
first found by DeWitt for the purely bosonic case by imposing the
Schr\"{o}dinger equation. Applying our results to anomalies in
$n$-dimensional quantum field theories, we note that the operator
ordering in the corresponding quantum mechanical Hamiltonians is fixed
in these cases. We present a
formula for the path integral action, which corresponds one to one
to any given covariant or noncovariant $\hat H$. We then evaluate these
path integrals through two loop order, and reobtain the same
propagators in all cases.

\end{narrow}

\vfil

\break

%********************* end titlepage *********************

\pagenumstyle{arabic}
\pagenum=0
{\bf\section{Introduction.}}

In this article we settle a long standing problem, namely the direct
computation of the phase space matrix element $\psprop$ for quantum
mechanical systems with a kinetic term of the form $a(\hat x) \hat p^2
+ b(\hat x) \hat p + c(\hat x)$. The standard example, on which we
concentrate, is the nonlinear sigma model describing a point particle
coupled to an external gravitational field. Another example is the
Hamiltonian for a point particle minimally coupled to an
electromagnetic field. We shall also consider the inclusion of
fermions, in particular supersymmetric quantum mechanics. Although our
results generalize straightforwardly to quantum field theory, we shall
restrict our discussion to quantum mechanical systems.

Through the work of Alvarez-Gaum\'{e} and Witten on chiral and
gravitational anomalies[\putref{AGWi},\putref{Alva}], and later the
work of Bastianelli and van Nieuwenhuizen on trace (Weyl)
anomalies[\putref{Bast},\putref{BaNi}], it is known that anomalies of
an $n$-dimensional quantum field theory can in general be written as
configuration space path integrals for nonlinear sigma models.
However, the precise relation between the quantum mechanical
Hamiltonian, which corresponds to the regulator in the quantum field
theory and whose operator ordering is unambiguously fixed by this
quantum field theory[\putref{textb}], and the corresponding action
$\Sco$ to be used in the configuration space integral, is unclear. For
chiral anomalies, the precise form of this action is not needed, as
due to their topological nature the naive classical action $\Scl$ is
sufficient.  However, for trace anomalies the extra terms by which
$\Sco$ differs from $\Scl$ are crucial. We therefore decided to settle
the relation between $\Sco$ and $\Scl$ in general.

Before going on, let us eliminate a possible source of confusion. We
are aware of an enormous amount of literature on configuration space
path integrals, in which geometric operators like the Laplace-Beltrami
operator and the Lichnerowicz operator play a role. In these articles
these operators are selected because of their nice mathematical
properties. In the present article we do not begin by selecting such
nice operators on mathematical grounds; rather, on physical grounds we
begin with certain Hamiltonians $\hat H$ whose operator orderings are
fixed a priori by corresponding quantum field theories, and our aim is
to find the corresponding $\Sco$ which produce exactly the same
propagator as $\hat H$. When it will so happen that these $\Sco$ do
not have nice mathematical properties, so be it. For us, the relevant
question is: which $\Sco$ corresponds to which $\hat H$? Of course,
this question only makes sense if we also give a precise definition of
how to evaluate the path integrals. We shall only consider perturbative
expansions, and the precise rules are given in
[\putref{AGWi},\putref{Bast},\putref{BaNi}]. They will be reviewed in
the appendix. (For the evaluation of anomalies, perturbation
theory gives exact results).

In the configuration space path integrals considered in
[\putref{AGWi},\putref{Alva},\putref{Bast},\putref{BaNi}], the
internal symmetry sector or the fermionic sector were still treated by
an operator formalism. We shall treat the bosonic and fermionic
sectors on equal footing, starting with an operator formalism and
deducing from it the corresponding path integral formalism. Our final
path integrals will be genuine path integrals, without sectors where
one still uses operators. To achieve
this, we need to evaluate the matrix element $\psprop$ for bosons or
$\langle x,\bar\eta| \exp \left( - {i\over\hbar} \hat H \Delta
t\right) | p,\xi\rangle$ for fermions where $\langle\bar\eta |$ and
$|\xi\rangle$ are fermionic coherent states.

At first sight the task seems appalling (which may account for the
absence of this result in the literature) because expanding the
exponent one finds momentum operators $\hat p$ in all possible places
between position operators $\hat x$ which appear in the arbitrary
metric $g_{ij}(\hat x)$. Pulling all $\hat p$ to the right and keeping
track of all possible commutators of $\hat p$ and $\hat x$ seems a
hopelessly complicated combinatorical problem. It seems universally
believed that it is sufficient to expand $\exp\left(-{i\over\hbar}\hat
H \Delta t\right)$ into $1-{i\over\hbar}\hat H \Delta t$, evaluate the
matrix element $\langle x|\hat H |p\rangle\equiv h(x,p)\langle
x|p\rangle$ or $\langle z^*|\hat H|z\rangle \equiv h(z^*,z)\langle
z^*|z\rangle$ and reexponentiate to obtain
$\exp\left(-{i\over\hbar}h(x,p)\Delta t\right)\langle x|p\rangle$ or
$\exp\left(-{i\over\hbar}h(z^*,z)\Delta t\right)\langle z^*|z\rangle$
(here $|z\rangle$ denote bosonic coherent
states)[\putref{Alva},\putref{textb}].  This we have found to be
incorrect: one must keep {\it all} terms in the expansion of the
exponent, but for each term one needs to take into account only a
finite number of commutators between $\hat p$ and $\hat x$ if one is
only interested in the result to a given order in $\Delta t$. This makes
the problem tractable after all. Including fermions, one obtains in
addition to $\hat x$ and $\hat p$ also operators $\hat\psi$ and
$\hat\psi^\dagger$. Again, expanding the exponent, all terms contibute
but only a finite number of anticommutators of $\hat\psi$ and
$\hat\psi^\dagger$ need be kept in each term, and the transition
element (also called propagator) $\langle
x,\bar\eta|\exp\left(-{i\over\hbar}\hat H \Delta t\right)
|p,\xi\rangle$ is obtained to any given order in $\Delta t$.

To avoid confusion, let us define what we mean by `to a given order
in $\Delta t$'. We say that the propagator is given up to order $l$ in
$\Delta t$ if it can be written as a product of some function
$f(x,y;\Delta t)$ and a factor $[1 + {\cal O}(\Delta t)^{l+1/2}]$. The
function $f$ itself will in general be singular in $\Delta t$.

The (incorrect) neglecting of terms of order $(\Delta t)^2$ in
nonlinear sigma models should not be confused with the (correct)
neglecting of terms of order $(\Delta t)^2$ in the approach of Feynman
and others for Hamiltonians of the form $T(\hat p) + V(\hat x)$, where
the Trotter formula[\putref{Trot}] actually proves the correctness of
this procedure.  Nonlinear sigma models are simply not in the class of
models to which the Trotter formula can be applied.

It is often claimed that after integrating out the momenta, one can
use the resulting propagators $\langle x_j| \exp \left( -
{i\over\hbar} \hat H \Delta t\right) | x_{j-1} \rangle$ to order
$\Delta t$ to obtain the configuration space path integrals by taking
products of these building blocks and integrating over the
intermediate points $x_1\ldots x_{N-1}$. However, there is a problem:
usually one wants to use {\it local} actions $\Sco$, and the matrix
elements $\langle x_j| \exp \left( - {i\over\hbar} \hat H \Delta t
\right) | x_{j-1} \rangle$ cannot in general be written as the
exponent of a local action. For example, for the nonlinear sigma model
we consider below, one finds terms $\Delta t R(x)$ and $R_{ij}(x)
(x-y)^i (x-y)^j$, and whereas the former can be written to order
$\Delta t$ as $\int\! dt\, R(x)$, the latter term cannot be written in
such a way. In fact, the nonlocal $\langle x_j| \exp \left( -
{i\over\hbar} \hat H \Delta t \right) | x_{j-1} \rangle$ satisfy the product
property of path integrals, and thus trivially produce in a path
integral the propagator $\langle x| \exp \left( - {i\over\hbar}
\hat H T \right) | y\rangle$. The problem
is rather to find a local action $\Sco$, which also produces in a path
integral this propagator. From the example of the point particle in
curved space it is known that $\Sco$ contains local terms of higher
order in $\hbar$, but in general the problem of constructing $\Sco$
from a given Hamiltonian (with given operator ordering) seems
unsolved. DeWitt has recently made great progress in this direction by
establishing a formal (unregularized) connection between $\hat H$ and
$\Sco$ using the concept of time-ordering[\putref{DeWib}]. Using his
results and the two and three-loop results of [\putref{BaNi}] on trace
anomalies, we shall give the exact local action $\Sco$ which
corresponds to any, covariant or noncovariant, Hamiltonian $\hat H$
with at most two $\hat p$ operators. We have not been able to give a
rigorous proof of this result, but we have checked our result through
two-loop order, and give general arguments which hold at higher loop
order.

Actually, Feynman already suspected that for nonlinear sigma models
integrating out the momenta would not lead to path integrals with the
classical action in the exponent[\putref{Feyn}]. In the literature an
indirect method has been used to obtain the propagator $\csprop $ in
configuration space. Namely, one requires that it satisfies the
Schr\"{o}dinger equation $H_x \csprop = i\hbar \left( \partial \bigl/
\partial t_2 \right) \csprop $, where $H_x$ is the Hamiltonian in the
$x$-representation, $\langle x | \hat H | x^\prime \rangle = H_x
\delta (x-x^\prime)$. In this way, DeWitt found that for small
$t_2-t_1$, this propagator does not approach Feynman's result, but an
extra term proportional to the scalar curvature is
present[\putref{DeWi}]. We shall be able to trace its origin by
following the direct calculation step by step. Bastianelli and one of
us[\putref{BaNi}] found the extension of DeWitt's result to
fermions. They took a very general ansatz for the action $\Sco$ in the
configuration space path integral, and after evaluating this
configuration space path integral by a loop expansion, they were able
to determine the propagator by again imposing the Schr\"{o}dinger
equation.

The reader may start wondering at this moment why we use at all the
direct but cumbersome method of evaluating $\int\! dp\, \langle x|
\exp \left( - {i\Delta t\over\hbar} \hat H \right) | p\rangle \langle p |
y\rangle$, rather than just following the heat kernel approach. We
have two reasons. First of all, since Feynman, most physicists derive
path integrals as we do, and it is desirable to understand this approach
in detail. Secondly, it will help us in determining the action
$\Sco$. In fact, in flat spacetime, one obtains directly $\Sco$ after
integrating out the momenta which are introduced by $\langle x|
\exp \left( - {i\Delta t\over\hbar} \hat H\right) | p\rangle$; only in curved
spacetime complications set in.

There are two distinct results one is interested in: the propagator
and its trace. Let us first consider the case of flat spacetime. Then
both results are obtained from a formula which is obtained in the
spirit of Feynman by inserting complete sets of $x$ and $p$ states or
coherent states, but in the former case the propagator can only be
written as a phase space path integral provided one adds boundary
terms depending on a {\it complex} classical trajectory, while for the
trace one directly obtains the path integral without the extra
boundary terms and without the complex classical trajectory. These
boundary terms can be found in textbooks[\putref{textb}], whereas the
results for the trace formula were used by Alvarez-Gaum\'{e} in
[\putref{Alva}]. In section 3 we explain the difference between both
approaches. In curved spacetime, as we already mentioned, even for
small $\Delta t$ the propagator may contain terms which cannot be
written as local actions, and in these cases the straightforward
Hamiltonian approach does not yield the action $\Sco$.  As we already
stated, we have a formula for the correct $\Sco$ in general, and it
contains the same kind of boundary conditions as found from the
Hamiltonian approach in flat spacetime.

Since it is sometimes claimed that phase space path integrals are less
well-defined than configuration space path integrals, and that path
integrals for fermions cannot be as rigorously defined as for bosons,
and because we were also ourselves confused for a long time, we decided
to give a detailed account of the resolution of these problems,
starting from scratch in section 2. In particular, we want to stress
here that path integrals for fermions are phase space path integrals,
and perfectly well defined in terms of coherent states.  (The
Grassmann integration makes them actually more convergent than their
bosonic counter parts). This is very well explained in some
textbooks[\putref{textb}], and we have nothing new to add, but for
completeness we also include these aspects in section 5. In reference
[\putref{textb}] the Bargmann-Wigner formalism of holomorphic functions
is used to define inner products, Hilbert spaces, and only afterwards
one tackles quantum mechanics. This we found an unnecessary detour;
one can directly define coherent states in quantum mechanics, and
use the inner product for these coherent states which follows
straightforwardly from the (anti)commutation relations of the
annihilation and creation operators.

The fermionic path integrals need Grassmann numbers $\bar\eta^a$ and
$\eta^a$ $(a=1\ldots n)$ which are independent. However, the
Hamiltonian which corresponds to the regulator $\Dsl\!\Dsl$ of
spin-${1\over 2}$ quantum field theory, contains only one kind of
fermionic operators, $\psi_1^a$. (One identifies the Dirac matrices
$\gamma^a$ with $\sqrt{2}\psi_1^a(t)$). This $\psi_1^a$ satisfies the
equal-time anticommutation relations
$\{\psi_1^a,\psi_1^b\}=\delta^{ab}$, but in order to obtain creation
and annihilation operators, one needs operators $\psi^{a\dagger}$ and
$\psi^a$ satisfying $\{\psi^a,\psi^{a\dagger}\}=\delta^{ab}$ and
$\{\psi^a,\psi^b\}=
\{\psi^{a\dagger},\psi^{b\dagger}\}=0$. It is suggested in the
literature that one may double the number of fermion species in the
Hamiltonian, and replace the spin connection
$\omega_{iab}\psi_1^a\psi_1^b$ by
$\omega_{iab}\psi^a_\alpha\psi^b_\alpha$ with $\alpha=1,2$. Afterwards
one is supposed to divide the final results by $2^{n/2}$ in order to
undo this doubling. This practise is well-known for one-loop problems
in quantum field theory[\putref{tHoo}], but it is incorrect to use it
in this context because the one-loop anomalies of a quantum field
theory correspond to {\it multiloop} corrections of the corresponding
quantum mechanical model[\putref{Bast},\putref{BaNi}]. We resolve this
problem by adding a second set of {\it free} fermions.

We begin in section 2 with a discussion of phase space path integrals
based on $x$ and $p$ eigenstates.  In section 3 we discuss the bosonic
phase space path integrals based on coherent states. Section 4
contains the heart of this paper: a direct evaluation of the
propagator through order $\Delta t$ of bosonic nonlinear sigma models.
We use here $x$ and $p$ eigenstates.  The result leads us to the
complex classical trajectories, and the trace formula. In section 5 we
study the $N{=}2$ supersymmetric nonlinear sigma model and discuss the
relation between operator ordering of the Hamiltonian and the
supersymmetry charge on the one hand, and the various supersymmetric
or non-supersymmetric actions on the other hand. In section 6 we
repeat the analysis of section 4 including fermions and using
fermionic coherent states. In both cases, the resulting propagator can
be written as a product of three factors: the classical action, the
(super) Van Vleck determinant, and a term which involves only the
scalar curvature and which is related to the trace anomaly in two
dimensions. In section 7 we discuss the $N{=}1$ supersymmetric
nonlinear sigma model, in particular the resolution of the problem of
fermion doubling alluded to above. Again the propagator can be written
as the product of the super Van Vleck determinant, the exponent of the
classical action, and a term involving the scalar curvature which
determines the trace anomaly of in this case a spin-${1\over 2}$
field. Section 8 contains the conclusions. In particular, we give
here, without proof, the one to one correspondence between $\hat H$
and $\Sco$ for any, covariant or noncovariant, $\hat H$. In the
appendix we perform the 2-loop calculations based on configuration
space path integrals, using covariant and noncovariant actions $\Sco$.

{\bf\section{Phase space path integrals in general.}}

Path integrals were (almost) introduced into quantum mechanis by Dirac
in 1933[\putref{Dira}]. In those days, quantum mechanics was obtained
from classical mechanics by replacing the canonical brackets of the
Hamiltonian formalism by corresponding quantum commutators. Dirac set
out to find a formulation which was based on the Lagrangian in order
to be able to retain manifest Lorentz invariance at the quantum
level. His starting point were the following relations of classical
mechanics between coordinates, conjugate momenta and action
$$\putequation{hameq}$$
where $\Sxy$ is the classical action for a point particle as a
function of its initial coordinate $y$ and its final coordinate $x$.
These relations follow immediately from the Euler-Lagrange variational
equations for $\int\limits_0^t \! dt^\prime \, L\bigl( x(t^\prime),
\dot x(t^\prime) \bigr) $ at fixed $t$. Dirac found a corresponding
operator equation in quantum mechanics as follows. Introducing
``moving frames''\footnote[$\ \dagger$]{These moving frames project a
ket $|\psi\rangle$ onto the corresponding Schr\"{o}dinger wave functon
$\psi(x,t)=\langle x,t|\psi\rangle$, and in this $x$ representation,
$\hat p(t)$ is represented on $\psi(x,t)$ as $p_x=-i\hbar
{\partial\over\partial x}$, as follows from $\langle x,t | \hat p(t) |
x^\prime,t \rangle = -i\hbar {\partial\over\partial x} \delta
(x-x^\prime)$. The overlap function in (2.2) gives then the
probability amplitude for the point particle to be at position $x$ at
time $t$, if it is initially at time $t=0$ at position $y$. In the
nonlinear sigma models we consider we shall use as inner product
$\sqrt{g(x)} \langle x|x^\prime \rangle = \delta (x-x^\prime)$ where
we define $\delta (x-x^\prime)$ by $\int \!  dx^\prime \, f(x^\prime)
\delta (x-x^\prime) = f(x)$. There is then an ambiguity in the
representation of $\hat p(t)$, namely $p_x = -i\hbar g^\alpha
{\partial\over\partial x} g^{-\alpha}$ with $g= \det g_{ij}(x)$, which
we fix by requiring that $\hat p(t)$ be hermitian with respect to this
inner product. The result is $\alpha = -1/4$.}  $|x,t\rangle =
\exp\left( {i\over\hbar} \hat H t \right) | x \rangle$, where
$|x\rangle$ are the eigenstates of the position operator $\hat x =
\hat x(t{=}0)$ at time $t=0$, and $|x,t\rangle$ are the eigenstates of
the position operator $\hat x(t) = \exp \left( {i\over\hbar} \hat H t
\right) \hat x \exp \left( - {i\over\hbar} \hat H t \right) $ at time
$t$, he noted that the inner product between these complete sets of
states, called ``transformation function'' by him, satisfies a
relation similar to (\puteqn{hameq}). Namely, defining a function
$U(x,y;t)$ by
$$\putequation{defU}$$
one has
$$\putequation{schr}$$
where $\hat p(t) $ is the momentum operator at time $t$. A similar
relation holds for $\hat p(0)$. Dirac then introduced an operator
$\hat U (\hat x(t),\hat x(0) )$ whose matrix elements are the overlap
functions in (\puteqn{defU})
$$\putequation{defUhat}$$
He called such operators ``well-ordered'', meaning that all $\hat
x(t)$ should be put on the left and all $\hat x(0)$ should be put on
the right, where they can be replaced by their eigenvalues $x$ and
$y$, respectively. Combining (\puteqn{schr}) and (\puteqn{defUhat}),
Dirac obtained operator equations corresponding to (\puteqn{hameq})
$$\putequation{hamopeq}$$
He then noted that according to the completeness postulate of quantum
mechanics one has
$$\putequation{compl}$$
and stated that for fixed $t_j$ and small $\hbar$ only stationary
points of $U$ contribute.

The relation (\puteqn{hamopeq}) suggests that the operator $\hat U$ is
the kind of quantum mechanical extension of the classical action Dirac
was looking for. In fact, he went further and carefully stated that
the propagator $\langle x_{j+1},t_{j+1} | x_j,t_j \rangle$
``corresponds to'' the classical action $S_{\rm
cl}(x_{j+1},x_j;t_{j+1}-t_j)$ for small $t_{j+1}-t_j$. He did not say
that they are equal but suggested that they were proportional, in
which case he obtained an equivalence principle for small $\hbar$,
provided the stationary trajectories of $U$ coincide with the
classical trajectories of $\Scl$.

In this article we consider quantum mechanical systems with a kinetic
term of the form $a(\hat x) \hat p^2 + b(\hat x)\hat p + c(\hat x)$,
where $\hat x$ and $\hat p$ have $n$ components ($\hat x^i$ and $\hat
p_i$ with $i=1\ldots n$), also called nonlinear sigma models. The
models we shall consider are point particles, or point spinors coupled
to an external gravitational field. An example of the kind of
Hamiltonians we study is given by
$$\putequation{Hferm}$$
where $g(x) = \det g_{ij}(x)$, $\alpha$ and $\beta$ run from $1$ to
$2$, and $a$,$b$ and $i$,$j$ run from $1$ to $n$. The Riemann tensor
is defined in (5.4), while
$$\putequation{defcovD}$$
This is the Hamiltonian for $N{=}2$ supersymmetric quantum mechanics.
For the $N{=}1$ case $\hat\psi^a_1 = \hat\psi^a_2 = {1\over\sqrt{2}}
\hat\psi^a_I$, and $\hat\pi_i$ can be identified with the covariant
derivative $\partial_i + {1\over 4}\omega_{iab} \gamma^a \gamma^b$ if
one identifies $\hat\psi^a_I$ with $\sqrt{\hbar} \gamma^a$. In fact,
for the $N{=}1$ case, and also for the purely bosonic case, the
particular ordering of all operators in $\hat
H$ (including the position operator $\hat x$ which we shall write from
now on for
notational simplicity without hat), is dictated by the $n$-dimensional
quantum field theory from which they come. We now explain this point;
however, readers not interested in quantum field theory may skip the
following paragraph and simply restrict their attention to the
particular orderings of $\hat H$ which we consider.

There is a general construction[\putref{DTNP}] for any quantum field
theory which yields the regulator $\hat {\cal R}$ one must use in
order to obtain consistent anomalies. If the Jacobian for an
infinitesimal transformation of a classical symmetry is given by
$1+J$, the anomaly is given by
$$\putequation{regan}$$
For a scalar field in $n$ dimensions this construction yields $\hat
{\cal R}_s = g^{-1/4} \partial_i g^{1/2} g^{ij} \partial_j g^{-1/4} $
for the consistent regulator which preserves general coordinate
invariance. On the other hand, the consistent regulator which
preserves Weyl (local scale) invariance but, as a consequence, breaks
Einstein (general coordinate) invariance, is given by $\partial_i
g^{1/2} g^{ij} \partial_j$. We call these regulators consistent
because they lead to anomalies which satisfy the consistency
relations. The corresponding quantum mechanical Hamiltonians are obtained by
replacing $\partial_i$ by ${i\over\hbar}\hat p_i$. The matrix elements
of $\hat p_i$ in the $x$-representation will not be needed (they are
given in the footnote above), since all we will use are the
commutation relations and the fact that $\hat p_i
|p\rangle = p_i |p\rangle$ on eigenfunctions $|p\rangle$. However, for
completeness we mention that in the $x$-representation with inner
product $(f_1,f_2)=\int\! dx \, \sqrt{g} f_1^\ast f_2$ one must
replace $\hat p_i$ by ${\hbar\over i}g^{-1/4}\partial_i g^{1/4}$. Then
the quantum
mechanical operator corresponding to the regulator becomes $g^{-1/2}
\partial_i g^{1/2} g^{ij} \partial_j$,
which is indeed hermitian with the inner product with $\sqrt{g}$.  For
spin-${1\over 2}$ fields $\hat {\cal R}_{fer} = g^{1/4} \Dsl \Dsl
g^{-1/4}$ which can be written as $\hat {\cal R}_{fer} = \hat {\cal
R}^0_{fer} + {1\over 4} R$ where ${\cal R}^0_{fer}$ is the first term
on the right hand side of (\puteqn{Hferm}) and $R$ is the scalar
curvature. The Dirac matrices $\gamma^a$ satisfy $\{
\gamma^a, \gamma^b \} = 2 \eta^{ab}$ and are represented in the
corresponding quantum mechanical model by operators $\sqrt{2}
\hat \psi^a_I$ satisfying $\{ \hat\psi^a_I, \hat\psi^b_I \} =
\eta^{ab}$. In this case there is thus
no ordering ambiguity in the Hamiltonian. For other
applications[\putref{Alva}] one may need an $N=2$ action with $(1,1)$
supersymmetry in which case $\alpha$ and $\beta$ range from $1$ to
$2$. In this case we shall fix the operator orderings in another way,
because there is no quantum field theory for which the regulator
corresponds to the $N{=}2$ Hamiltonian. The result is that the last
term in the Hamiltonian in (\puteqn{Hferm}) is then the
same (up to a sign) as the 4-fermion term in the classical action.
Even though there are no ordering ambiguities in the Hamiltonian which
corresponds to $\hat {\cal R}_s$ or $\hat {\cal R}_{fer}$, it remains
to be seen whether the
corresponding action we will produce in the configuration space
path integrals is the invariant classical action. We shall obtain the
anomalies of the corresponding $n$-dimensional quantum field theory
from the propagator $\cspro0$ by multiplying with the Jacobian $J(x)$,
putting $x=y$, integrating over $x$ {\it and taking the limit $t$
tending to zero}, retaining only the finite part. ($t^{-1}$ plays the
role of the square of the regulator mass, $M^2$). As a result the
classical terms (of order $\hbar^{-1}$) in the propagator as well as
the one-loop ($\hbar$-independent) terms will not contribute, and only
the two-loop terms of order $\hbar$ will contribute for small $\hbar$
in $n=2$ dimensions. This is not in disagreement with Dirac's results,
since he only made a statement about the propagator for
$\hbar\rightarrow 0$ at fixed $t$.

Dirac did not write down an equation relating the propagator $\cspro0$
for small $t$ to the classical action $\Scl(x,y;t)$ because he
probably did not know the normalization factor between them. Before
him, it was already known to some mathematicians (for example Wiener
in 1923[\putref{Wien}]) that one could define functionals of functions
$x(t)$ by time discretization and taking {\it ratios} of two such
functionals. For example, for the functional $F\bigl[ V[x] \bigr] =
\int \! d\mu \, \exp \left( \int_0^t V(x) dt\right)$ with $V(x)$ a
local smooth function of $x(t)$ such as $x^2(t)$, and $\int\! d\mu =
\int \! Dx \, \exp \left( - {1\over 2} m \int_0^t \dot x^2(t) dt
\right)$, the following limit exists and defines the symbol $\int\!
Dx$
$$\putequation{ratfunc}$$
Here $F\bigr[ V,N \bigr]$ contains the time discretization
$$\putequation{tdisc}$$
with $N\epsilon =t$ and $x_0 = x(t{=}0) = y$ is fixed. The path
integral in the denominator can be viewed as a measure for the path
integration. In the Hamiltonian approach, to which we now turn, one
{\it derives} the measure from first principles. This has the
advantage that one need not worry whether the measure in
(\puteqn{ratfunc}) is the only reasonable choice of measure.

Path integrals as such, and their normalization, were found by Feynman
in 1948[\putref{Feyn}], who went back to Dirac's work, but inserted
not only the $N\!-\!1$ complete sets of intermediate coordinate
eigenstates, but also $N$ complete sets of momentum eigenstates. He
considered quantum mechanical systems with a Hamiltonian of the form
$\hat H = T(\hat p) + V(\hat x)$, and using the relation
$$\putequation{Trotf}$$
he obtained phase space path integrals for a point particle with $n$
components by substituting the following expression into
(\puteqn{compl})
$$\putequation{intpj}$$
where $\Delta t = t_{j+1}-t_j$. For $T(p) = {1\over 2m} p^2$, he could
integrate over $p$ and obtain the configuration space path integrals
of Dirac, but with their normalization
$$\putequation{cspi}$$
Here $S=T-V$, but in Euclidean space one may start with $\exp \left( -
{1\over \hbar} H \Delta t \right)$ in (\puteqn{Trotf}) and then one finds
$\exp \left( -{1\over\hbar} S_E \right) $ in (\puteqn{cspi}) with $S_E
= T+V$.

In the actual loop by loop calculations we perform in the appendix, we
do not use (\puteqn{compl}) and (\puteqn{cspi}), but rather go over to
so-called mode-variables in the continuum theory $a_k$. One can
carefully follow the steps leading from the discrete-time variables to
the variables $a_k$, and then take the limit $N\rightarrow\infty$. The
result is that one obtains a nontrivial measure $d\mu =
(2\pi\epsilon\hbar)^{-n/2} \prod \left( {\pi k^2\over 4\epsilon\hbar}
\right)^{1/2} da_k$. Using this measure, propagators and vertices
follow, and loops can be calculated. For a detailed derivation of this
measure see [\putref{Bast},\putref{BaNi},\putref{Sken}].

Since Feynman, path integrals have taken an enormous flight. There
are two approaches: the phase space approach and the
configuration space approach. In the phase space approach there always
comes a moment where $\langle x_j| \exp \left( - {i\over\hbar}
\hat H \Delta t \right) | p_j \rangle$ has to be evaluated. It is
our contention that the evaluation of this basic matrix element is
always done incorrectly for nonlinear sigma models, even though for
models in flat space the
usual procedure gives the correct result.  Namely, without exception
(to our knowledge), it is always assumed that the following is a
correct approximation if one is interested in a result which is
corrrect up to order $\Delta t$
$$\putequation{pspi}$$
We have evaluated this matrix element exactly for nonlinear sigma
models, and found, as conjectured by Feynman, that one cannot neglect
terms of order $(\Delta t)^k$ in the expansion of the exponent for any
$k$. Most of these $\Delta t$
will be re-exponentiated and appear as $-{i\over 2m\hbar} p^2 \Delta
t$ in the exponent; they yield the normalization factor in
(\puteqn{cspi}) after integrating over $p$.  However, some terms
remain, proportional to the same exponent times a polynomial in
$p$. One should really keep all terms in the expansion of the
exponent, move all $\hat p$ to the right and all $\hat x$ to the left,
and each time a $\hat p$ passes a $\hat x$ it produces a factor
$-i\hbar$. In terms of rescaled variables $q=p \sqrt{\epsilon/\hbar}$
where $\epsilon = \Delta t$ one then obtains expressions of the form
$$\putequation{rescpi}$$
where $A$ contains one or three factors of $q$ and contains the terms
due to one $p$,$x$ commutator, while $B$ contains zero, two, four or
six $q$ momenta and contains the terms due to two $p$,$x$
commutators. Integration over $q$ replaces $q$ by $m
(x-y)/\sqrt{\epsilon\hbar}$, and since $x-y$ is of order
$\sqrt{\epsilon}$, we can stop the expansion in (\puteqn{rescpi}) at
$B$ for a two-dimensional quantum field theory.  The factors
$g^{-1/4}(x)$ and $g^{-1/4}(y)$ come from the inner product
$$\putequation{inpr}$$
which is a direct consequence of the orthogonality relations
$\sqrt{g(x)} \langle x | x^\prime \rangle = \delta(x-x^\prime)$ and
$\langle p | p^\prime \rangle = \delta(p-p^\prime )$. Putting $A=B=0$
amounts to the approximation in (\puteqn{pspi}), but is is clear from
the way we have displayed the further terms in (\puteqn{rescpi}) that
after integration over $q_i$ the terms with $A$ and $B$ will
contribute corrections to the propagator of order
$\sqrt{\epsilon\hbar}$ and $\epsilon\hbar$. This last term will give
the trace anomaly for $n=2$. For a
quantum field theory in $n_0$ dimensional space (or spacetime), one
must determine the expression within square brackets in
(\puteqn{rescpi}) to order $(\epsilon\hbar)^{n_0/2}$, which is
cumbersome. In the configuration space
approach[\putref{AGWi},\putref{Alva},\putref{Bast},\putref{BaNi}]
which uses the Schr\"{o}dinger equation, one must compute ``graphs''
on the worldline with $n_0$ loops, which is also cumbersome. Only for
chiral anomalies where the Grassmann integration over the fermionic
zero modes brings in factors of $\epsilon\hbar$ one need only expand the
configuration space path integral to second order in quantum
fluctuations[\putref{AGWi},\putref{Alva}]. There are then already
enough factors of $\hbar$ to ensure that it makes no difference
whether one takes the classical action or another action which
corresponds to a different operator ordering of the Hamiltonian.

{\bf\section{Path integrals with coherent states.}}

In this section we want to discuss phase space path
integrals based on coherent states. These are the path integrals one
uses for fermions (because for fermions $\psi$ and $\psi^\dagger$ are
more like $a$ and $a^\dagger$ then like $p$ and $x$). There are some
misconceptions about such path integrals, having to do with the
observation that a path connecting a point in phase space to another
point in phase space cannot in general be a classical path. We begin by
quoting Schulman[\putref{Schu}] who in his very readable
textbook on path integrals already anticipated the extra terms
in (\puteqn{rescpi}):``$\ldots$ No one (to my knowledge) has made a
serious investigation of the neglected terms $\epsilon (z_j-z_{j-1})$
[he refers here to the approximation in (\puteqn{pspi}) and the extra
terms in (\puteqn{rescpi})]. My own guess is that they can contribute
and that this contribution will be related to the operator ordering
problems in quantum mechanics $\ldots$''. We agree with his observation
on operator ordering: if the exponent of the Hamiltonian of the
nonlinear sigma model
would have been well-ordered in Dirac's sense, no extra terms would
have appeared in (\puteqn{rescpi}), but physics (the $n$-dimensional
quantum field theory) has given us another ordering. However, Schulman
considers at this point coherent states $|z,t\rangle$ (see below),
rather than the eigenstates $|x,t\rangle$ and $|p,t\rangle$ we
considered so far, and claims that the extra terms are produced
because the curves $z(t)$ are more singular than the Brownian motion
paths that enter the usual path integrals (for which $(\Delta x)^2
\sim \Delta t$). We do not agree with him on this point; already in
(\puteqn{rescpi}) which does not use coherent states one encounters
the extra terms. And he goes on[\putref{Schu}]:``$\ldots$ A well known
feature (see Section 31) of the phase space path integral is that
paths in phase space are discontinuous so that a term $\epsilon \Delta
p$ need not go to zero any faster than $\epsilon$ $\ldots$''. Indeed,
if $p(t)$ is the time derivative of $x(t)$, and the paths $x(t)$ are
continuous but not differentiable, then the paths $p(t)$ will be
discontinuous. However, looking up this Section 31, one discovers that
he considers the following path integrals
$$\putequation{phsppi}$$
He then states[\putref{Schu}]:``$\ldots$ we interpret the formula
(\puteqn{phsppi}) as a sum over paths [correct]. The integration
variables suggest that one is adding trajectories in phase space
[incorrect in our opinion] $\ldots$ However, $\ldots$ the fact that a
single point in phase space determines the classical path, causes some
difficulty [indeed]. Here are two interpretations $\ldots$''. And he
goes on to consider either nonclassical paths in phase space for which
$x_j$ is continuous but $p_j$ jumps at the end of each time segment,
or classical paths for which $p(t_j + {1\over 2} \Delta t) =p_j$ and
$x(t_j+\Delta t)=x_{j+1}$ with $t_{j+1}-t_j=\Delta t$. We claim that
in (\puteqn{phsppi}) one is considering ordinary paths in
configuration space, with beginpoint $x_0=y$ and endpoint $x_N=x$. For
coherent states things are more subtle as we now discuss.

We define coherent states for bosonic point particles by
$$\putequation{defcoh}$$
where $z^* = (z)^*$ is the complex conjugate of $z$, and
$z=(x+ip)/\sqrt{2\hbar}$ while $z^* = (x-ip)/\sqrt{2\hbar}$. (This
corresponds to a harmonic oscillator with $\hat H = {1\over 2} \hat
p^2 + {1\over 2} \hat x^2 = \hbar\bigl( a^\dagger a + {1\over
2}\bigr)$. For $\hat H = {1\over 2m}
\hat p^2 + {1\over 2} m \omega^2 \hat x^2$, one must use $\hat x
\sqrt{m\omega}$ and $\hat p / \sqrt{m\omega}$ in these
definitions). The vacuum is annihilated by $a$, and one has the
following inner product and decomposition of unity
$$\putequation{inpcoh}$$
The symbol ${dz dz^* \over 2\pi i}$ always denotes ${dxdp\over
2\pi\hbar}$, and we stress that $z$ and $z^*$ are not independent
complex variables. The propagator $\procoh \equiv \langle w^* | \exp
\left( - {i\over\hbar} \hat H T \right) | z \rangle $ is then equal to
$$\putequation{defprocoh}$$
with $\Delta t = T/N$. Note that this is an exact result for any $N$.
In section 4 we shall compute for small
$\Delta t$ the matrix elements of $\exp\left( -{i\over\hbar} \hat H
\Delta t \right)$. Anticipating this calculation, let us define a
function $h(z^*_{j+1},z_j)$ by
$$\putequation{defscrham}$$
Then the integrand in the propagator can be written as
$$\putequation{intprop}$$
where $z_0=z$ and $z^*_N=w^*$. This is the formula referred to in the
introduction, from which the propagator and its trace are obtained.

For the trace anomaly one should take the trace of the propagator and
evaluate
$$\putequation{trace}$$
where $w^*$ must be equated to $(z)^*$. The integration in
(\puteqn{trace}) is then again an ordinary Gaussian integral. To prove
(\puteqn{trace}), one may either use (\puteqn{inpcoh}) or expand the
coherent states into harmonic
oscillator eigenfunctions $|n\rangle =
(n!)^{-1/2}(a^\dagger)^n|0\rangle$. The result reads then indeed
$\sum_n \langle n | \exp \left( - {i\over\hbar} H \Delta t \right) | n
\rangle $. The extra term $e^{-w^*z}$ in (\puteqn{trace}) completes
the exponent of (\puteqn{intprop}) to an action provided we identify
$z=z_0$ with $z_N$, hence $z=z_0=z_N$
$$\putequation{pipbc}$$
This path integral has then periodic boundary conditions. In terms of
$x$ and $p$ the kinetic term reduces to the familiar $-{i\over\hbar} p
\dot x $. For the computation of other anomalies one must evaluate the
trace of $\langle w^* | \hat J \exp\left( - {i\over\hbar} \hat H
\Delta t \right) | z \rangle$, where $\hat J$ is the quantum
mechanical operator which corresponds to the Jacobian of an
infinitesimal symmetry transformation of the quantum field
theory. Depending on the problem, the presence of $\hat J$ can be
translated into a change in boundary conditions (periodic,
antiperiodic, etc.[\putref{Alva}]).

For other problems than anomalies, one may need the propagator itself.
In order to still be able to write (\puteqn{intprop}) as a path
integral, we no longer have the extra exponent due to the trace
operation but rather we must add and substract a suitable term by
hand. In principle, any term will do, but for practical calculations
it is very useful to decompose $z(t)$ and $z^*(t)$ into a classical
part which satisfies the equation of motion, and a quantum part over
which one still integrates with the measure ${dz^*_j dz_j\over 2\pi i}
e^{-z^* z}$. Terms linear in the quantum field then cancel from the
action, and the corrections begin with the harmonic (quadratic)
approximation, and then higher terms. To achieve this one considers
{\it two independent} classical trajectories $z_{\rm cl}(t)$ and
$w_{\rm cl}^*(t)$, uniquely specified by $z_{\rm cl}(0)=z$ and $w_{\rm
cl}^*(T)=w^*$. Of course, the value of $z_{\rm cl}(T)$ at $t=T$ is
then not $w$, nor is $w_{\rm cl}^*(t{=}0)$ equal to $(z)^*$. Rather, to
avoid confusion, we keep denoting them by $z_{\rm cl}(T)$ and
$w^*_{\rm cl}(0)$. We then
complete (\puteqn{intprop}) by adding and substracting a term $w^*
z_{\rm cl}(T)$. This yields as path integral
$$\putequation{pathcoh}$$
One can also treat the beginpoint and endpoint more symmetrically, by
adding one-half the above term, and another term given by ${1\over 2}
w^*_{\rm cl}(0) z$. In this case one finds the path integral[\putref{textb}]
$$\putequation{pathcohsym}$$
The boundary conditions are now: $z^*(T)=w^*$, $z^*(0)=w^*_{\rm
cl}(0)$, $z(T)=z_{\rm cl}(T)$, $z(0)=z$.
In the action one is then to replace $z(t)$ by $z_{\rm cl}(t) + z_{\rm
qu}(t)$ and similarly $z^*(t)$ by $w^*_{\rm cl}(t) + z^*_{\rm qu}(t)$.
The quantum fields vanish at the boundaries
$$\putequation{qufi}$$
and we still have $z = (x+ip)/\sqrt{2\hbar}$, $z^* =
(x-ip)/\sqrt{2\hbar}$, with real $x$ and $p$. The measure ${Dz^*
Dz\over 2\pi i} = {Dx Dp\over 2\pi\hbar}$ defines an ordinary Gaussian
integral.  One could define a classical $p$ and $x$ by
$$\putequation{xpcl}$$
In that case, $x_{\rm cl}(t) = (w^*_{\rm cl}(t)+ z_{\rm
cl}(t))\sqrt{\hbar /2} $ and in particular $x_{\rm cl}(t{=}0),x_{\rm
cl}(t{=}T)$ are
{\it complex} (because $w^*_{\rm cl}(t{=}0) \neq z^*$). Similarly for
$p_{\rm cl}(t)$.  There is no problem with classical paths connecting
two points in phase space because we have here {\it two} classical
paths which do not connect $w^*$ and $z$, but one of which starts at
$z$ and ends wherever the dynamics may take it, while the other starts
at $w^*$ and ends somewhere else. No discontinuity or jumps of paths
need be considered.

To evaluate (\puteqn{pathcohsym}) for the harmonic oscillator is
trivial [\putref{textb}]. Terms linear in quantum fields cancel
whereas the terms quadratic in quantum fields are independent of the
classical fields and yield an overall constant, so that the classical
factor in front of the integral in (\puteqn{pathcohsym}) contains the
whole result up to a constant. This constant can be fixed by
considering the limit $T\rightarrow 0$, or by doing the path integral
over $z$ and $z^*$ explicitly. For more complicated systems, one can
evaluate the path
integral by expanding the action in terms of classical and quantum
fields, but then $z^*_{\rm qu}(t)$ is not an independent complex
variable, but rather expressed in terms of $\left( z_{\rm qu}(t)
\right)^*$ by $z^*_{\rm qu}(t) = \left( z_{\rm cl}(t) \right)^* -
w^*_{\rm cl}(t) + \left( z_{\rm qu} (t) \right)^*$. We can define
$x_{\rm qu}$ and $p_{\rm qu}$ by $z_{\rm qu} = \left( x_{\rm qu} + i
p_{\rm qu} \right) / \sqrt{2\hbar}$ and $z^*_{\rm qu} = \left( x_{\rm
qu} -i p_{\rm qu} \right) / \sqrt{2\hbar}$, but then also $x_{\rm
qu}(t)$ and $p_{\rm qu}(t)$ are complex. Using
$$\putequation{defxpqu}$$
it is clear that the Jacobian for the transformation of the
integration variables $x(t)$ and $p(t)$ to $x_{\rm qu}(t)$ and $p_{\rm
qu}(t)$ is unity, so that the measure becomes
$$\putequation{meas}$$
where $a_j = {1\over2} \left( z_{\rm cl}(t_j) + w^*_{\rm cl}(t_j)
\right)$ and $b_j = {1\over 2i} \left( z_{\rm cl}(t_j) - w^*_{\rm
cl}(t_j) \right)$ are complex constants for each $t_j$. In other
words, $x_{\rm qu}(t)$ and $p_{\rm qu}(t)$ lie on a line in the
complex plane parallel to the real axis. In a loop expansion, the
integrand is analytic in all variables, and we may simplify the
integrand to $z_{\rm qu}(t) = \left( x_{\rm qu}(t) + i p_{\rm qu}(t)
\right) / \sqrt{2\hbar}$ and $z^*_{\rm qu}(t) = \left( x_{\rm qu}(t) -
i p_{\rm qu}(t) \right) / \sqrt{2\hbar}$ with real $x_{\rm qu}(t)$ and
$p_{\rm qu}(t)$, and the integration measure becomes
$$\putequation{newmeas}$$

For fermions we consider operators $\hat\psi^\dagger$ and $\hat\psi$
satisfying $\{ \hat\psi, \hat\psi^\dagger \} = 1$, and {\it
independent} Grassmann variables $\bar\eta$ and $\eta$. This will
force us to double the number of fermionic operators, from
$\hat\psi^a_1$ to $\hat\psi^a_\alpha$ as in (\puteqn{Hferm}). The
operators $\hat\psi^a$ and $\hat\psi^{a\dagger}$ are the defined by
$(\hat\psi^a_1 + i \hat\psi^a_2)/\sqrt{2}$ and $(\hat\psi^a_1
-i\hat\psi^a_2)/\sqrt{2}$, respectively. We continue in this section
with the case $n=1$, omitting the superscript $a$. The coherent states
are now defined by
$$\putequation{cohfer}$$
satisfying $\hat\psi |\eta\rangle = \eta |\eta\rangle$ and $\langle
\bar\eta | \hat\psi^\dagger = \langle \bar\eta | \bar\eta $. The inner
product and decomposition of unity read formally the same as in the
bosonic case
$$\putequation{inpcofer}$$
but now the ordering of $d\bar\eta$ and $d\xi$ as well as the ordering
of $\bar\eta$ and $\xi$ in $\bar\eta \xi$ is important, while the
factor $(2\pi i)^{-1}$ is absent from the measure because we now have
Grassmann integration according to which $\int\! d\xi \, \xi = \int\!
d\bar\eta \, \bar\eta =1$. The results in (\puteqn{inpcofer}) follow
directly by expanding the exponent; one finds $\unit = |0\rangle
\langle 0| + |1\rangle \langle 1|$ with $\hat\psi^\dagger |0\rangle =
|1\rangle$, which is evidently correct. We can repeat the same steps
as before, leading to (\puteqn{intprop}). The propagator is then given
by
$$\putequation{propcohfer}$$
or, in the symmetric case, by
$$\putequation{prchfersym}$$
Again we obtain the well known extra boundary terms in these path
integrals with fermionic coherent states[\putref{textb}].
On the other hand, the trace formula is given by
$$\putequation{fertra}$$
where now $\xi$ and $\bar\eta$ are independent variables. The proof
follows directly from Grassmann integration and shows that $d\xi$
stands to the left of $d\bar\eta$. The trace of the
propagator becomes then
$$\putequation{traprofer}$$
This can be interpreted as a phase space path integral with
$\int\!dt\, (-\bar\eta \dot\xi - {i\over\hbar} h)$, provided one
interprets $\xi(t{=}0)\equiv \xi_0 = -\xi(t{=}T)$. By putting the term
$\bar\eta \xi$ in the exponent to the far right, one again finds a
path integral, now with $\int\!dt\, (\dot{\bar\eta} \xi - {i\over\hbar}
h)$, provided $\bar\eta (t{=}0) \equiv -\bar\eta = -\bar\eta (t{=}T)$.
Taking the symmetric case, we recover (\puteqn{pipbc}), but now with
antiperiodic boundary conditions for $\xi$ and $\bar\eta$. For other
anomalies with $\hat J$ not equal to unity, the boundary conditons on
$\xi$ and $\bar\eta$ may be different. For example, for chiral
anomalies one obtains periodic boundary conditions for $\xi$ and
$\bar\eta$ [\putref{Alva}].

{\bf\section{Transition amplitude for the bosonic case.}}

We will consider the matrix element
$$\putequation{Matrelt}$$
with
$$\putequation{defham}$$
We are working from now on in Euclidean space; to obtain the results
for Minkowski
spacetime one should replace $\epsilon$ by $i\epsilon$. For $\xi =
{1\over 4} {n-2\over n-1}$, $\hat H$ is the regulator for a classically
conformal invariant scalar quantum field theory in $n$
dimensions[\putref{Bast}]. Expanding the exponent in
(\puteqn{Matrelt}), we define
$$\putequation{defAkl}$$
where $A^k_l(x)$ is a $c$-number function and $p^l$ denotes a
homogeneous polynomial of order $l$ in the momenta.

In order to compute the transition amplitude $\Ampl$ to order
$\epsilon$ compared to the leading terms, it will turn out that we
only need terms on the right hand side of (\puteqn{defAkl}) with
$l=2k$, $2k-1$, and $2k-2$. We find, defining $p^2 = g^{ij}(x) p_i p_j
$,
$$\putequation{A2k}$$
Since this is the term containing the maximal number of $p$'s, it can
be easily computed because all $\hat p $ operators are just replaced
by the corresponding $c$-numbers when acting on $| p \rangle $.

The next term is
$$\putequation{A2kmin1}$$
In this expression one of the $\hat p$'s acts as a derivative, whereas
the other $2k-1$ are replaced by the corresponding $c$-numbers. The
first term in (\puteqn{A2kmin1}) comes about when the derivative acts
within the same factor $\hat H$ in which it appears, and is multiplied
by $k$ since there are $k$ factors of $\hat H$. The second term arises
if this derivative acts on a different factor of $\hat H$. For this to
occur there are $k\choose 2$ possible combinations, and taking into
account that there are two $\hat p$'s in each factor of $\hat H$ we
get an extra factor $2$. Notice that in both cases the terms involving
a derivative acting on $g$ cancel.

The last term we have to calculate is
$$\putequation{A2kmin2}$$
In this expression two of the $\hat p$'s act as derivatives, except
for the last term which is already of order $p^{2k-2}$.

The first set of terms appears when both derivatives act within the
same factor $\hat H$; again there are $k$ terms of this kind.

The next set of terms arises when only two of the factors $\hat H$
play a r\^{o}le. There are four possibilities: $(i)$ one $\hat p$ from
the left factor acts on the right factor, while another $\hat p$ from
the right factor acts within the right factor, $(ii)$ the first $\hat
p$ acts within the first $\hat H$, while the second $\hat p$ acts
within the second $\hat H$, $(iii)$ both $\hat p$'s come from the left
$\hat H$, but one of them acts inside the left $\hat H$ while the
other acts on the right $\hat H$, and $(iv)$ both $\hat p$'s from the
left $\hat H$ act on the right $\hat H$. In all cases it is easy to
see that again the derivatives on $g$ cancel.

The following set of terms comes from combinations using three factors
$\hat H$, hence its overall factor $k\choose 3$. There are again four
cases: $(i)$ a $\hat p$ from the first $\hat H$ and a $\hat p$ from
the second $\hat H$ hit the third $\hat H$, $(ii)$ one $\hat p$ acts
inside the factor $\hat H$ in which it appears whereas a $\hat p$ from
another $\hat H$ hits the remaining $\hat H$ (there are $3$ terms of
this kind), $(iii)$ a $\hat p$ from the first $\hat H$ hits the second
$\hat H$, and a $\hat p$ from the second $\hat H$ hits the third $\hat
H$, and $(iv)$ of the two $\hat p$'s from the first $\hat H$ one acts
on the second, and one on the third $\hat H$.

Finally, the term with $k\choose 4$ involves four factors $\hat H$,
such that one $\hat p$ from one $\hat H$ hits another $\hat H$, and
the other $\hat p$ from one of the remaining factors $\hat H$ hits the
last $\hat H$.

The reason further terms do not contribute can be most easily seen if
we rescale $q = \sqrt{{\epsilon\over\hbar}}p$. Then the transition
amplitude becomes
$$\putequation{ordEps}$$
where we have used (\puteqn{inpr}). So only the $A^k_{2k-1}$ and
$A^k_{2k-2}$ terms contribute through order $\epsilon$ compared to the
leading term $A^k_{2k}$.

The sum over $k$ in (\puteqn{ordEps}) can be performed, leading to
$$\putequation{momint}$$
We can now complete the square in the exponent and integrate out the
momenta $q_i$, since the integral becomes just a sum of Gaussian
integrals which can easily be evaluated. The problem is now to
factorize the result such that it is manifestly a scalar both in $x$
and in $y$ (a `bi-scalar') under general coordinate transformations.
We expect, of course, to find at least the classical action integrated
along a geodesic (see (4.11)). In the expansion of this
functional around $x(0)=x$, we recognize many of the terms in
(\puteqn{momint}). However, there are terms left over. They combine
into $R$ or $R_{ij}$, while expansion of $g(y)$ yields terms with
$\partial \log g$ or derivatives thereof. With this in mind, we write
the result in a factorized form, where in one factor we put all terms
which possibly can come from expanding some power of $g(y)$, while
into another factor we put the expanded action and curvature terms. It
is quite nontrivial, and an excellent check on the results obtained so
far, that this is at all possible. The resulting expression is
$$\putequation{factTra}$$
where the Ricci tensor is defined by $R_{ij}=R_i{}^k\!{}_{kj}$ and
$R_{ijkl}$ is given in (5.4).  The terms within the
first pair of square brackets are, through order $\epsilon$, equal to
$g^{1/4}(x) g^{1/4}(y) $ and cancel the factors $g^{-1/4}(x)
g^{-1/4}(y) $ in front of the whole expression. Note that from the
term in the exponent in (\puteqn{factTra}) it follows that the
difference $(y-x)$ is of order $\sqrt{\epsilon}$, thus one indeed
finds this expansion through order $\epsilon$. Similarly the terms
within the second pair of square brackets are of order $\epsilon$ or
less. The terms with $\partial_kg_{ij}$ and its square are the first
two terms in the expansion of an exponent. This suggests to
exponentiate all terms, yielding
$$\putequation{infampl}$$
All terms in the exponent except the last two just correspond to an
expansion around $x$ of the classical action, which is equal to the
integral along the geodesic joining $x$ and $y$ of the invariant line
element (cf. [\putref{DeWi}])
$$\putequation{expAct}$$
when boundary conditions $x(-\epsilon )=y$, $x(0)=x$ are imposed.

Our final result for the transition amplitude can thus be written as
$$\putequation{final}$$
This shows that (\puteqn{infampl}) is, to the order we are
expanding, symmetric under the exchange of $x$ and $y$. One may check
that it satisfies the composition rule
$$\putequation{comp}$$
A quick way to check this is to use normal coordinates, in which case
one only needs to retain the leading term in the classical action
(since $\partial_ig_{jk}=\partial_{(i} \partial_j g_{kl)}=0$), while
$g_{ij}(z) = g_{ij}(x)+ {1\over 3} R_{kijl}(x)(z-x)^k(z-x)^l $ through
order $\epsilon$. Taking the opposite point of view, we can impose
(\puteqn{comp}) as a consistency condition on the amplitude. This
fixes the higher order terms in the expansion of (\puteqn{final}),
with exception of the terms proportional to the scalar curvature,
which are related to the trace anomaly. One should expect this to be
the case since we can always add such a term to the Lagrangian and
interpret it as an external potential, which should not be fixed by
the requirement (\puteqn{comp}).

The transition element can of course also be evaluated using path
integral methods. One should be careful however in choosing the
correct action in the configuration space path integral, since our
Hamiltonian (\puteqn{defham}) is not Weyl-ordered[\putref{Weyl}] or
time-ordered[\putref{DeWib}]. The
corresponding Weyl-ordered or time-ordered Hamiltonian is (see e.g.
[\putref{Mizr},\putref{DeWib}]), dropping a noncovariant term
proportional to the product of two Christoffel symbols for reasons
given in the conclusions,
$$\putequation{HWeyl}$$
Since it is $\hat H_{\rm Weyl}$ that corresponds to the classical
action in the path integral,
the (Euclidean) action that corresponds to $\hat H$ in
(\puteqn{defham}) is
$\Sco = \Scl + {1\over 8} \hbar^2 \int \! dt \, R $, in agreement with
the results in [\putref{Bast}].

We have obtained the result for $\Ampl$ from a direct computation; the
extra $R$ and $R_{ij}$ terms are due to ordering the $\hat p$
operators in all terms of the expansion of $\exp \left( -
{\epsilon\over\hbar} \hat H \right)$. The terms with $R_{ij}$ can be
expressed in terms of the classical action as follows. By using
$$\putequation{defD}$$
where $D(x,y;\epsilon ) = \det D_{ij}(x,y;\epsilon)$ is the Van Vleck
determinant [\putref{Vlec},\putref{More}]
$$\putequation{defDij}$$
we can rewrite (\puteqn{final}) as
$$\putequation{rewr}$$
where $\tilde D = g^{-1/2}(x) D(x,y;\epsilon) g^{-1/2}(y)$ and we have
neglected terms of higher order in $\epsilon$. Note that $\tilde D $
as well as $S_{{\rm cl}}(x,y;\epsilon)$ transform as biscalars under
general coordinate transformations ($D$ itself is a bi-density
[\putref{DeWia}]), hence also the infinitesimal transition amplitude
$\Ampl$ does not depend on the coordinates chosen. Obviously this
result is not affected by the terms proportional to the scalar
curvature.

The Van Vleck determinant gives the harmonic approximation of the
quantum terms in the path integrals[\putref{Schu}]. For example, the
Feynman factor $\left( {m\over \Delta t} \right)^{n/2}$ in
(\puteqn{cspi}) is a Van Vleck
determinant. Since ${\partial \over \partial y^j} S_{{\rm
cl}}(x,y;\epsilon ) = -p_j $ with $p_j$ the momentum conjugate to
$y^j$, $D$ is the Jacobian for the transition of the phase space
measure $dx_2dp_1dx_1$ to the configuration space measure $dx_2dydx_1$
where $y$ divides the interval $(x_1,x_2)$ into two smaller intervals
of size ${1\over 2} \Delta t$. To the order we working in, $\tilde D$
for the half-interval is equivalent to $\tilde D^{1/2}$ over the whole
interval and reproduces the $R_{ij}$ terms. One might think that one
could explain the $R_{ij}$ terms in the propagator by taking as
$\exp\bigl(\Sco\bigr)$ the expression $\tilde
D^{1/2}\exp\bigl(\Scl\bigr)$. However, this is incorrect because the
path integral for this $S_{{\rm conf}}$ produces its own $R_{ij}$
terms, on top of those contained in $\tilde D$, as was already noted
in [\putref{Bast}].

We conclude that the final result can be written as the product of
three factors: one term which is only related to the trace anomaly,
the exponent of the classical action, and the square root of the Van
Vleck determinant.

As one would expect, the transition amplitude is the Green function
associated with the Schr\"{o}dinger equation (diffusion equation for
Euclidean time)
$$\putequation{Schrod}$$
where
$$\putequation{defHamy}$$
To prove (\puteqn{Schrod}) we begin with the identity
$$\putequation{ddeps}$$
If we would expand the propagator around $y$, and act on it with $\hat
H(x)$ as in (\puteqn{ddeps}), we obtain the same result as acting with
$\hat H(y)$ on our propagator (which is expanded around $x$), because
the right hand side of (\puteqn{Schrod}) is symmetric in $x$ and $y$.
Using the footnote in section 2 and (\puteqn{Schrod}) we may check
(\puteqn{infampl}).

We can now simply compute the anomaly in $n=2$ by taking the trace in
(\puteqn{final}) and singling out the $\epsilon $ independent part.
The result is[\putref{Bast},\putref{BaNi}]
$$\putequation{An2}$$

{\bf\section{The classical and quantum Hamiltonians for the $N{=}2$
supersymmetric nonlinear \\ sigma model.}}

As an example of a case where the operator orderings in the
Hamiltonian are unambiguously fixed by requiring that the
corresponding configuration space path integral has the original
classical action in its exponent, consider $N{=}2$ supersymmetric
quantum mechanics. The classical action in Minkowski time is given by
$$\putequation{Lsusy}$$
and is invariant under the following supersymmetry transformations
$$\putequation{susytra}$$
This action can either be obtained by dimensional reduction of the
$N{=}(1,1)$ supersymmetric nonlinear $\sigma $ model from $1+1$
dimensions to one time dimension[\putref{AGWi}], or directly by
requiring invariance
under $(1,1)$ supersymmetry transformations with parameter
$\varepsilon_\alpha $. One finds either way the classical supersymmetry
transformations, and the classical supersymmetry charge. The latter is
given by
$$\putequation{Qcl}$$
To define the corresponding quantum generator, it is helpful to
introduce fermions with tangent-space indices, because their brackets
are field-independent. Therefore define $\psi^a_\alpha = e_i^a(\phi)
\chi^i_\alpha $ where $e^a_i e^b_j \delta_{ab} = g_{ij} $.
Then the action in Minkowski time becomes
$$\putequation{Lsusyflat}$$
The supersymmetry charge becomes
$$\putequation{Qclflat}$$
So far, this is standard supergravity.

To quantize this system, we first observe that there are second class
constraints
$$\putequation{constr}$$
The Poisson brackets are given by $ \bigl\{ p(\psi^a_\alpha ),
\psi^b_\beta \bigr\}_P = - \delta^{ab} \delta_{\alpha\beta} $, where
$ p(\psi^a_\alpha ) = {\partial \over \partial \dot\psi^a_\alpha }
{\cal L} $ (the minus sign in the Poisson bracket is not a matter of
convention but follows from requiring the Heisenberg field equations
to hold). Using $ \bigl\{ C^a_\alpha , C^b_\beta \bigr\}_P = -i
\delta^{ab} \delta_{\alpha\beta} $, we find the Dirac brackets
$$\putequation{DiracB}$$
which is as usual half the Poisson brackets. Thus the quantum
anticommutator reads (we set $\hbar =1$ in this section)
$$\putequation{Anticom}$$
Furthermore, the conjugate momentum of $\phi^i$ is given by
$$\putequation{momentum}$$

We now choose as quantum supersymmetry charge
$$\putequation{Qqu}$$
We choose this particular ordering because then $Q_\alpha^{({\rm
qu})}$ is hermitian, and hence also $H^{({\rm qu})}$ will be
hermitian. To prove that this $Q_\alpha^{({\rm qu})}$ is hermitian,
one may show that after hermitian conjugation one gets three extra
terms which cancel due to the ``vielbein postulate'' that the total
covariant derivative of the vielbein (with Cristoffel and spin
connection) cancels. We evaluate the
quantum anticommutator $ \bigl\{ Q_\alpha^{({\rm qu})}, Q_\beta^{({\rm
qu})} \bigr\} $ and show that the result is equal to $
\delta_{\alpha\beta} H^{({\rm qu})} $ with $H^{({\rm qu})}$ by
definition the quantum Hamiltonian. Classically, the Hamiltonian is
given by ${1\over 2} g^{ij} \pi_i \pi_j - {1\over 8} R_{abcd}(\omega)
\bigl( \psi^a_\alpha \psi^b_\alpha \bigr) \bigl( \psi^c_\beta
\psi^d_\beta \bigr) $, but quantum mechanically there are many
possibilities for the Hamiltonian, because $ \bigl\{ \psi^a_\alpha ,
\psi^b_\beta \bigr\} X_{ab}^{\alpha\beta} $ vanishes classically for
any $X$, but equals $\delta^{ab} \delta_{\alpha\beta}
X^{\alpha\beta}_{ab}$ at the quantum level. We define the Hamiltonian
as the square of the supersymmetry charge because, as we shall discuss
in section 6, when this $H^{({\rm qu})}$ is used in $\Samhis1 $ it
gives the same propagator as the path integral in which the original
supersymmetric action appears. Hence, we have chosen what might be
called supersymmetric ordering for $Q_\alpha^{({\rm qu})}$ and
$H^{({\rm qu})}$.  Also,
with this ordering presciption, $Q_\alpha^{({\rm qu})}$ and
$H^{({\rm qu})}$ transform as scalars under
general coordinate transformations applied to operators (cf.
[\putref{DeWib}]).

The direct evaluation of $\bigl\{ Q_\alpha^{({\rm qu})},
Q_\beta^{({\rm qu})} \bigr\} $ is tedious, but straightforward. Using
$$\putequation{commut}$$
one finds, keeping all operators in the order
they come,
$$\putequation{QQ}$$
(We used the vielbein postulate $\partial_j e^i_a = - \Gamma^i_{jk}
e_a^k - \omega_{ja}{}^b e_b^i $ and found that all $\Gamma $ terms
cancel). The last term vanishes classically, but at the quantum level
it is nonvanishing; in fact, it just covariantizes the $\pi^2$ term
to the complete d'Alembertian for a scalar particle! Using
$$\putequation{pipi}$$
and the cyclic identity
$$\putequation{cycl}$$
we see that also the four-fermion term is proportional to
$\delta_{\alpha\beta} $. Our final result reads
$$\putequation{QQisHqu}$$

Thus, imposing the quantum algebra $\bigl\{ Q_\alpha^{({\rm qu})},
Q_\beta^{({\rm qu})} \bigr\} = \delta_{\alpha\beta} H^{({\rm qu})} $
with given $Q_\alpha^{({\rm qu})}$ has fixed the operator ordering in
$H^{({\rm qu})}$ as given in (\puteqn{QQisHqu}). Note that this
quantum Hamiltonian is already Weyl-ordered\footnote[$\
\ddagger$]{We obtain this covariant result if we use a Weyl ordering
in which we consider $\pi$
as an independent operator, and not $p$.  Note that DeWitt in
[\putref{DeWib}] at this point switches notation, and denotes
$g_{ij}\dot x^j$ by $p_i$ (his equation 6.7.26).  Weyl ordering (which he
discusses when $p$ and $\pi$ coincide) in terms of his new $p$ and $x$
then indeed agrees with our observation that (\puteqn{QQisHqu}) is
already Weyl-ordered. However, from a canocical point of view, $\pi$
is not an independent operator. We discuss this point further in the
conclusions.}: the term with $\pi^2$ yields upon
Weyl-ordering again a term $+{1\over 8} R$, see (\puteqn{HWeyl}),
whereas the $R\psi^4$ term contributes $-{1\over
8}R$, cancelling the bosonic contribution. To show this latter result,
write the Weyl-ordered operator corresponding to ${1\over 8}
R_{abcd}\psi^a_\alpha \psi^b_\alpha
\psi^c_\beta \psi^d_\beta = {1\over 2} R_{abcd}\bar\psi^a \psi^b
\bar\psi^c \psi^d$ (using (6.3)) as
$$\putequation{HSWeyl}$$
where $t_a$ denotes the time associated with $\psi^a$, and the limit
$t_a,t_b,t_c,t_d \rightarrow t$ is understood. We do not have to
include the curvature term in the Weyl ordering because it commutes
with all the fermions. To compute the difference between
(\puteqn{QQisHqu}) and (\puteqn{HSWeyl}), we decompose unity in a sum
of products of $\theta$-functions, and write the operator in
(\puteqn{QQisHqu}) as follows
$$\putequation{1isth}$$
Now substract (\puteqn{1isth}) from (\puteqn{HSWeyl}), using the
anticommutation relations (\puteqn{Anticom}). When we make use of
identities such as
$$\putequation{thiden}$$
then, in the limit that all times coincide, we find that the
difference equals $- {1\over 8} R$. So the two extra terms due to Weyl
ordering the two terms in $H^{({\rm qu})}$ in (\puteqn{QQisHqu})
cancel, and indeed the corresponding configuration space path integral
which follows from this particular $\hat H^{({\rm qu})}$ has the
classical supersymmetric action in its exponent.

{\bf\section{Transition amplitude for $N{=}2$ supersymmetric sigma model.}}

We will now compute the matrix element
$$\putequation{SMatelt}$$
where
$$\putequation{defSham}$$
As shown in (\puteqn{QQisHqu}), this is the quantum Hamiltonian for
the $N{=}2$ supersymmetric nonlinear sigma model.  The states $| \chi
\rangle $ and $ \langle \bar\eta | $ are eigenstates of the operators
$\psi^a$ and $\bar\psi^b $ respectively, $\psi^a | \chi \rangle =
\chi^a | \chi \rangle$ and $\langle \bar\eta | \bar\psi^b = \langle
\bar\eta | \bar\eta^b $. They are coherent states, $| \chi \rangle =
\exp\bigl( \bar\psi^a \chi^a \bigr) | 0 \rangle$ and $\langle \bar\eta
| = \langle 0 | \exp\bigl( \bar\eta^a \psi^a\bigr)$, and satisfy the
completeness relation $\unit = \int \! \bigl( d\bar\xi^1 d\xi^1 \ldots
d\bar\xi^n d\xi^n \bigr) \, \exp \bigl( - \bar\xi^a \xi^a \bigr)
|\bar\xi \rangle \langle \xi | $, and inner product $ \langle\bar\eta
| \xi \rangle = \exp \bigl( \bar\eta \xi \bigr) $.

Compared to the $D_i$ appearing in the regulator for the trace anomaly
of a spin-${1\over 2}$ field ($\hat {\cal R}_{fer} = g^{-1/4}\Dsl \Dsl
g^{1/4} $), we have twice as many fermionic degrees of freedom
$(\alpha = 1,2)$.  The case with half as many fermionic degrees of
freedom (with $\alpha=1$ in (\puteqn{defcovD})), does correspond to a
quantum field theory for spin-${1\over 2}$ fields and will be
discussed in section 7. We define
$$\putequation{defferm}$$
Analogously to the bosonic case we expand the exponent in
(\puteqn{SMatelt}) and define
$$\putequation{defBkl}$$
Again we will only need the terms with $l=2k$, $2k-1$, and $2k-2$.
It will be convenient to express the Hamiltonian as $\hat H =
\hat\alpha + \hat\beta + \hat\gamma$, with
$$\putequation{Hparts}$$
The leading term is the same as in the bosonic case
$$\putequation{B2k}$$
The next term equals
$$\putequation{B2kmin1}$$
The first term comes from $k$ factors $\hat\alpha$, and is identical
to what the purely bosonic case yields, whereas the second term arises
when we combine $k-1$ factors $\hat\alpha$ with one factor
$\hat\beta$, and replace all operators by their corresponding c-number
or Grassmann number values. The last term we have to compute is
$$\putequation{B2kmin2}$$
There are four possible combinations that give a contribution: $(i)$
$k$ factors $\hat\alpha$, yielding the same expression as in the
bosonic case, $(ii)$ $k-1$ factors $\hat\alpha$ and one factor
$\hat\beta$; now one $\hat p$, either from one of the $\hat\alpha$'s
or from $\hat\beta$ acts as a derivative, again either on one of the
$\hat\alpha$'s or on $\hat\beta$, yielding the next four terms,
$(iii)$ $k-1$ factors $\hat\alpha$ and one factor $\hat\gamma$ and
replacing all $\hat p$'s by their c-number values, and $(iv)$ $k-2$
factors $\hat\alpha$ and two factors $\hat\beta$, which gives rise to
the last term. Notice that to evaluate the last two terms we first
needed to normal order the fermionic operators which produced extra
terms proportional to $\delta^{ab}$ according to (\puteqn{defferm}).
In detail, we write $\bar\psi^a \psi^b \bar\psi^c \psi^d$ as
$\bar\psi^a \psi^d \delta^{bc} - \bar\psi^a \bar\psi^c \psi^b \psi^d$.
We can now proceed in the same way as in the bosonic case, by doing
the sum over $k$, extracting a factor $\exp \bigl( - {1\over
2\epsilon\hbar} p^2 \bigr) $, and performing the Gaussian integral
over the momenta. In the last term of (\puteqn{B2kmin2}), the integral
over $p_jp_l$ contributes a term with $g_{jl}$ which cancels the
complete $\omega^2$ term in the line above it, plus a term with
$(y-x)_j(y-x)_l$. In this result, the contribution proportional to
$\omega^2 \bar\eta^2 \xi^2$ is the square of the second term on the
right hand side of (\puteqn{B2kmin1}) and is removed by
exponentiating. The two-fermion part of the last term in
(\puteqn{B2kmin2}) survives and is also exponentiated. We then compute
the transition amplitude
$$\putequation{Smomint}$$
Using the inner product
$$\putequation{defSinp}$$
we find for the amplitude
$$\putequation{STraAmp}$$
where
$$\putequation{SBos}$$
is the expansion through order $\epsilon$ of the length of the
geodesic joining $x$ and $y$ (cf. (\puteqn{expAct})), and
$$\putequation{SFer}$$
If we add an extra boundary term $\hbar \delta_{ab}\bar\psi^a_{\rm cl}(0)
\psi^b_{\rm cl}(0) = \hbar \delta_{ab} \bar\eta^a \psi^b_{\rm cl}(0)$ to
(\puteqn{SFer}), then the sum of the first term in (\puteqn{SFer})
together with this extra term becomes the leading term in the
fermionic action
$$\putequation{SFercon}$$
We can easily check that the expansion through order $\epsilon$ of
(\puteqn{SFercon}) indeed equals the expression in (\puteqn{SFer})
when the equations of motion are imposed. These read (of course we
also need the bosonic part of the action to find the full equations of
motion)
$$\putequation{eom}$$
We can now expand the Lagrangian in a Taylor series around its value
at $t=0$, and then do the trivial time integrations. This yields
$$\putequation{TayL}$$
We thus expand all fields in the Lagrangian around their
values at $t=0$, making use of the equations of motion (\puteqn{eom}).
The expansions up to the order we need are given by
$$\putequation{expter}$$
Inserting these expansions into $\tilde S_F$ in (\puteqn{SFercon})
yields the expression (\puteqn{SFer}).

Again, it is easy to check that the composition rule holds
$$\putequation{Scompo}$$
Imposing this rule as a consistency condition fixes the
term proportional to the Ricci curvature times $(y-x)^i (y-x)^j$ in
the expansion of (\puteqn{STraAmp}), whereas it leaves arbitrary
all curvature terms which have a factor $\epsilon$ in front.

Also for this $N=2$ supersymmetric case one can derive the resulting
transition element from a path integral. As we explained in section 5,
the Hamiltonian to be used in the path integral is the supersymmetric
one, leading to the following configuration space path integral
$$\putequation{SconFer}$$
where we have now rescaled the fermions by a factor $\sqrt{\hbar}$, in
order to obtain the usual anticommutation relation
$\{\bar\psi^a,\psi^b\} = \hbar \delta^{ab}$. The last term in
(\puteqn{SconFer}) is the extra term one always needs when writing
down path integrals for coherent states as is discussed in
[\putref{textb}]. We indeed find that it must be included to obtain
the correct propagator. The actual computation of this path integral
is worked out in the appendix.

We will now show that the final result for the transition amplitude
can again be written as the product of three factors: a term
containing only the scalar curvature which is related to the trace
anomaly, the exponent of the classical action, and the square root of
in this case the supersymmetric generalisation of the Van Vleck
determinant. The latter is defined by[\putref{DeWib},\putref{Nieu}]
$$\putequation{defDS}$$
where $\Phi^A = (x^i, \bar\eta^a)$ and $\Phi^B =(y^j, \chi^b)$, and
for $S_B$ and $\tilde S_F$ we substitute the expressions
(\puteqn{SBos}) and (\puteqn{SFer}), with the fermions rescaled by a
factor $\sqrt{\hbar}$. To evaluate $D_S$ write
$$\putequation{DisABCD}$$
We find, expanding in normal coordinates around $x$ to simplify the
expressions,
$$\putequation{ABCD}$$
We do not need terms of order $\epsilon$ in $B$ and $C$, since $D_S =
\det A \det^{-1} \bigl( D - C A^{-1} B \bigr)$ and $A^{-1}$ is already
of order $\epsilon$. Writing $A_{ij} = {1\over\epsilon} g_{ik} \bigl(
\delta^k{}_j + \epsilon a^k{}_j \bigr)$ and $D_{ab} = (\delta_{ab} +
d_{ab})$, we can write the expansion of the super Van Vleck
determinant as
$$\putequation{expSdet}$$
Multiplying by $g^{-1/4}(x) g^{-1/4}(y)$ to transform $D_S^{1/2}$ into
a bi-scalar, we obtain
$$\putequation{DSis}$$
So indeed we can write
$$\putequation{proprew}$$
similarly to (\puteqn{rewr}). All terms involving the Ricci curvature
in (\puteqn{STraAmp}) are thus completely accounted for in the super
Van Vleck determinant! This would not have been the case if we would
have used the ordinary determinant. Note that it is $\tilde S_F$ and not $S_F$
which appears in the exponent, for reasons explained in
(\puteqn{propcohfer}). As we already mentioned, also in the path
integral we shall need $\tilde S_F$.

{\bf\section{The $N{=}1$ supersymmetric sigma model.}}

For the case of $N{=}(0,1)$ supersymmetry we start again with the
supersymmetric sigma model as defined in (\puteqn{Lsusyflat}), but now
truncate the theory by putting in the action $\psi_1^a=\psi_2^a =
{1\over\sqrt{2}} \psi^a_I$.  This then requires $\epsilon_1 +
\epsilon_2 =0$, see (\puteqn{susytra}). Canonical quantization then
yields $\{ \psi^a_I, \psi^b_I \} = \delta^{ab}$, and $p_i = g_{ij}
\dot \phi^j + {i\over 2} \omega_{iab} \psi^a_I\psi^b_I$. The
truncation of $H^{({\rm qu})}$ would yield again (\puteqn{QQisHqu})
but now with $\psi^a_I\psi^b_I\psi^c_I\psi^d_I$ instead of $\left(
\psi^a_\alpha \psi^b_\alpha \right) \left( \psi^c_\beta
\psi^d_\beta \right)$. Using the cyclic identity of the Riemann
tensor, one obtains
$$\putequation{Htrun}$$
However, this is {\it not} the Hamiltonian we are interested in for
two reasons: $(i)$ we want a Hamiltonian which corresponds to the
regulator of a spin-${1\over 2}$ field in quantum field theory. The
latter is given by
$-{1\over 2} \hbar^2 g^{1/4} \Dsl \! \Dsl g^{-1/4}$ which differs from
the truncated Hamiltonian because it has a term $-{1\over 8}\hbar^2 R$
instead of $-{1\over 16}\hbar^2 R$, and $(ii)$, we want the $N{=}1$
supersymmetric quantum Hamiltonian, given by the square of the $N{=}1$
supersymmetry charge. Truncating the $N{=}2$ supersymmetry charge in
(\puteqn{Qqu}) to the $N{=}1$ case, one obtains the operator $Q^{({\rm
qu})}_{N=1} = g^{1/4} e^i_a(\phi) \psi^a_I \pi_i g^{-1/4}$. Since this
operator corresponds to the Dirac operator $g^{1/4} \Dsl g^{-1/4}$
whose square contains a term ${1\over 4} R$ we clearly obtain the
Hamiltonian with $-{1\over 8} \hbar^2 R$.

The truncated Hamiltonian is thus no longer supersymmetric. The reason
is that the field $\psi_1 - \psi_2$ can consistently be set to zero at
the classical level, as its classical supersymmetry variation vanishes,
but at the quantum level the anticommutator of this field with itself
is nonvanishing. So instead of the Hamiltonian in (\puteqn{Htrun}) we
should use the Hamiltonian
$$\putequation{FerReg}$$
We can now repeat the computation of the transition element of section
6. Note that in contrast with (\puteqn{defSham}) we have only one
species of fermions present, namely $\psi_I^a$. In order to compute
the transition element (and subsequently if needed its trace), we need
two operators $\psi^a$ and $\bar\psi^a$ as in (\puteqn{cohfer}). Hence
we enlarge our Hilbert space by introducing a second set of free fermions,
$\psi^a_{II}$, satisfying $\{ \psi^a_{II},\psi^b_{II} \}
=\delta^{ab}$ and $\{ \psi^a_I,\psi^b_{II} \} = 0$. We then again
define $\psi^a = {1\over\sqrt{2}}\bigl(
\psi^a_I + i\psi^a_{II}\bigr)$ and $\bar\psi^a =
{1\over\sqrt{2}}\bigl( \psi^a_I - i\psi^a_{II}\bigr)$. We rewrite the
operators $\psi^a_I$ appearing in the Hamiltonian (\puteqn{FerReg}) in
terms of the operators $\psi^a$ and $\bar\psi^a$ as $\psi^a_I =
{1\over\sqrt{2}}( \psi^a + \bar\psi^a)$, and proceed analogously to the
computation in section 6 to obtain the transition element. Compared to
this section the only difference is that we replace the operators
${1\over 2}\omega_{iab}\psi^a_\alpha\psi^b_\alpha =
\omega_{iab}\bar\psi^a\psi^b$ of section 6 by ${1\over
2}\omega_{iab}\psi^a_I\psi^b_I={1\over
4}\omega_{iab}(\bar\psi^a+\psi^a)(\bar\psi^b+\psi^b)$, and
$R_{abcd}\psi^a_\alpha\psi^b_\alpha\psi^c_\beta\psi^d_\beta$ by $R$.
The net result is that $(i)$ all $\bar\eta^a\xi^b$ terms in
(\puteqn{B2kmin1}) and (\puteqn{B2kmin2}) are replaced by ${1\over 4}
(\bar\eta^a + \xi^a) (\bar\eta^b + \xi^b)$, $(ii)$ in
(\puteqn{B2kmin2}) the coefficient $\bigl(\bar\eta^a \xi^b \bar\eta^c
\xi^d + \delta^{bc} \bar\eta^a \xi^d \bigr)$ of the $\omega^2$ term is
replaced by ${1\over 16} \left[ (\bar\eta^a + \xi^a) (\bar\eta^b +
\xi^b) (\bar\eta^c + \xi^c) (\bar\eta^d + \xi^d) + 2 \delta^{bc}
\delta^{ad} \right]$ (we have used that the expressions are
antisymmetric under $a\leftrightarrow b$ or $c\leftrightarrow d$, and
symmetric under $ab\leftrightarrow cd$), and $(iii)$ all contributions
from the $R\psi^4$ term in the Hamiltonian disappear and are replaced
by ${1\over 8} \epsilon\hbar R$. One finds
$$\putequation{FerTrans}$$
where
$$\putequation{Soverh}$$
and $S_B$ is defined in (\puteqn{SBos}).

The various terms in (\puteqn{FerTrans}) and (\puteqn{Soverh}) can be
understood as follows. The $\bar\eta \chi$ term comes from the inner
product of coherent states in (\puteqn{defSinp}). The rest of the
terms in (\puteqn{Soverh}) can be obtained from (\puteqn{SFer}) by
making the replacement $\bar\eta^a \chi^b \rightarrow {1\over 4}
(\bar\eta^a + \chi^a )(\bar\eta^b + \chi^b)$. Since this substitution
yields a result antisymmetric in $a\leftrightarrow b$, the
$(x-y)^2\omega^2
\bar\eta \chi$ term can not have a counterpart in (\puteqn{Soverh}).
However, there is a $(x-y)^2\omega^2$ term in (\puteqn{FerTrans}), and
this term is the counterpart of the $(x-y)^2\omega^2\bar\eta \chi$ in
(\puteqn{SFer}) as one may trace by recalling that $\bigl(\bar\eta^a
\xi^b \bar\eta^c \xi^d + \delta^{bc}
\bar\eta^a \xi^d \bigr)$ was replaced by ${1\over
16} \left[ (\bar\eta^a + \xi^a) (\bar\eta^b + \xi^b) (\bar\eta^c +
\xi^c) (\bar\eta^d + \xi^d) + 2 \delta^{bc} \delta^{ad} \right]$. The
first part of this term is the square of the $(x-y)\omega (\bar\eta
+\chi)^2$ term in $\tilde S_F$, whereas the second part explains the
presence of the $(x-y)^2\omega^2$ term in (\puteqn{FerTrans}).
Finally, the $-{1\over 12} \epsilon\hbar R$ from the bosonic case in
(\puteqn{infampl}) combines with the ${1\over 8} \epsilon\hbar R$ from
(\puteqn{FerReg}) into the term ${1\over 24}\epsilon\hbar R$.

We will now show that the expression for the propagator can again be
written as the product of the super Van Vleck determinant, the
exponent of the classical action, and a term involving the scalar
curvature which, as shown at the end of this section, determines the
trace anomaly of a spin-${1\over 2}$ field. Defining the super Van
Vleck determinant as in (\puteqn{defDS}) and (\puteqn{DisABCD}), we
find from (\puteqn{Soverh})
$$\putequation{1ABCD}$$
which yields, using again (\puteqn{expSdet}),
$$\putequation{DS1is}$$
So we can indeed write
$$\putequation{prop1re}$$
similarly to (\puteqn{rewr}) and (\puteqn{proprew}).

Again, the same result for the propagator can also be obtained from a
configuration space path integral with a suitable action. As exlained
in the appendix, also in this case we have to introduce an extra set of
free fermions, $\psi^a_2$, in order to make the path integral
well-defined. The action one should use in the configuration space
path integral is (we again rescaled the fermions by a factor
$\sqrt{\hbar}$)
$$\putequation{SferAn}$$
where compared to the naive action corresponding to the Hamiltonian in
(\puteqn{FerReg}) the ${1\over8}\hbar^2 R$ term is cancelled because
of Weyl ordering of the bosonic term (in this $N{=}1$ case there is no
$R\psi^4$ term which would yield another $- {1\over 8} R$ term in the
action, upsetting the cancellation of $R$ terms in (\puteqn{SferAn})).
We have introduced an extra set of fermions that do not couple to any
of the other fields, this way making certain that we do not alter the
dynamics. In order to keep local Lorentz invariance we require that
the fermions $\psi^a_2$ are inert under local Lorentz transformations.

We can now easily compute the trace anomaly for a spin-${1\over 2}$
field by taking the trace of the transition element
(\puteqn{FerTrans}) and singling out the $\epsilon$-independent part.
Since by introducing the free fermions $\psi^a_2$ ($a=1\ldots n$), we
have (for even $n$) $2^{n/2}$ states in the $\psi_1$ sector and
$2^{n/2}$ states in the $\psi_2$ sector (combining the $n$ $\psi^a_1$
into half as many pairs of creation and absorbtion operators, and
similarly for $\psi^a_2$). Hence, we must divide the trace over
$\psi_1$ and $\psi_2$ by a factor $2^{n/2}$, since we really should
only take the trace in the $\psi_1$ sector. The trace anomaly for a
spin-${1\over 2}$ field in $n$ dimensions is therefore given
by (cf. [\putref{BaNi}], equation (2.9))
$$\putequation{SAnn}$$
For $n=2$ the trace anomaly becomes ${-\hbar\over 24}R$, which is
indeed the result for a Dirac fermion; for the anomaly of a Majorana
fermion we have to divide this expression by two.

{\bf\section{Conclusions.}}

We have considered the evaluation of the propagator in quantum
mechanics for a point particle in curved space, and for a point
particle with its fermionic extension in curved space. In the latter
case we considered both $N{=}1$ and $N{=}2$ supersymmetric quantum
mechanics. We first used the Hamiltonian approach and obtained the
propagators through order $\Delta t=\epsilon$. In this approach there
are no amibiguities at all, but the calculations are somewhat
tedious. Afterwards we considered the problem of finding an action for
a path integral, together with a presciription for evaluating this
path integral, which reproduces the propagator. This problem has been
studied at great length in the literature, but in our opinion no
complete solution has yet been found. We now summarize the
contribution of this paper to these problems.

In the Hamiltonian approach, the propagator is defined by $\langle x |
\exp \left( - {\epsilon\over\hbar} \hat H \right) | y\rangle$ where we
assume $\hat H$ to be an operator with a given, a priori fixed,
ordering of the operators $\hat x^i$ and $\hat p_j$ but without an
explicit time dependence. We evaluated this matrix element, following
Feynman, by inserting a complete set of states $\int\! d^np\,
|p\rangle\langle p|$, expanding the exponent, and moving in each term
all $\hat x$ to the left and all $\hat p$ to the right, where we
replace them by their eigenvalues $x$ and $p$, and finally
reexponentiating the result.  It seems widely believed that for small
$\epsilon$ it is sufficient to only retain the term linear in $\hat H$
in the expansion, to define $\langle x | \hat H | p \rangle = h(x,p)
\langle x | p \rangle$, and to replace $\langle x| \left( 1 -
{\epsilon\over\hbar} \hat H \right)| p\rangle$ by $\exp \left( -
{\epsilon\over\hbar} h(x,p) \right) \langle x|p\rangle$. If $\hat
H(\hat x,\hat p)$ is of the form $T(\hat p)
+ V(\hat x)$, this procedure is correct, and proven by the Trotter
formula (see e.g. [\putref{Schu}]). However, for more general systems,
such as a particle in curved space with Hamiltonian
$$\putequation{partcs}$$
where $\int\!d^nx \, \sqrt{g(x)} |x\rangle\langle x| = \unit$,
or a point particle in flat space coupled to electromagnetism with
Hamiltonian
$$\putequation{ppMaxw}$$
where $\int\!d^nx \, |x\rangle\langle x| = \unit$, this linear
approximation is incorrect. In both cases, the commutator of $\hat
p_i$ with a function $f(\hat x)$ is given by ${\hbar\over i} \left(
{\partial f\over \partial x^i} \right) (\hat x)$ (we never need to use
the fact that for
hermiticity $\hat p_i$ must be represented in curved space by
${\hbar\over i} g^{-1/4}(\hat x) \partial_i g^{1/4}(\hat x)$ if one
uses an inner product defined by $\int\!d^n x\, \sqrt{g(x)}
|x\rangle\langle x|=\unit$, see section 2). The reason that terms
linear in $\epsilon$ in the propagator also come from terms nonlinear
in $\epsilon$ in the expansion of $\exp \left( - {\epsilon\over\hbar}
\hat H \right)$ is that the $p$-integration uses a Gaussian
integrand $\exp\left( - \epsilon p^2 \right)$, so that under the
integral $\hat p$ is of order $\epsilon^{-1/2}$.  Hence each
commutator $[\hat x,\hat p]$ which removes an operator $\hat p$,
contributes a factor $\epsilon^{1/2}$ to the final answer.
Consequently, to obtain the terms of order $\epsilon^{m/2}$ in the
propagator for (\puteqn{partcs}) it is sufficient and necessary to
retain all terms with at most $m$ commutators in the evaluation of the
matrix element $\langle x| \sum {1\over k!}\left( -
{\epsilon\over\hbar} \right)^k \hat H^k | p\rangle $. In particular,
terms proportional to $\epsilon$ (which are needed to check that the
propagator satisfies the Schr\"{o}dinger equation to lowest order in
$\epsilon$) come from terms containing none, one or two commuators in
the evaluation of $\langle x| \hat H^k |p\rangle$ {\it for any} $k$!
Forgetting these commutators leads to an incorrect result.

We have been able to resum this infinite series both for the bosonic
and for the supersymmetric cases. It should be stressed that all terms
in $\langle x| \hat H^k |p\rangle$ are well defined, and no
amibiguities exist in the evaluation of the propagator in the
Hamiltonian approach. Only the commutation relations between $\hat p$
and $\hat x$ were needed, and no regularization of products of
operators at the same time was used. Our final answer for the bosonic
point particle
in curved space agrees with the literature, proving that it was indeed
necessary to keep terms beyond the linear approximation. The result
through order $\epsilon$ is a product of $(i)$ the classical action for a
solution of the classical equation of motion for $x^i(t)$, describing
a particle moving from $y$ to $x$ in time $\epsilon$, $(ii)$ the
square root
of the Van Vleck determinant, which is anyhow needed to make the
propagator a general coordinate scalar both in $x$ and in $y$ (a
bi-scalar [\putref{DeWia}]), and $(iii)$ the trace anomaly. Namely
$$\putequation{respro}$$

We would like to remark that the extension of this result to include
coupling to an electromagnetic field can easily be deduced from its
supersymmetric extensions we considered in section 6 and 7. One simply
identifies ${e\over c}A_i(x)$ with ${i\hbar\over 2}\omega_{iab}
\psi^a_\alpha \psi^b_\alpha$ in section 6, or with ${i\hbar\over 2}
\omega_{iab} \psi^a_I \psi^b_I$ in section 7, and retains only the terms
which are not coming from either anticommuting the fermions or from
the inner product of the fermionic coherent states. This yields again
the expression (\puteqn{respro}), where now $S_{\rm cl}$ is the
expansion through order $\epsilon$ of the classical action of a point
particle in curved space coupled to electromagnetism.

Although in (\puteqn{respro}) the term $R(x)\epsilon$ can be rewritten
as an integral of a
local funtional of $x(t)$, $R(x)\epsilon = \int\!dt\, R(x_{\rm cl}(t))
$ to linear order in $\epsilon$, the term with
$R_{ij}(x)(x-y)^i(x-y)^j$ can not be obtained by expanding the
integral of a {\it local} functional. For this reason, one cannot
simply string a set of matrix elements $\langle x_i| \exp \left( -
{\epsilon\over N\hbar} \hat H \right) | x_{i+1} \rangle$ together and,
by integrating over the intermediate points $x_1\ldots x_{N-1}$,
obtain a path integral. The path integrals we are interested in have,
by definition, a {\it local} action, and the problem one is faced with
is to find a local functional $\Sco[x(t)]$ which produces the
propagator $\langle x| \exp\left( - {\epsilon\over\hbar} \hat H
\right) |y\rangle$ when inserted into a path integral of the form
$\int \! Dx(t) \, \exp \left( - {1\over\hbar} \Sco[x(t)] \right)$. The
paths all run from $x(t{=}{-}\epsilon) = y$ to $x(t{=}0)=x$. Of course, one
should also specify how to evaluate the path integral, and this we do
by ``mode-expansion''[\putref{AGWi},\putref{Bast},\putref{BaNi}],
according to which $x(t)$ is expanded in terms of a complete set:
$x(t) = \sum a_n \phi_n(t)$. The obvious and most convenient choice
for the complete set of $\phi_n(t)$ are the eigenfunctions of the
kinetic operator, because in that case the kinetic terms become a
diagonal quadratic expression in terms of the $a_n$. The problem is
therefore to find a local Lagrangian $L_{\rm conf}[x(t)]$, which can
be split into a kinetic part $L_{\rm conf}^0$ and the rest, denoted by
$L_{\rm conf}^{int}$, which will provide the vertices. As shown in
[\putref{BaNi}] the correct trace anomaly, in $d=2$ due to the terms
proportional to $\epsilon$, and in $d=4$ due to the terms proportional
to $\epsilon^2$, is obtained by using the covariant action
$$\putequation{Scova}$$
{\it provided one uses normal coordinates}. In this paper for
supersymmetric systems, and in [\putref{Bast}] for a scalar particle,
it was found that the complete propagator through order $\epsilon$
is produced by the covariant Lagrangians we have discussed, again
provided one uses normal coordinates.

It is not sufficient only to know how to calculate in normal
coordinates, because some Hamiltonians one encounters are {\it not}
generally covariant. We have in mind the consistent anomalies in
quantum field theories which maintain local scale (Weyl) invariance
and consequently break general coordinate (Einstein) invariance. For
these systems, the regulator can be constructed using the algorithm of
reference [\putref{DTNP}], and, as expected, the corresponding
Hamiltonians are not generally covariant. Thus, one must know the
action $\Sco$ and the rules to evaluate path integrals in arbitrary
coordinates. Of course, for generally covariant Hamiltonians, the
propagator is a coordinate scalar, hence one may evaluate it in normal
coordinates to obtain the correct answer.

For general coordinates, the form of $L_{\rm conf}$ is unknown. In the
literature, some attempts have been made to deduce $H_{\rm conf}$
(related to $L_{\rm conf}$ by a Legendre transformation, both are
functions, not operators) from $\hat H$ by a functional differential
equation. In particular, DeWitt has pursued this problem in the latest
version of his book on supermanifolds[\putref{DeWib}], but he runs
into the well-known problem how to define products of operators at the
same point (i.e., at the same time). He proposes to use
``time-ordering'', according to which one first defines all operators
at different times, then considers all permutations of these
operators, and finally takes the limit that all times become
equal. Following this procedure for the bosonic case, one obtains not
only the ${\hbar^2\over 8} R$ term mentioned above, but also an extra
noncovariant term
$$\putequation{Hconf}$$
Concluding that time-ordering breaks general coordinate covariance,
he chooses a pseudo-Euclidean frame in which $\Gamma^i{}_{jk}(x(t{=}0))
=0$ and computes (for the supersymmetric case) some two loop diagrams
and finds that they are in agreement with the result for the
propagator (which he obtained long ago by heat kernel methods).

We believe that this is not the whole story. First of all, the
presence of a $\Gamma^2$ term at all $t$ will show up in calculations
using normal coordinates at order $\epsilon^2$ (where 3-loop diagrams
contribute). Although these contributions do look covariant in normal
coordinates, one obtains an incorrect result for the $d=4$ trace
anomaly. Hence the ${\hbar^2\over 8}\Gamma^2$ term without further
modifications should not be present in $L_{\rm conf}$ when one
uses normal coordinates. Of course, once the correct action $\Sco$ is
known in normal coordinates, one can find the corresponding action in
general coordinates by making a change of integration variables. The
Jacobian will contribute new vertices, and although we have not
worked out the path integrals for general metrics, we have a remark.
The background decompositon $x^i(t)
= x^i_0(t) + x_{\rm qu}^i(t)$ seems too linear to us for such a
nonabelian theory as gravity. Let us recall that in superspace
Yang-Mills theory, the Yang-Mills field is contained in $\exp (V)$,
and a background field approach replaces this by $\exp(V_{\rm back})
\exp(V_{\rm qu})$. This is clearly an infinitely nonlinear
split. Analogously, we expect also in gravity a nonlinear split of the form
$x^i(t) = x^i_0(t) + x^i_{\rm qu}(t) + (x^i_{\rm qu}(t))^2 {\rm terms}
+ \ldots$. We do not claim to have exhaustively studied this problem ,
and intend to return to it in the future. However, we have made some
introductory calculations which we now describe.

In our explicit computation of the path integral to order $\epsilon$
(2 loops), we took $L_0= {1\over 2} g_{ij}(x) \dot x^i \dot x^j$ where
the $x$ in $g_{ij}(x)$ is the endpoint of the interval. Decomposing
the action into the sum of this kinetic term and vertices we computed
loops. We then proceeded to evaluate all diagrams up to 2 loops {\it
in an arbitrary coordinate frame}
$$\putequation{Spluspro}$$
where $\tau = t/\epsilon$. Here $a$, $b$, $c$ are anticommuting ghosts
due to exponentiating the measure. They were first introduced by
Bastianelli in [\putref{Bast}]. The present form was obtained in
[\putref{BaNi}]. Due to these ghosts all loop contributions become
completely finite. We contract all quantum fields, using (see
(A.7))
$$\putequation{onlypro}$$
where
$$\putequation{propmod}$$
The great advantage of the formulation in (\puteqn{Spluspro}) is that
it allows a regularization which treats $x$ and the ghosts in a uniform
manner, namely by mode cut-off: in all graphs one first uses
$\Delta(\sigma,\tau)$ in (\puteqn{propmod}) where the sum runs up to
$N$, and only at the end takes $N$ to infinity.
Of course, $x_0(t) = x + (y-x) t/\epsilon$ is only a solution for
$S_0$ and not a
solution of the full equation of motion, hence there are terms linear
in $x_{\rm qu}(t)$ in the action. The path integral expectation value
of a single $x_{\rm qu}(t)$ vanishes (since $L_0$ is symmetric under
$x_{\rm qu}(t) \leftrightarrow - x_{\rm qu}(t)$), but contractions of
the form $x_{\rm qu}(s) x_{\rm qu}(t)$ contribute.

We found
that the total set of $2$ loop diagrams in the bosonic case precisely
reproduces all terms in the propagator through order $\epsilon$, except
for one term (which vanishes in normal coordinates). Namely, if one
uses the covariant action (omitting the ${1\over
8}\Gamma^2$ term), the path integral result is equal to the propagator
plus the following remainder
$$\putequation{restpath}$$
If one adds the ${1\over 8}\Gamma^2$ term, part of this remainder is
canceled, but we are still left with another extra term
$$\putequation{pathres}$$
We believe that a proper treatment of the Jacobian for the
transformation from normal to general coordinates will cancel
(\puteqn{restpath}).
Our conclusion for the bosonic case is that in the Hamiltonian sector
we have a complete understanding, but in the path integral sector, we
believe we know the correct $\Sco$ (namely the covariant action with
the ${1\over 8} R$ term but without $\Gamma^2$ terms) but we have no
proof of this rule.

We now turn to the fermionic sector where we have not only found
similar results, but also new results for fermionic phase space path
integrals. In the $N{=}2$ supersymmetric case, there are also fermionic
variables $\psi^a_\alpha(t)$ present, with $\alpha = 1,2$ and
$a=1\ldots n$. The Hamiltonian for this system is not obtained from
the regulator of a corresponding quantum field theory, but we fixed
its ordering by requiring the supersymmetry relation
$2\delta_{\alpha\beta} \hat H = \{ \hat Q_\alpha , \hat Q_\beta \}$
where the operator orderings in the supersymmetry charge operator
$\hat Q_\alpha$ in (\puteqn{Qqu}) were fixed by requiring
hermiticity. We then computed the propagator between states
$|y,\xi\rangle$ and $\langle x,\bar\eta |$. The states $|\xi\rangle$
and $\langle\bar\eta |$ are fermionic
coherent states, and using general theory of coherent states, suitably
adapted to the fermionic case, we straightforwardly obtained the
propagator to order $\epsilon$. Of course, we must now also take into
account the commutators $\{ \hat\psi^{\dagger a}, \hat\psi^b
\} = \delta^{ab}$. The final result in (\puteqn{STraAmp}) was then
written as $\exp\left( -{1\over\hbar} \bigl( S_B + S_F \bigr) -
\delta_{ab} \bar\eta^a \psi^b_{\rm cl}(0) \right)$ times a factor containing
only terms with curvatures. The extra term $-\delta_{ab}\bar\eta^a
\psi^b_{\rm cl}(0)$ (where $\psi^b_{\rm cl}(t)$ is a solution of the
field equations
with boundary value $\psi^b_{\rm cl}(t{=}{-}\epsilon) = \chi^b$) is
well-known from
the theory of path integrals for coherent states, and can be found in
textbooks, see section 3. The curvature terms should describe the Van
Vleck determinant and a term with $R$ which would have been the trace
anomaly of a quantum field theory if $\hat H$ could have been
identified with its regulator. However, in supersymmetric theories it
is more natural to use superdeterminants, and hence we considered the
double superderivative of the action. The super Van Vleck determinant
absorbed then all terms with Ricci curvatures, just as in the bosonic
case. We were left with a term involving the scalar curvature, which
is exactly the same as in the bosonic case. (Using the ordinary Van
Vleck determinant, one is left with a term containing the Ricci
curvature). Thus, the propagator
factorizes exactly as in the bosonic case, but now with a super Van
Vleck determinant.

To reproduce these operator results for the $N{=}2$ supersymmetric case
from a path integral, we took as action $\Sco$ the classical $N{=}2$
action in (\puteqn{SconFer}) with the boundary term, including the
$R\psi^4$ term but without any further term proportional to
$R$. The reason we did not include an $R$ term is that, applying the
time-ordering prescription to the $N{=}2$ Hamiltonian,
we have found that the term $-{\hbar^2\over 8} R$ of the bosonic
action is precisely canceled by a similar term obtained by
time-ordering the $R\psi^4$ term! (In the $N{=}1$ case a similar
cancellation of ${1\over 8} R$ terms occurs for different
reasons). Time-ordering however, also produces the $\Gamma^2$ term of
the bosonic
sector discussed above, together with an $\omega^2$ term if one treats
$\hat p$, $\hat x$, $\hat{\bar\psi}$ and $\hat\psi$ as the independent
variables. (However, with $\hat p_i - \omega_{iab}(\hat x)
\hat{\bar\psi}^a \hat\psi^b$, $\hat x$, $\hat{\bar\psi}$ and
$\hat\psi$ as independent variables, no $\omega^2$ term is produced).

With the fermionic propagator given by the mode expansion (and with
the propagator for the bosonic and ghost fields mentioned before) we
then may compute all 2-loop graphs
using (\puteqn{Spluspro}) and (\puteqn{onlypro}) together with
(A.15) and (A.16). The result of these calculations can be
summarized as follows: to order $\epsilon$, the path integral produces
the propagator plus two extra terms. The first one comes from the
bosonic sector and is equal to (\puteqn{restpath}), the second one is
due to contracting two $\int\!
d\tau\, \dot x^i_{\rm qu} \omega_{iab}(x) \bar\psi^a_{\rm qu}
\psi^b_{\rm qu}$ vertices. Writing the three propagators in a mode
expansion, and truncating the bosonic and fermionic propagators at the
same $n=N$, we have found (by a numerical evaluation) a finite but
nonzero result
$$\putequation{Nis2r}$$
In normal coordinates $\omega_{iab}(x)$ vanishes, and hence in normal
coordinates we obtain complete agreement between the operator approach
and path integrals. In a future extension to general coordinates, the
contribution in (\puteqn{Nis2r}) should be taken into account.

Finally, we considered the supersymmetric $N{=}1$ case. This case is
more complicated than the $N{=}2$ case, because there are only real
fermions present, whereas one needs complex
fermions, both in the Hamiltonian approach (for operators $\psi$ and
$\psi^\dagger$) and in the path integrals (to define separate boundary
conditions on the left and the right). We resolved this problem by
adding a set of free fermions in both cases; this should not alter the
model. To preserve local Lorentz invariance, we required that these
free fermions are inert under these transformations.

In their fundamental paper on chiral and gravitational
anomalies[\putref{AGWi}], Alvarez-Gaum\'{e} and Witten did not need to
introduce these extra fermions because they only considered traces of
the propagator, so they did not need to specify separate boundary
conditions on the left and the right. Moreover, the Jacobian for
chiral transformations of the quantum fields led in the corresponding
quantum mechanical model to periodic boundary conditions for the
$N{=}1$ case, and half-periodic and half-antiperiodic or completely
periodic for the $N{=}2$ case. In all these cases, there is a zero
mode $\psi_0$ in the expansion of $\psi(t)$ in terms of eigenfuncions
of the operator ${\partial\over\partial t}$ (corresponding to our
$S_0$). The Grassmann integration then requires a minimal number of
$\epsilon R \psi^2\psi_0^2$ vertices to saturate the integration over
these zero modes, and this already produces enough powers of $\epsilon$ to
cancel the $\epsilon^{-n/2}$ in the measure of the path integral.
Consequently, for them the harmonic approximation is sufficient. For
us, however, this is not sufficient. For example, for the trace
anomaly we need completely antiperiodic boundary conditions for the
fermions, so that no $\epsilon$-producing zero modes $\psi_0$ are
present. In addition, we considered the full propagator, and not only
its traces.

The actual computations in the $N{=}1$ case became very similar to
those of the $N{=}2$ case, once the free fermions were added. The
Hamiltonian which corresponds to the general coordinate invariant
regulator of a spin-${1\over 2}$ field, was the sum of the
Laplace-Beltrami operator for fermions plus a term $-{1\over 8}
\hbar^2 R$ (twice as much as truncation of the $N{=}2$ Hamiltonian
would give). This result was consistent with the requirement that
$\hat H$ be equal to the square of a hermitian supersymmetry charge.
Again we found that the propagator factorized into the classical action
(with boundary term), the super Van Vleck determinant, and a scalar
curvature term which yields now the trace anomaly for the quantum
field theory of a (Dirac or Majorana) fermion minimally coupled to
gravity.

Then we considered the path integral approach to the propagator of the
$N{=}1$ case. In this case we took as action $\Sco$ the action in
(\puteqn{SferAn}), which is the classical supersymmetric action (with
boundary term). No extra $R$ term was present because we showed that
the $R$ term due to time-ordering in the bosonic sector cancels the
explicit $R$ present in the Hamiltonian. Hence, for different reasons,
both in the $N{=}1$ and $N{=}2$ cases, the action $\Sco$ equals the
classical supersymmetric action. This result one might have taken
anyhow on the basis of symmetry arguments, but it is gratifying that
if follows rigorously from our Hamiltonian approach. We computed all
graphs through two loops which contribute through order $\epsilon$. In
normal coordinates we found again complete agreement with the operator
approach, but in general coordinates we found again an extra term
which a future extension of the path integral approach using arbitrary
coordinates should explain.

We conclude this paper with a general solution for the path integral
in terms of an action $\Sco$, which corresponds to any, covariant or
noncovariant, Hamiltonian of the form $\hat H = a(\hat x) \hat p^2 +
b(\hat x) \hat p + c(\hat x) $ with $n$ operators $\hat x^i$ and $\hat
p_i$. First rewrite $\hat H$ into the form
$$\putequation{Hamgen}$$
where $g=\det g_{ij}$. Then use the metric $g_{ij}(x)$ to construct
normal coordinates. In these normal coordinates one has
$$\putequation{Scofigu}$$
The path integral is evaluated by splitting off a kinetic term
$g_{ij}(x(0)) \dot x^i(t) \dot x^j(t)$, and adding the ghost action
in (\puteqn{Spluspro}). Propagators and vertices are now
defined, and the loop expansion of the propagator in normal
coordinates is obtained. Using mode cut-off
[\putref{Bast},\putref{BaNi}] to regulate divergent graphs, it is
found that at any loop level the final result is completely finite. For
general coordinate invariant $\Sco$, this
result also holds in general coordinates. In other cases one must
transform back from normal to general coordinates. We have checked
that this procedure reproduces the propagator
obtained from $\hat H$ through order $\epsilon$, and trace anomalies
through order $\epsilon^2$, for the point particle, and $N{=}1$ and
$N{=}2$ supersymmetry. Summarizing: whereas the heat kernel or the
Hamiltonian approach yield the propagator directly, we have found the
corresponding action $\Sco$ which exactly reproduces this propagator
from a path integral.

\sectionnumstyle{blank}
{\bf\section{Acknowledgements.}}

We would like to thank F. Bastianelli, E. Br\'{e}zin, B. DeWitt, M.
Ro\v{c}ek, K. Skenderis and J. Zuber for discussions.

\sectionnum=0
\sectionnumstyle{Alphabetic}
{\bf\section{Appendix: Evaluation of path integrals.}}

The path integral in the bosonic case (section 4) has already been
evaluated in normal coordinates in [\putref{Bast},\putref{BaNi}], by
expanding fields in Fourier series and converting the path integral to
an integral over the Fourier coefficients. An important detail is that
we must introduce a ghost system to exponentiate the factors
$\sqrt{g(x_j)}$ at the $N-1$ intermediate points $x_j$ in the measure,
hereby making the remainder of the
measure translationally invariant. This makes loop corrections
finite. We will now first evaluate the bosonic path integral in an
arbitrary coordinate frame, and afterwards we will compute the
fermionic path integrals.

We follow the approach of [\putref{Bast},\putref{BaNi}], and start
with the action in configuration space (note that extra terms
of the type $R+\Gamma^2$ are already of order $\epsilon$, and can, to
the order we are considering, simply be replaced by their classical
values)
$$\putequation{Scogen}$$
where we have rescaled $\tau=t/\epsilon$, and introduced a ghost
system to exponentiate the factors $\sqrt{g(x(t))}$ appearing in the
measure. The boundary conditions are given by $x^i(-1) =y^i$,
$x^i(0)=x^i$. All ghosts vanish at both endpoints,
$b^i(-1)=b^i(0)=c^i(-1)-c^i(0) =a^i(-1)=a^i(0) =0$, because there are
only $\sqrt{g}$ factors at the intermediate points. We
will now make a decomposition in background and quantum fields, where
we choose the background fields to satisfy the boundary conditions. We
take vanishing background fields in the ghost sector, whereas
for the bosonic fields we take
$$\putequation{bgdec}$$
We now evaluate the path integral by expanding the term
$g_{ij}(x(\tau))$ around its value at $x(\tau)=x$, and write
$S=S^{kin} + S^{int}$, where
$$\putequation{skin}$$
and
$$\putequation{sint}$$
Now all functions $g$, $\partial g$ are evaluated at the endpoint $x$.
Since all
quantum fields vanish at $\tau=0$ and at $\tau=-1$, we can decompose
them in a mode expansion, yielding a natural regularization scheme
for the propagators. The propagator was obtained in reference
[\putref{BaNi}] by
using the mode expansion, but one can also obtain it (more easily)
from canonical methods. Expanding the Heisenberg operator $\hat x$ in
a complete set of zero modes of the kinetic operator $-g_{ij}(x)
{d^2\over dt^2}$ we get $\hat x^i(t) = \hat x^i + g^{ij}(x)\hat p_j
t$, with $[\hat x^i,\hat p_j] = i\hbar \delta^i_j$. Imposing the
boundary conditions $x_{\rm qu}(t{=}{-}\epsilon) = x_{\rm qu}(t{=}0) = 0$ as
conditions on the bra and ket vacua, $\hat x(t{=}{-}\epsilon) |0\rangle
=0$ and $\langle 0| \hat x(t{=}0) = 0$, we find $\langle 0| \hat x^i =0$
and $(\hat x^i - \epsilon \hat p^i) |0\rangle =0$. Hence
$$\putequation{canprop}$$
The Euclidean version of this result, after we make the rescaling
$\tau=t/\epsilon$, is given by
$$\putequation{GreenBos}$$
Using the mode expansion and (\puteqn{skin}) we find the same
propagator, but now with a regularized expression for
$\Delta(\sigma,\tau)$ (see [\putref{BaNi}])
$$\putequation{modpro}$$

The simplest term comes about when we only consider the background
fields $x_0(\tau)$. This yields
$$\putequation{Sbackg}$$
which already reproduces most terms in the classical action (compare
(\puteqn{expAct})); note that since we are not expanding around the
full solution to the equation of motion, we cannot expect to reproduce
the full classical action in $S_0$. We will see that the remainder of
the classical action is produced by quantum corrections.

Next we consider contributions from terms in which we contract one pair
of quantum fields. These come about either when we just take $S^{int}$
by itself, or when we take $\left( S^{int} \right)^2$. The terms of
the first type yield
$$\putequation{lS1}$$
which is part of the $R_{ij} (x-y)^i (x-y)^j$ term.
Furthermore, the terms of the second type yield
$$\putequation{qS1}$$
which we recognize as the remainder of the classical action.

The next set of terms comes from contracting two pairs of quantum
fields. Again there are two types of contributions, from $S^{int}$ by
itself and from its square. The first terms are
$$\putequation{lS2}$$
which we can identify as part of the scalar curvature term, whereas
the terms coming from $\left( S^{int} \right)^2$ combine with the
terms in (\puteqn{lS1}) into $-{1\over 12} R_{ij} (x-y)^i (x-y)^j$.

Finally, we should contract three pairs of quantum fields, now only in
$\left( S^{int} \right)^2$. Part of these terms we need to
covariantize the expression in (\puteqn{lS2}) into ${1\over 24}
\epsilon\hbar R$. The remaining terms are given by
$$\putequation{qS3}$$
which is equal to
$$\putequation{qS3p}$$
The significance of this term is discussed in the conclusions.

We now evaluate the fermionic (supersymmetric) path integrals. In
these cases, the bosonic and ghost sector of the path integral can be
handled in exactly the same way as before, again yielding the action
$S_B$ together with the contribution ${1\over 24}
\epsilon\hbar R - {1\over 12} R_{ij} (x-y)^i (x-y)^j$. We now consider
the fermionic sector.

Similarly to [\putref{Bast},\putref{BaNi}], we rescale
$\tau=t/\epsilon$ to make the $\epsilon$-dependence more explicit and
facilitate keeping track of the order in the expansion in $\epsilon$.
We now first note that we can write
$$\putequation{Ssplit}$$
Here $S_0$ contains all terms with only bosonic or ghost fields. The
other two terms are given by
$$\putequation{Skinfer}$$
and
$$\putequation{Sintfer2}$$
for the $N{=}2$ supersymmetric nonlinear sigma model considered in
section 6, whereas for the $N{=}1$ model discussed in section 7 we have
$$\putequation{Sintfer1}$$

In a background field approach we decompose $\psi(\tau) = \psi_0(\tau) +
\psi_{\rm qu}(\tau)$, and $\bar\psi(\tau) = \bar\psi_0(\tau) +
\bar\psi_{\rm qu}(\tau)$. We could have used fermionic normal
coordinates and added a term proportional to $x_{\rm qu}$ as in
[\putref{Alva}], but the result should not depend on the split between
classical and quantum fields, and our choice is quite simple. We
choose $\bar\psi_0(\tau)$ and $\psi_0(\tau)$ to be the solutions
the equations of motion of the kinetic part of the action,
(\puteqn{Skinfer}), which satisfy the boundary conditions, i.e. we
take $\bar\psi_0^a(\tau) =\bar\eta^a$ and $\psi_0^a(\tau) =\chi^a$.
Since we
require $\bar\psi^a(0)=\bar\eta^a$ and $\psi^a(-1) = \chi^a$, this
implies that the quantum fields need to satisfy the boundary
conditions $\bar\psi^a_{\rm qu}(0) = 0$ and $\psi^a_{\rm qu}(-1) =0$.
The Green function for the action (\puteqn{Skinfer}) of the quantum
fields then reads
$$\putequation{ferGreen}$$
where $\epsilon_{12} = +1$. Or, in terms of $\bar\psi$ and $\psi$,
$$\putequation{Greenfer}$$
One can easily check that this Green function is a solution to the
equations of motion of (\puteqn{Skinfer}) and that the boundary
conditions are indeed satisfied. In particular, at $\sigma=\tau$ we
define $< A(\tau) B(\tau) > = \lim_{\delta\rightarrow 0} {1\over 2} \bigl[
A(\tau+\delta) B(\tau-\delta) + A(\tau-\delta) B(\tau+\delta) \bigr]$,
and at $\sigma=\tau=0$ or $\sigma=\tau=-1$ one gets the correct result
using $\theta(0) = {1\over 2}$.

The propagator for the fields
$\bar\psi(t)$ and $\psi(t)$ can also
easily be found from the mode expansion or from canonical methods. In
the canonical approach
$\psi(t) =\hat\psi$ and $\hat{\bar\psi}(t) =\hat{\bar\psi}$ with
boundary conditions on the vacua given by $\langle 0|
\bar\psi(t{=}0) = 0$ and $\psi(t{=}{-}\epsilon) |0\rangle =0$. In the mode
expansion, we write
$$\putequation{modexp}$$
and the kinetic matrix $\int\limits_{-\epsilon}^0 \! dt \, \bar\psi
\dot\psi $ becomes $\sum {1\over 2} \pi (n+ {1\over 2}) \bar b_n b_n$,
yielding the propagator
$$\putequation{femopr}$$
This is the $\theta$-function $< T \bar\psi^a(s) \psi^b(t) > =
\delta^{ab} \theta(s-t)$ as follows from a Fourier analysis. For our
purposes we prefer the expression (\puteqn{femopr}) for the
propagators in terms of modes, since this yields a convenient regularization.

We can now straightforwardly compute the transition element by
expanding $\exp \left( - {1\over\hbar}  S^{int}_{\rm fer} \right)$
and contracting the quantum fields, using the above Green
functions for the fermions. For the bosonic and ghost fields, the
Green functions have been computed in [\putref{BaNi}]. The propagator
for the bosonic fields is given in (\puteqn{GreenBos}).

We will first consider the $N{=}2$ case, with the interaction given in
(\puteqn{Sintfer2}). When we expand $\exp \left( -{1\over\hbar}
S^{int}_{\rm fer} \right)$ we will for the first term in this
expansion only need the contraction at equal time  $<\bar\psi^a(\tau)
\psi^b(\tau) > = - {1\over 2}\hbar\delta^{ab}$. Since this is symmetric
in $a$ and $b$, the first term in (\puteqn{Sintfer2}) will yield no
contribution, whereas the second term contributes ${1\over 2} \epsilon
R_{ab}\bar\eta^a\chi^b-{1\over 8}\epsilon\hbar R$. Next consider the
term ${1\over 2} \left( {1\over\hbar} S^{int}_{\rm fer} \right)^2$.
Terms involving the $R\psi^4$ term are of higher order in $\epsilon$
(for example, the cross term is of order
$(x-y)\epsilon\sim\epsilon^{3/2}$),
so we only need to find the contribution from the square of the $\dot
x \omega \psi^2$ term. When we contract only the four fermionic fields,
using (\puteqn{Greenfer}), one finds zero. The contraction of the
two $\dot x$ fields and two
fermionic fields (using (\puteqn{Greenfer}) and (\puteqn{GreenBos}))
vanishes because the fermionic contractions yield
$\theta(\sigma-\tau)+\theta(\tau-\sigma)=1$, and the integral of the
bosonic part
vanishes. Finally, we can contract four fermionic and two bosonic
fields, which yields a contribution $-{1\over 12}\epsilon\hbar g^{ij}
\omega_{ia}{}^b \omega_{jb}{}^a$. In order to obtain this result, we
used the regularized version of the propagator from the mode
expansion, evaluated the integrals using the same number of
bosonic and fermionic modes, and only afterwards let the number of
modes go to infinity. Contractions involving the other bosonic
fields are again of higher order and need not be considered.

We should be careful since we are in fact expanding the fermionic
fields $\bar\psi^a(\tau)$ and $\psi^a(\tau)$ around the constant fields
$\bar\eta^a$ and $\chi^a$ respectively, which are {\it not} solutions
to the equations of motion (\puteqn{eom}) of the full action. Since
$<\psi_{\rm qu}(\tau)> = <\bar\psi_{\rm qu}(\tau)> =0$ due to the fact
that the action is even in the number of Grassmann fields, the term
from $S^{int}_{\rm fer}$ which is linear in quantum fields vanishes.
However, we do have to take into account the contribution when we
contract only one pair of fermions, each field from a
different factor $S^{int}_{\rm fer}$ in ${1\over 2} \left(
{1\over\hbar} S^{int}_{\rm fer} \right)^2$.  This yields the
contribution ${1\over 2} (x^i-y^i) (x^j-y^j) \omega_{iab}{}^c
\omega_{jcb} \bar\eta^a \chi^b$. When we were expanding around the
solutions to the equations of motion, this term was part of the
classical action (\puteqn{SFer}). Since we are now expanding around
the constant fields $\bar\eta^a$ and $\chi^a$, this term is not part
of the expansion of the action but instead appears as an additional
quantum correction.

Taking the contributions from the bosonic and ghost sector, together
with the above computed contributions from the fermionic sector, we
arrive at (\puteqn{STraAmp}), plus a remainder $-{1\over 12}
\epsilon\hbar g^{ij} \omega_{ia}{}^b \omega_{jb}{}^a - {1\over 24}
\epsilon\hbar
g_{kl} \Gamma^k{}_i{}^j \Gamma^l{}_j{}^i$. Similarly to
the bosonic case, this term vanishes in normal coordinates, so that
we obtain complete agreement between the Hamiltonian and
path integral approach. In general coordinates, we again expect that a
proper treatment of the Jacobian generated by a change of variables
from normal to general coordinates will lead to cancellation of this
term.

Finally we consider the $N{=}1$ case. Our aim is to reproduce the result
in (\puteqn{FerTrans}) from a path integral with (\puteqn{SferAn}) as
action, and with boundary conditions $\bar\psi^a(0) = \bar\eta^a$ and
$\psi^a(-1) = \chi^a$. As explained in the beginning of this appendix
we should obtain from the fermionic sector the terms in $\tilde S_F$ in
(\puteqn{Soverh}) but also the last term in (\puteqn{FerTrans}). In
this case it is easier to work with the $\psi_1$ propagator given in
(\puteqn{ferGreen}) then with the $\bar\psi$, $\psi$ propagators,
since the fermionic sector of the interactions, given in
(\puteqn{Sintfer1}), only depends on $\psi_1$. We again decompose
$\psi^a_1$ into a sum $\psi^a_{1,0} + \psi^a_{1,{\rm qu}}$ with
$\psi^a_{1,0} = {1\over\sqrt{2}}\bigl( \bar\eta^a + \chi^a \bigr)$ and
$\psi^a_{1,{\rm qu}}$ appearing in the propagators. The first term in
$\tilde S_F$ is of course due to the last term in (\puteqn{SferAn}).
The next set of contributions comes from the expectation value of
$-{1\over\hbar}S^{int}_{\rm fer}$. The classical part is given by
$$\putequation{Sintcl1}$$
where $x^i_0(\tau)$ is a solution of the bosonic part of the action.
Expanding $x^i_0(\tau) = x^i + (x-y)^i \tau + {\cal O} (x-y)^2$, one
recovers the last two terms in (\puteqn{Soverh}). The equal time
contraction of $\psi^a_1 \psi^b_1$ in (\puteqn{Sintfer1}) vanishes by
itself (it anyhow is proportional to $\delta^{ab}$). Also the
contribution from the contraction in $\dot x^i \omega_{iab}$ in
(\puteqn{Sintfer1}) vanishes, as it is proportional to
$\int\limits_{-1}^0 d\tau \, < \dot x^i(\tau) x^j(\tau) >$ which is
zero according to (\puteqn{GreenBos}). We are left with the last term
in (\puteqn{FerTrans}), which should come from the term  ${1\over 2}
\left( {1\over\hbar} S^{int}_{\rm fer} \right)^2$. The classical part
of this term is of course contained in the expansion of $\exp \left( -
{1\over\hbar} \tilde S_F \right)$. The contraction of $\dot
x^i(\sigma)$ with $\dot x^j(\tau)$ is of order $\epsilon$, and would
contribute, but its integral over $\sigma$ and $\tau$ vanishes. One
$\dot x^i(\sigma)$ with $x^j(\tau)$ contraction is also of order
$\epsilon$, but would leave a factor $\dot x^i_0$ which is of order
$\epsilon^{1/2}$, so this term is of higher order. The
contraction of all four fermions is nonvanishing, and indeed
reproduces the term ${1\over 16} (x^i-y^i) (x^j-y^j) \omega_{ia}{}^b
\omega_{jb}{}^a$, in agreement with (\puteqn{FerTrans}).
Finally, we can contract four fermionic fields and two bosonic fields
$\dot x^i$. This yields $-{1\over 24}\epsilon\hbar g^{ij}
\omega_{ia}{}^b \omega_{jb}{}^a$, where, as in the $N{=}2$ case, we take
the same number of bosonic and fermionic modes to regularize the
integrals. Adding all contributions we find the propagator for the
$N{=}1$ case as given in (\puteqn{FerTrans}), plus a remainder which
now equals $- {1\over 24} \epsilon\hbar g^{ij} \omega_{ia}{}^b
\omega_{jb}{}^a- {1\over 24} \epsilon\hbar
g_{kl} \Gamma^k{}_i{}^j \Gamma^l{}_j{}^i$. Again, for general
coordinates we expect this term
to cancel against a Jacobian; in normal coordinates we find complete
agreement with the Hamiltonian result.

\sectionnumstyle{blank}
{\bf\section{References}}
\begin{putreferences}
\reference{AGWi}{L. Alvarez-Gaum\'{e} and E. Witten, Nucl. Phys. {\bf
B234} (1984) 269.}
\reference{Alva}{L. Alvarez-Gaum\'{e}, in `Supersymmetry', eds. K.
Dietz et al., Plenum Press, 1984.}
\reference{Bast}{F. Bastianelli, Nucl. Phys. {\bf B376} (1992) 113.}
\reference{BaNi}{F. Bastianelli and P. van Nieuwenhuizen,
Nucl. Phys. {\bf B389} (1993) 53.}
\reference{textb}{L. Faddeev and A. Slavnov, `Gauge Fields: an
Introduction to Quantum Theory', $2^{\rm nd}$ ed., Addison-Wesley, Redwood
City, 1991; C. Itzykson and J.-B. Zuber, `Quantum Field Theory',
McGraw-Hill, New York, 1980; T.D. Lee, `Particle Physics and
Introduction to Field Theory', Harwood Academic Publishers, Chur,
1981.}
\reference{Trot}{H. Trotter, Proc. Amer. Math. Soc. {\bf 10} (1959)
545.}
\reference{DeWib}{B. DeWitt, `Supermanifolds', $2^{\rm nd}$ ed.,
Cambridge University Press, 1992.}
\reference{Feyn}{R. Feynman, Rev. Mod. Phys. {\bf 20} (1948) 367.}
\reference{DeWi}{B. DeWitt, Rev. Mod. Phys. {\bf 29} (1957) 377.}
\reference{tHoo}{G. 't Hooft and M. Veltman, Ann. Poincar\'{e} {\bf
20} (1974) 69.}
\reference{Dira}{P. Dirac, Phys. Z. Sowj. {\bf 3} (1933) 64.}
\reference{DTNP}{A. Diaz, W. Troost, P. van Nieuwenhuizen and A. van
Proeyen, Int. J. Mod. Phys. {\bf A4} (1989) 3959.}
\reference{Wien}{N. Wiener, J. Math. and Phys. Sci. {\bf 2} (1923) 132.}
\reference{Sken}{K. Skenderis and P. van Nieuwenhuizen, ITP-SB-93-86.}
\reference{Schu}{L. Schulman, `Techniques and Applications of Path
Integration', John Wiley and Sons, New York, 1981.}
\reference{Weyl}{H. Weyl, `Theory of Groups and Quantum Mechanics',
Dover, New York, 1950.}
\reference{Mizr}{M. Mizrahi, J. Math. Phys. {\bf 16} (1975) 2201;
Nuovo Cim. {\bf 61B} (1981) 81.}
\reference{Vlec}{J. Van Vleck, Proc. Natl. Acad. Sci. {\bf 24} (1928) 178.}
\reference{More}{C. Morette, Phys. Rev. {\bf 81} (1951) 848.}
\reference{DeWia}{B. DeWitt, in `Relativity, Groups and Topology',
eds. C. DeWitt and B. DeWitt, Les Houches 1963; Phys. Rep. {\bf 19}
(1975) 295.}
\reference{Nieu}{P. van Nieuwenhuizen, Phys. Rep. {\bf 68} (1981) 189.}
\end{putreferences}

\bye